\pdfoutput=1
\documentclass[aps,pre,twocolumn,amsmath,amssymb,notitlepage,superscriptaddress,longbibliography,nofootinbib,10pt]{revtex4-1}
\usepackage[T1]{fontenc}
\usepackage[left]{lineno}
\usepackage{amsmath}
\usepackage{amssymb}
\usepackage{bm}
\usepackage{graphicx}
\usepackage{comment}
\usepackage{xcolor}
\definecolor{bleuf}{rgb}{0,0,0}
\definecolor{redf}{rgb}{0.61,0.24,0.38}
\usepackage[%
  colorlinks=true,
  urlcolor=redf,
  linkcolor=redf,
  citecolor=redf
]{hyperref}
\usepackage{url}
\usepackage{soul}
\usepackage{upgreek}

\begin{document}
\title{
 Phase coherence and disorder-induced wave propagation in micromotor arrays 
}

\author{Romane Braun}
\affiliation{ENS de Lyon, CNRS, LPENSL, UMR5672, 69342, Lyon cedex 07, France.}
\author{Alexis Poncet}
\affiliation{ENS de Lyon, CNRS, LPENSL, UMR5672, 69342, Lyon cedex 07, France.}
\author{Alexandre Morin}
\affiliation{Huygens-Kamerlingh Onnes Laboratory, Universiteit Leiden,
PO Box 9504, 2300 RA Leiden, the Netherlands}
\author{Denis Bartolo}
\affiliation{ENS de Lyon, CNRS, LPENSL, UMR5672, 69342, Lyon cedex 07, France.}
%
\begin{abstract}
Machines are designed, assembled, and programmed to convert power into predetermined dynamics and functions. 
In contrast, living systems such as interacting cells and animal groups self-organize, synchronize, and perform complex tasks without predefined patterns. 
Inspired by these decentralized architectures, experiments have shown that small assemblies of elastically coupled  self-propelled robots can achieve two fundamental functionalities observed in nature: collective motion and oscillatory deformations ~\cite{Ferrante2013,Woodhouse2018,boudet2021,baconnier2022,Zheng2023,Hernandez2024,xi2024,Xia2024,martinet2025}.
{\color{bleuf}{However, biological inspiration has steered research toward translational self-propulsion, while active rotation remains an underexplored route to designing broader animate materials~\cite{ball2021}.}
Here, we study the self-organization 
of microscopic metamachines~\cite{aubret2021} composed of thousands of 3D-printed rotary motors. 
We first demonstrate and explain how motors precessing in unspecified directions collectively arrange their dynamics into a pristine antiferromagnetic phase.
Next, we elucidate the emergence of spatiotemporal order in the form of phase coherence in the rotors’ precession. 
Finally, we show how quenched disorder initiates the free propagation of phase waves across self-organized regions with mismatched rotation speeds.
Our results suggest that spinner-based metamachines could illuminate metachronal-wave formation in living systems~\cite{gilpin2020multiscale,byron2021metachronal}, and signal propagation in synthetic animate materials~\cite{ball2021,volpe2025roadmap}}. 
\end{abstract}
\maketitle

\noindent From motile cilia to cell tissues and human crowds, coordinated motions emerge across scales when living systems are physically coupled~\cite{gilpin2020multiscale,kruse2011spontaneous,Gu2025}.
In contrast to 
 most man-made machines, these collective dynamics are spontaneous: interacting living units  achieve spatial and temporal organization without relying on programmed instructions.
{\color{bleuf} {Organizing self-propelled robots into mechanical lattices, a recent surge of experiments and theoretical studies have set the stage for artificial machines capable of self-organization~\cite{Ferrante2013,Woodhouse2018,boudet2021,baconnier2022,Zheng2023,Hernandez2024,xi2024,Xia2024}.}
These  active structures feature emergent collective motions reminiscent of living matter, and can be generally defined as metamachines~\cite{aubret2021, volpe2025roadmap}: an artificial assembly of machines that interact to produce emergent behaviors beyond those programmed into any single component.
Until now, metamachines have primarily relied on translational self-propulsion. It is however not the only form of mechanical activity.}
Rotary motors, for example, are among the most fundamental autonomous machines.
Yet, despite progress in active spinning matter~\cite{petroff2015,aubret2018,soni2019,liu2020,Tierno2021,han2021,Massana2021,tan2022,liebchen2022,chen2024,chao2024}, and pioneering studies on non-reciprocally coupled motors~\cite{brandenbourger2019,veenstra2024,veenstra2025}, the collective dynamics of metamachines composed of independently driven  rotary units remain largely unexplored.
To address this gap, we leverage advances in 3D nanoprinting and colloidal motorization to miniaturize, assemble, and power large-scale arrays of rotary motors originally conceived in the late 19th century~\cite{quincke1896}. 
Through a combination of experiments, simulations, and theory, we uncover  the basic physical mechanisms that enable unprogrammed metamachines to self-organize their rotatory dynamics in time and space across system-spanning scales.

\begin{figure*}
\includegraphics[width=\textwidth]{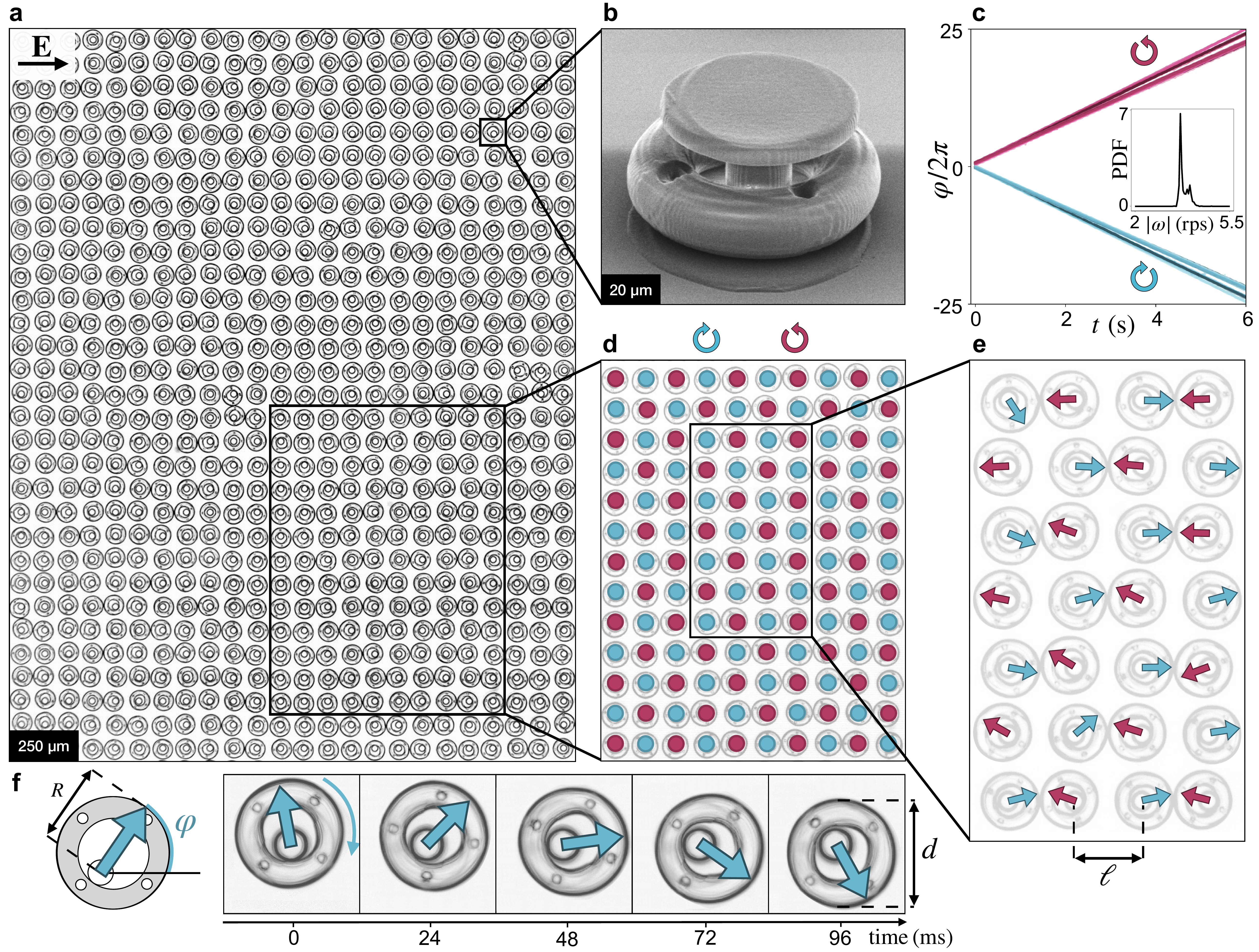}
\caption{{\bf Metamachines assembled from interacting Quincke motors.}
    {\bf a.} Optical micrograph of  3D printed micromotors arranged on a square lattice inside a microfluidic chip. 
    {\color{bleuf}The image shows 50\% of the whole lattice. All motors are identical both in the bulk and at the edges of the lattice.} $\ell=100\,\rm \mu m$, scale bar 250 $\mu$m.
    The direction of the $\mathbf E$ field aligns with one of the principal axes thereby promoting the Quincke rotation of the toroidal rotors ($E=0.65\,\rm V/\mu m$).
    {\bf b.} Scanning Electron Microscope image of an isolated motor. The rotor has a toroidal shape and the stator is attached to the bottom wall of the microchannel.
    {\bf c.} The phase $\varphi$ of a collection of non-interacting motors increases linearly with time as they are  activated by the Quincke mechanism. 
    The two rotation directions are equiprobable and all the rotors spin at a well defined rotation rate, which we define as the maximum of the probability distribution function (PDF) of the rotation speed $|\omega|\equiv|\dot\varphi|$.
    {\bf d.} When the motors are close enough -- as in {\bf a} -- their rotation directions self-organize into a perfect antiferromagnetic order. The color indicates the direction of rotation of the motors.
    {\bf e.} The rotor dynamics features strong phase coherence. Even though the rotation direction alternates, the instantaneous orientations of the rotors are spatially ordered.
    {\bf f.}  Past a critical field amplitude, the rotors are off-centered and undergo a steady orbital (hula-hoop-like) motion. 
    Their instantaneous configurations are determined by the rotation velocity $\omega$ and instantaneous phase $\varphi(t)$.
}
    \label{Fig1}
\end{figure*}
{\bf }
Our experiments consist in 3D printing  square lattices of microscopic motors, and studying their emergent dynamics (Figure~\ref{Fig1}a and Supplementary Movie 1). 
All motors are identical, they consist of a cylindrical stator and of a toroidal rotor made of photoresist resin (see Figure~\ref{Fig1}b and  Supplementary Information). 
{\color{bleuf}{We stress that unlike e.g. in Refs.~\cite{scholz2018,soni2019,Tierno2021,tan2022,ceron2023}, the motors are fixed on a substrate and arranged  on fixed square  lattices}.}
We power the motors using Quincke electrorotation~\cite{quincke1896,Pannacci2007,jakli2008,Bricard2013}.
In short, we place the motor lattices in a microfluidic channel filled with a conducting oil and apply a DC electric field $E$ along one of the principal axes of the lattice to induce rotation around the stator axis, see Figure~ \ref{Fig1}a.
As illustrated in Figure~\ref{Fig1}f and Supplementary Video 1,
past a critical field amplitude, the rotors do not merely spin but display a  hula-hoop motion: they roll and slip on the stator  following orbital trajectories (see SI for a thorough discussion of the single-motor dynamics). 
The instantaneous configuration of the motor located on the $n^{\rm th}$ node of the lattice is  determined both by  its instantaneous orientation $\mathbf u_n= (\cos \varphi_n(t),\sin \varphi_n(t))$ and
angular velocity $\omega_n(t)=\dot{\varphi}_n(t)$.
Crucially, neither the sign of $\omega$ nor the instantaneous phase   $\varphi(t)$ are prescribed by the applied electric field. 
The Quincke mechanism  only guarantees that the  rotors operate at a well defined rotation speed  $|\omega|$ narrowly distributed around a value $\omega_0$ set by the magnitude of the $E$ field (see Figure~\ref{Fig1}c and Supplementary Information).

\noindent{\bf Self-organization of interacting micromotors.}
Our goal is to understand how the speed and phase of the motors self-organize in space and time as they interact.
When the lattice spacing $\ell$ is large, the rotors orbit with random phases in random directions as shown in~\ref{extended_data_1}.
However, Supplementary Videos 1 and 2, and Figure~\ref{Fig1} reveal that when $\ell$ is sufficiently small  order emerges in the metamachine: 
(i) the directions of rotation 
exhibit pristine antiferromagnetic (AF) order -- the signs of the $\omega_n$ are staggered across the whole device (Figure~\ref{Fig1}d);
(ii) the instantaneous phases of the motors vary periodically in space with  (Figures \ref{Fig1}e).
This organization is commonly referred to as phase coherence and leads to spatiotemporal oscillations of the displacement field $\mathbf u$ at frequency $\omega_0$ and wavelength $2\ell$; 
(iii) However, in this rich dynamical steady state, the motors do not rotate in perfect unison, and SI videos 1 and 2 show that phase waves freely propagate over macroscopic scales.

{\color{bleuf}
To explain these  spatiotemporal organizations, we first identify the emergence of AF order as a robust feature of interacting active hysteretic units.
We then single out the microscopic interactions capable of sustaining  phase-coherent dynamics in our metamachine.
We finally show how quenched disorder, and designed speed heterogeneities, power and guide phase waves in collection of interacting motors.}

\begin{figure*}
\includegraphics[width=\textwidth]{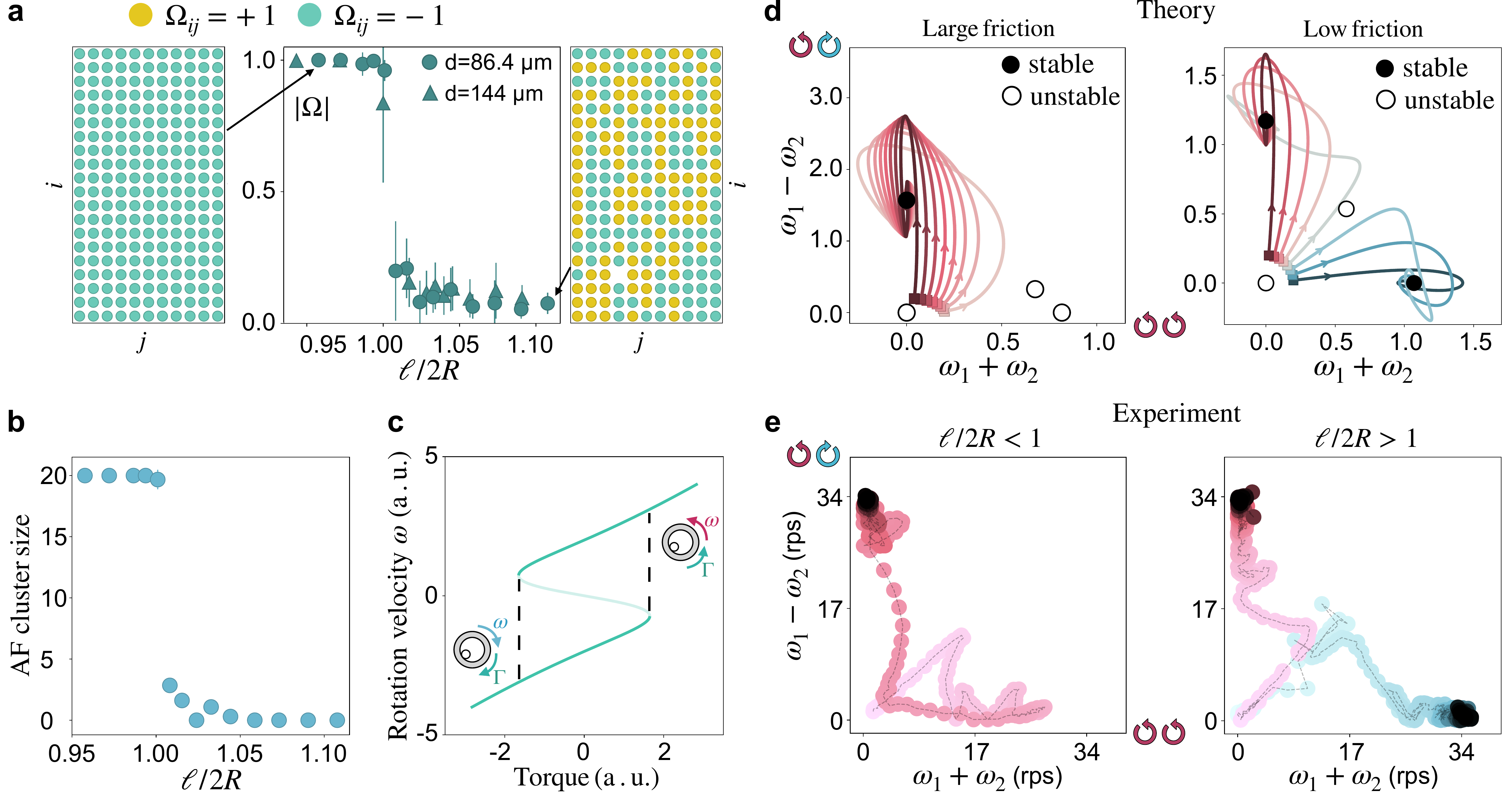}
\caption{{\bf Antiferromagnetic ordering.}
    {\bf a.} (Left and right) Local AF order parameter $\Omega_{ij}$ in the ordered and disordered phases, respectively. (Center) The global AF order parameter $\Omega$ rises sharply as $\ell$ approaches $2R$.
    Triangles: rotors of diameter 144  $\mu$m, $E=0.75\,\rm V/\mu m$. Circles: rotors of diameter 86.4 $\mu$m, $E=1.1\,\rm V/\mu m$. 
    {\bf b.} 
     {\color{bleuf}The AF cluster size  increases sharply from near one to the system size at the onset of the order-to-disorder transition. We define the cluster size as the correlation length associated to the AF order parameter normalized by the lattice spacing.
     %
      {\bf c.} The torque-velocity relationship of a Quincke rotor shows a clear hysteretic behavior (see SI).} 
    {\bf d.} Theory: Phase space trajectories of two Quincke rotors coupled through transverse friction (see SI). $\omega_1(t)$ and $\omega_2(t)$ denote the instantaneous rotation rates of the two rotors.
    For strong coupling, the system exhibits a single stable fixed point corresponding to counter-rotation. 
    Conversely, below a critical coupling strength, a new stable fixed point emerges, allowing both co- and counter-rotation states.
    The dark circles indicate stable fixed points. The open circles correspond to unstable fixed points.
    See SI for numerical details.
    {\bf e.} Experiments: We track the trajectories of isolated pairs of rotors in the $(\omega_1 - \omega_2,\ \omega_1 + \omega_2)$ plane ($E=0.75\,\rm V/\mu m$).
When $\ell/2R \lesssim 1$, all trajectories converge towards a counter-rotation state, even when the rotors initially rotate in the same direction: interactions select exclusively the AF state.
By contrast, when $\ell/2R \gtrsim 1$, the trajectories can converge to any of the four possible final rotation states.
}
    \label{Fig2}
\end{figure*}
{\color{bleuf}
\noindent {\bf  Antiferromagnetic ordering.} We classically quantify the AF order of the motors, with the order parameter $\Omega$ defined as the space and time average of the staggered rotation  $\Omega_{ij}\equiv(-1)^{i+j}{\rm sign}( \omega_{ij})$, where $(i,j)$ are the motor coordinates (Figures~\ref{Fig2}a). 
We repeat two series of experiments using rotors with different radii, and denote by $R$ the radius of the circle that encloses their hula-hoop trajectories (Figure~\ref{Fig1}f).
At large distances ($\ell/2R\gg 1$), we consistently find that  the directions of rotation are randomly distributed.
Upon decreasing $\ell/2R$, small AF domains form thereby leading to a continuous increase of the order parameter $\Omega$ (Figures~\ref{Fig2}a and~\ref{Fig2}b).   
This smooth evolution ends when $\ell/2R$ reaches a value close to unity. 
Below this sharp threshold, the whole metamachine self-organizes into a pristine antiferromagnetic phase where $\Omega=1$ (Figures~\ref{Fig2}a and~\ref{Fig2}b).}

{\color{bleuf}To understand how order builds up, we first note that both viscous  and solid frictions generate torques that favor counter-rotation, and hence AF interactions, between neighboring rotors. 
An analogy with AF Ising spin is therefore tempting. 
However, our motors operate at zero temperature, and within this analogy the metamachine would be expected to order at arbitrarily weak interactions, in contrast with our observations. 
To resolve this apparent paradox, we note that Quincke motors are intrinsically bistable, realizing dynamical analogues of Preisach hysterons~\cite{keim2019}.
To see this, we compute  their torque-velocity relationship in SI and show that it displays a clear hysteresis in Figure~\ref{Fig2}c. 
Unlike isolated Ising spins, which flip upon  application of a vanishingly small field, reversing the direction of our motors requires a finite torque.
This multivalued response already hints to a discontinuous transition towards AF order at finite coupling strength.}

{\color{bleuf} 
To confirm this prediction, we  introduce in SI a minimal model where two Quincke motors are coupled through transverse frictional interactions.
We find that at low friction both the co-rotating and counter-rotating states are stable fixed points of the motors' dynamics (Figure~\ref{Fig2}d).
It is only past  a critical coupling  that friction destabilizes the corotating states and uniquely selects AF configurations.
These predictions are further confirmed by the phase-space trajectories of isolated pairs of 3D-printed motors (Figure~\ref{Fig2}e and SI). 
At large distances (low friction), the motors converge to either co-rotating or counter-rotating states.
Conversely, at short distances, while initial conditions can transiently guide the Quincke motors toward co-rotation, this state is unstable and always relaxes to a permanent antiferromagnetic configuration.

Beyond the specifics of our experiments, our results reveal that, even in the absence of thermal fluctuations, bistable active units require strong interactions to self-organize their dynamics at all scales. }\\

\begin{figure*}
\includegraphics[width=\textwidth]{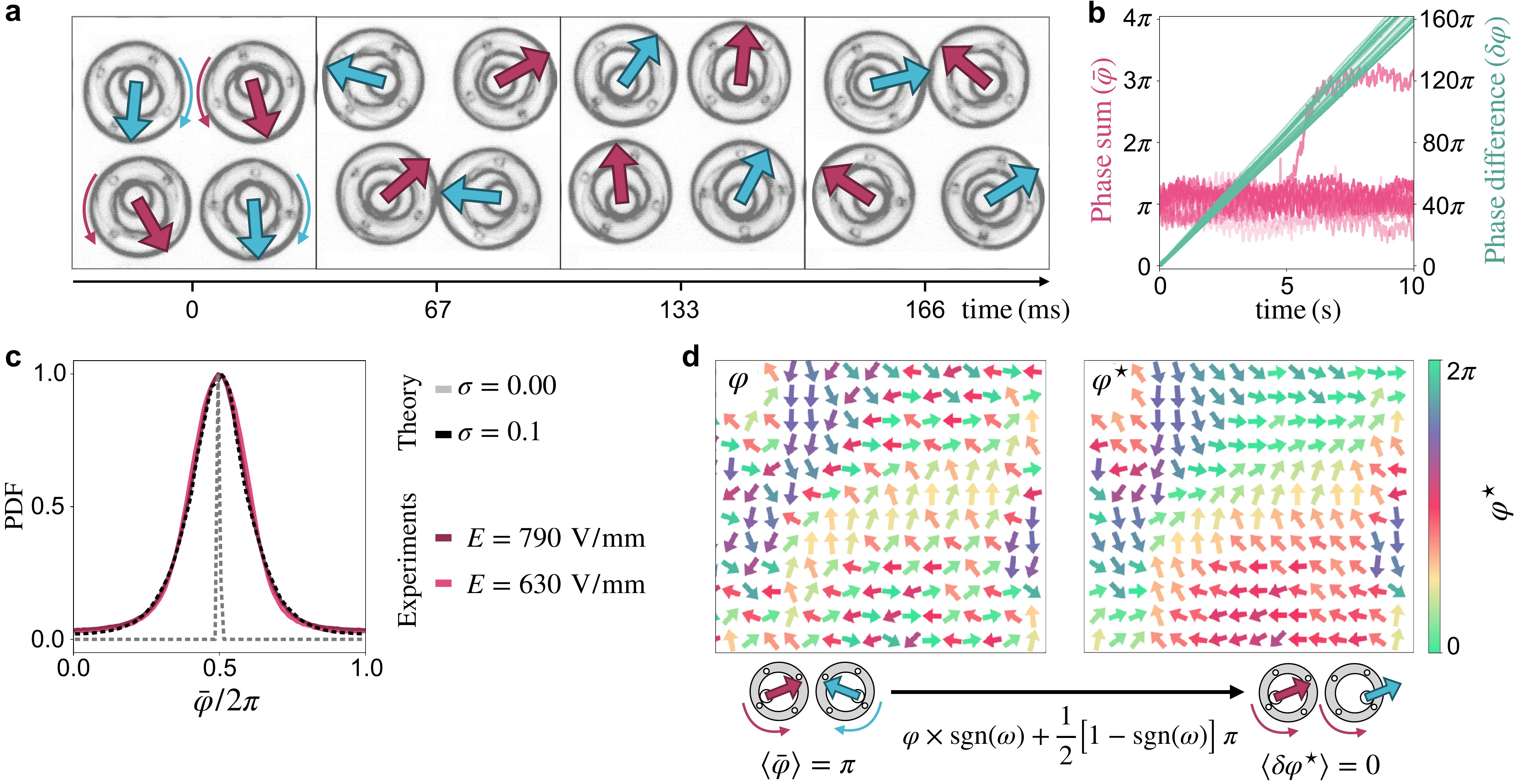}
\caption{{\bf Phase coherence}
    {\bf a.} Four subsequent snapshots of four rotors located in the bulk of a metamachine with $d=86.4\,\rm \mu m$, $\ell=99.6\,\rm \mu m$, and $E=0.628\,\rm V\mu m^{-1}$. Adjacent rotors rotate in opposite directions. However, the phase sums $\bar \varphi$ defined on the edges of the lattice are constant and equal to $\pi$. As a result, the rotors are synchronized along the $(1,1)$ and $(1,-1)$ directions.
  {\bf b.} Time evolution of the phase differences, $\delta \varphi(t)$, and phase sums, $\bar \varphi(t)$, on the four edges connecting the motors shown in {\bf a}. 
  {\color{bleuf}
  The phase difference  increases linearly in time. Conversely, the phase sum $\bar \varphi(t)$ oscillates at $|\omega|$ around a constant values equal to $\pi$ for all rotor pairs, in agreement with our theoretical prediction (see SI).}
    {\bf c.} 
    {\color{bleuf} Probability Distribution Function (PDF) of the phase sum over the whole lattice for two different electric field amplitudes (experiments, solid line) and two values of sigma (simulations, dashed lines).
    }
    {\color{bleuf}{\bf d.} In the ordered metamachine the bare phase $\varphi_n$ of the motors varies strongly over one lattice spacing.  We perform the gauge transformation sketched in the lower panel to define an equivalent  smooth phase field $\varphi^\star$.}}
    \label{Fig3}
\end{figure*}
%



\noindent {\bf Phase-coherent metamachines.}
We now focus on the antiferromagnetic phase and quantify the degree of phase coherence among the motors.
Figure~\ref{Fig3}a shows four subsequent snapshots of four rotors located in the bulk of the metamachine.
As expected from their opposite rotation directions, the phase difference $\delta\varphi$ between neighboring rotors increases linearly with time (Figure~\ref{Fig3}b).
However, the sum of their instantaneous phases, $\bar\varphi_{\langle n,m\rangle} = \varphi_n(t) + \varphi_m(t)$, remains nearly constant on each edge $\langle n,m\rangle$ of the square lattice (Figure~\ref{Fig3}b).
This locking of phase sums is a defining signature of phase coherence for coupled oscillators with opposite rotation frequencies~\cite{Acebron2005,box2015}.
Here, the motors spontaneously organize their dynamics to maintain  coherence across the entire system: the distribution of $\bar\varphi_{\langle n,m\rangle}$ peaks sharply around $\pi$, uniformly in both principal directions.
Varying the magnitude of the $E$ field, the value of  $\bar \varphi$ remains unchanged (Figure~\ref{Fig3}c).
{\color{bleuf}This self-organization results in a counterintuitive dynamics in which the  motor kinematics are synchronized and phase locked along the diagonal of the square lattice, but markedly different   along the $x$ direction, where rotor pairs come into contact, and the $y$ direction where they never touch (Figure~\ref{Fig3}a).}

To elucidate the mechanisms underlying this  spatiotemporal order, we introduce a minimal model capturing the phase dynamics.
In the overdamped limit, and assuming that the motors interact only through pairwise interactions, the phase $\varphi_n$ of rotor $n$ evolves as:
\begin{equation}
\partial_t \varphi_n(t) = \tau_n + \sum_m T_{nm}(\varphi_n, \varphi_m),
\label{eq:equationofmotion}
\end{equation}
where $\tau_n$ denotes the active torque, and $T_{nm}$  the interaction torque exerted by rotor $m$ on rotor $n$.

Guided by our measurements, we simplify the model by assuming constant driving torques of equal magnitudes and alternating signs ($|\tau_n| = \tau$).
In practice,  Quincke motors are coupled via a number of interactions: contact forces, viscous hydrodynamic flows,  electrostatic repulsion and dipolar forces. 
{\color{bleuf}Because contact and hydrodynamic interactions are isotropic (when $\tau$ is constant), they cannot account on their own for the anisotropic kinematics seen in Figure~\ref{Fig3}a.
We  elaborate on this argument in detail in the SI, where we  develop a minimal theory, inspired by models of motile cilia. It allows us to independently estimate the contributions of the various interactions on phase coherence~\cite{Vilfan2006,guirao2007,Goldstein2009,Uchida2010,bruot2016,meng2021}.
Unlike previous findings on driven colloids~\cite{bruot2016}, we confirm that repulsion forces as well as far field hydrodynamics are insufficient to explain the  self-organization  of  the motors, even when including rotation-speed modulations.  
Near-field hydrodynamic flows are more challenging to model but, in their simplest forms, cannot explain our findings either. 
Ultimately, we find that only the action of dipolar electrostatic interactions between the Quincke motors can yield a uniform phase coherence state where $\bar \varphi = \pi$.

Given this analysis we can now forge more intuition on the motor dynamics. 
The electric dipoles that power  Quincke rotation are oriented predominantly opposite to $\bf E$. 
As a result, the dipolar forces that couple the rotors have opposite signs along the $x$ and $y$ directions. 
This anisotropy rationalizes why the rotors organize their phases so as to come in contact  along the $x$ direction and remain at a distance along the $y$ direction. 
}\\

\noindent {\bf Disorder-induced wave propagation.} 
It is tempting to test  the predictive power of our theory beyond the average phase coherence.  
Unlike in our experiments, however,  the simulated dynamics of motor lattice always converges towards a quiet steady state and fails to capture the most striking feature of our experiments:  the spontaneous emission and propagation of phase waves through the metamachines (see  Supplementary Videos 2
and 3). 
We show in the following that disorder is the last crucial ingredient required to explain the emergence and propagation of phase waves

%
\begin{figure*}
\includegraphics[width=\textwidth]{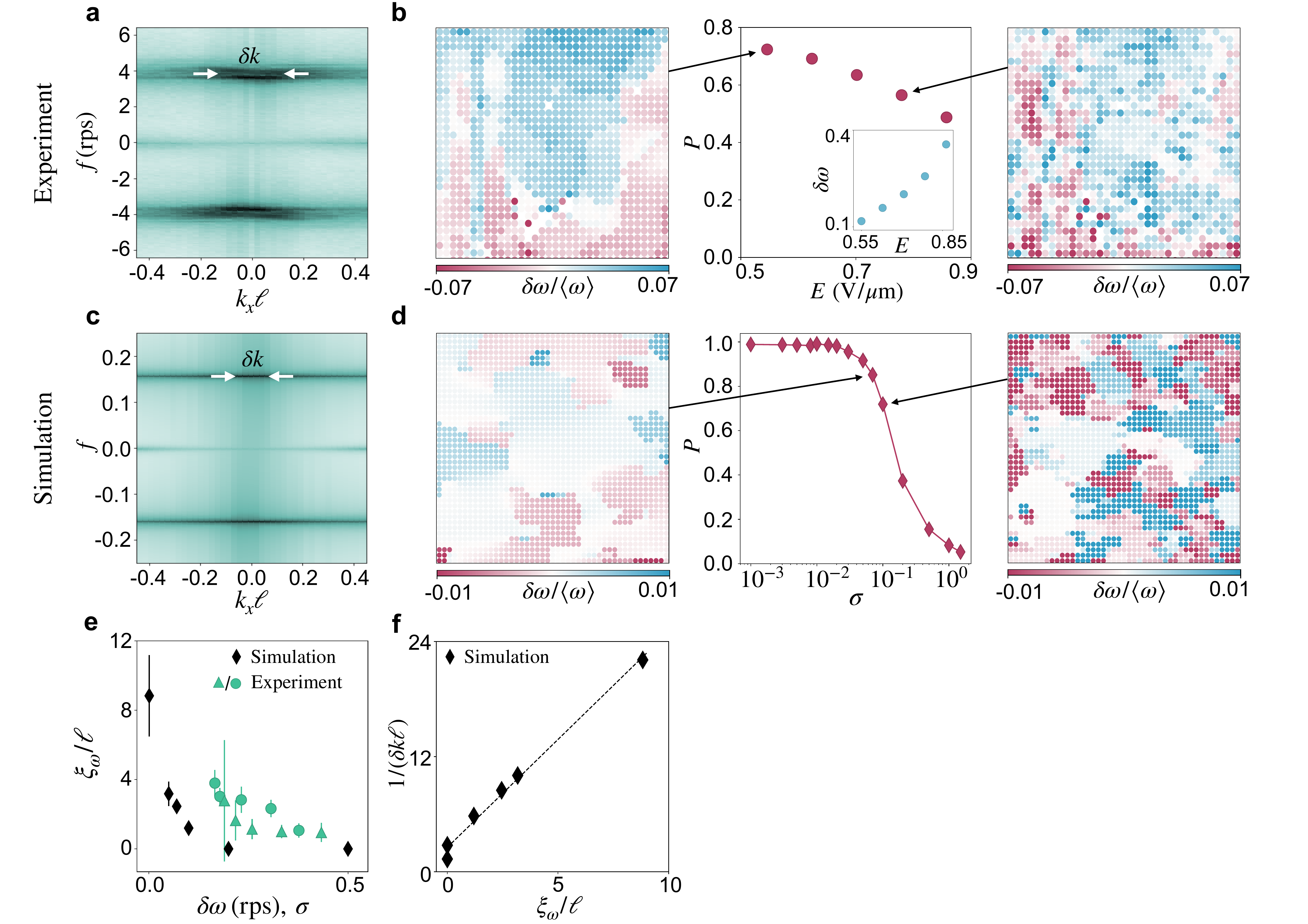}
\caption{{\bf Disorder-induced phase waves.}
  {\color{bleuf} We focus on a  $36\times 33$ region in the bulk of the metamachine composed in total of $48\times44$ motors.}
    {\bf a.} We Fourier transform the displacement field in space and time in the bulk,  and plot its corresponding power spectrum $\langle |\mathbf u_n^\star (k_x,k_y=0,f)|^2\rangle$ for wave vectors $(k_x,0)$.
    Two flat bands at frequencies $\pm \omega_0$ dominate the spectrum. We note $\delta k$ their spectral width.
    {\bf b.} Maps of the motor rotation speed at high and low $E$ values, and variations of the synchronization order parameter $P$ with E. Inset: The standard deviation of the rotation frequency increases with $E$.
    $d=86.4\,\rm \mu m$, $\ell=10099.6 \,\rm \mu m$.
    {\color{bleuf}The white spots corresponds to defects in the print process: rotors detached or broken.}
    {\bf c.} {\color{bleuf} Same plot as in {\bf a} for our numerical simulations. As observed experimentally, the power spectrum features two flat bands, whose width $\delta k$ does not depend on the boundary conditions (see SI).}
    {\bf d.} {\color{bleuf}Same plots as in {\bf b} for our numerical simulations.} The oscillators undergo a synchronization transition as the level of disorder decreases. 
    {\bf e.} Correlation length associated to the rotation speed field plotted as a function of $\delta \omega$ (experiments) and $\sigma$ (simulations).
     {\bf f.} The inverse of the spectral width $\delta k$, i.e. the correlation length of the phase fluctuations, and the correlation length of the $|\omega_n|$ are proportional.
}
    \label{Fig4}
\end{figure*}
To build our reasoning, we first need to evidence and quantify the phase-wave dynamics.
The task is not simple due to the fast spatial variations of the rotor orientations, Figure~\ref{Fig3}a. 
To investigate their long-wavelength dynamics, we therefore perform a gauge transformation inspired by antiferromagnetism, and illustrated in Figure~\ref{Fig3}d and Supplementary Video 4. 
As adjacent  rotors orbit in opposite directions, we redefine the sign of the phase according to the active torque: $\varphi_n\to  {\rm sign}(\omega_n)\varphi_n$. 
We then use the average phase coherence to define a slowly varying field as $\varphi^\star_n\equiv{\rm sign}(\omega_{n})\varphi_{n} + \frac 1 2(1-{\rm sign}(\omega_{n}))\pi$ (Figure~\ref{Fig3}d). 
Within this representation, an ideal  phase coherent state where $\bar \varphi=\pi$ reduces to a classical synchronized state where  the phases $\varphi^\star_n$ of all the motors are equal.  

We are now equipped to quantify the fluctuations in the motor dynamics. 
We compute the power spectrum of $\mathbf u_n^\star=(\cos \varphi_n^\star,\sin\varphi_n^\star)$ shown in Figure~\ref{Fig4}a. 
We find that 
it is peaked on two flat bands at frequencies $\pm \omega_0$. 
The power distribution within the two bands indicates that multiple waves can indeed propagate through our collection of coupled rotors.
They are associated with a distribution of wave vectors of width $\delta k$ (Figure~\ref{Fig4}a), which provides a direct measurement of the  correlation length of the displacement field $\mathbf u_n^\star$.
{\color{bleuf}This situation contrasts with experiments and simulations on beating cilia and colloidal oscillators, where hydrodynamic interactions produce constant phase shifts between neighboring elements, giving rise to so-called metachronal waves associated to a single wave vector~\cite{brumley2015,bruot2016,gilpin2020multiscale,byron2021metachronal,meng2021,Metti2026}}.

{\color{bleuf}
To explain this correlated dynamic we  make a simple yet crucial observation. 
Even when coupled, the motors do not  all rotate at the exact same rate. 
As we increase the strength of the $E$ field, we find  that the spreading of the frequencies $\delta \omega$ increases (Figure~\ref{Fig4}b inset). 
This spreading reflects the shrinkage of the compact regions where the rotation frequency is locally homogeneous, and  alters the level of phase coherence, which we quantify 
with the  classical order parameter
$P=\langle \exp[i(\varphi^\star_n(t)-\varphi^\star_m(t))]\rangle_{\langle n,m\rangle,t}$ (Figures~\ref{Fig4}b and~\ref{Fig4}e).}

{\color{bleuf}Guided by our observations, we refine our model. 
We focus exclusively on dipolar interactions  and
 classically add disorder to the driving torques (Eq.~\ref{eq:equationofmotion}).
 The $\tau_n$ are now uniformly distributed random variables of width~$\sigma$.
Our simulations are consistent with our experiments.
They account for the broadening of the distribution of the phase differences (Figure~\ref{Fig3}c), and reveal a synchronization physics akin to short range Kuramoto models (Figure~\ref{Fig4}d).  
When oscillators are coupled through short-range interactions, synchronization proceeds continuously and extends over finite domains whose size decreases as the distribution of natural frequencies broadens (Figure~\ref{Fig4}e)~\cite{sakaguchi1987local,Acebron2005}}.

In addition to explaining the motor synchronization,
our simulations  explain how  disorder  induces the propagation of phase waves sharing the same flat-band spectrum as in our experiments (Figure~\ref{Fig4}c and Supplementary Video 5).  
Plotting the extent $\xi_{\omega}$ of the {\color{bleuf} synchronized} domains versus the inverse of the spectral width $1/\delta k$ in Figure~\ref{Fig4}f, we find that they are linearly related. 
This simple observation implies that the phase waves freely propagate within the domains. 
Randomness in natural rotation speed is not necessary to yield wave propagation, what  matters is spatial heterogeneities in the rotation velocities.

\begin{figure*}
\includegraphics[width=\textwidth]{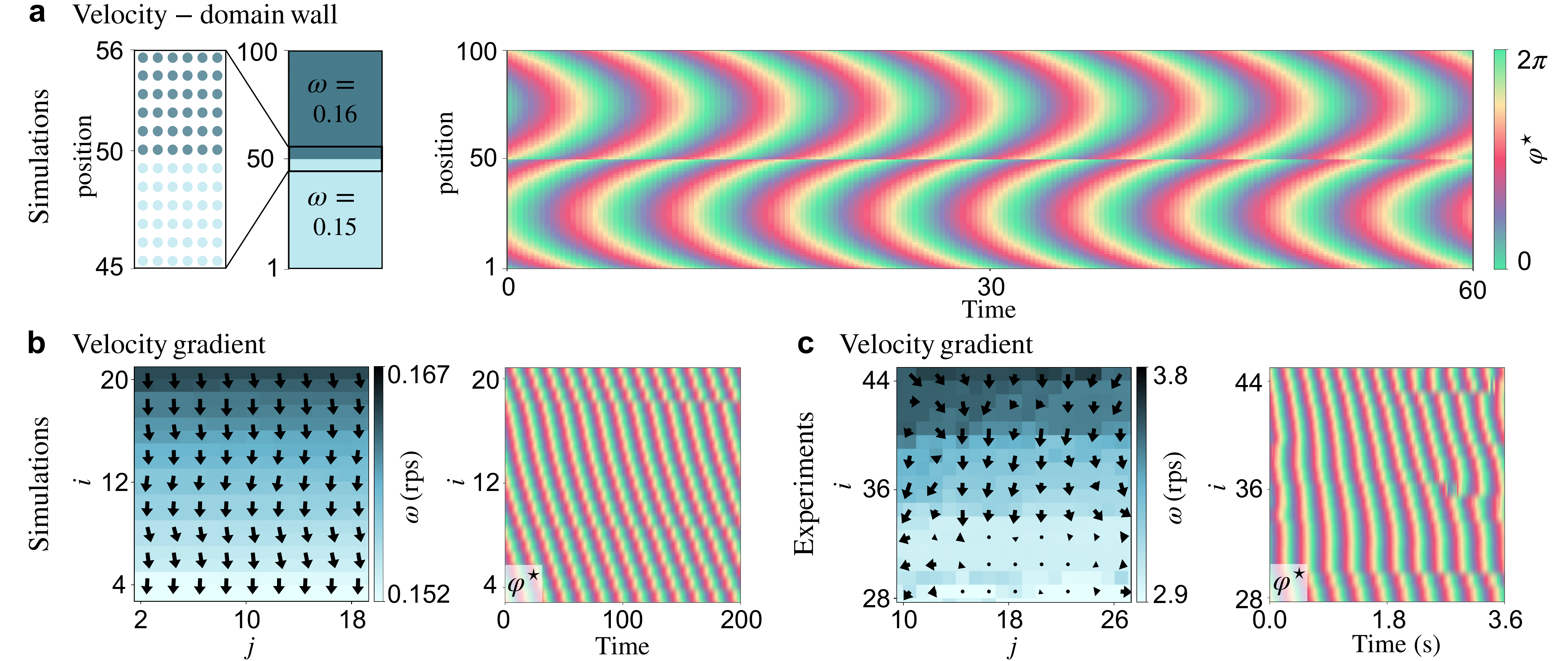}
\caption{
{\color{bleuf}{\bf Control of the phase waves.}
{\bf a.}
Numerical simulations. 
(left) Sketch of the geometry. We prepare a heterogeneous system with periodic boundary conditions along the vertical direction. It is composed of two domains where the natural frequencies are different. 
(right) Spatiotemporal plot of the modified phase field ($\varphi^\star$).
In steady state, phase waves emanate from the domain wall in the form of parabolic phase profiles. 
Simulation parameters in SI.
{\bf b.}
(left) Simulation of a square lattice of motors driven by heterogeneous torques. We impose a vertical gradient of natural frequencies. The arrows indicate the gradient of the resulting phase field $\bm\nabla \varphi^\star$. The phase gradient points in the direction opposite to the speed gradient.
(right) The spatiotemporal plot of the average phase $\langle\varphi^\star(i,j)\rangle_j$ clearly shows the control of the phase-wave  in the $i$ direction.
{\bf c.}  Experiments. 
When a speed gradient of the motors points in the $i$ direction the measured phase gradient $\bm\nabla \varphi^\star$ points on average in the $-i$ direction. 
As a consequence, we  control  the propagation of the phase-wave  in the $i$ direction as clearly seen in the spatiotemporal plot of  $\langle\varphi^\star(i,j)\rangle_j$. 
}}
\label{Fig5}
\end{figure*}

{\color{bleuf}To explain this counterintuitive effect where spatial heterogeneities are required to observe wave propagation,} we  consider an even simpler situation:  domain walls separating two regions where the driving torques differ by $\delta\tau$. 
{\color{bleuf}Figure~\ref{Fig5}a shows that phase waves emanate continuously from the  walls and propagate along their normal directions (see also Supplementary Video~6).}
To gain a more quantitative understanding, we  solve analytically the long-wavelength dynamics of $\varphi^\star$ in SI. 
We show that the phase gradient obeys a Laplace equation. 
Therefore,  perturbations created at a domain wall are transmitted throughout the system in the form of parabolic phase patterns, in agreement with our numerical observations (see Figure~\ref{Fig5}a).

\noindent{\bf Controlling phase waves in metamachines.} 
{\color{bleuf}
Our results point to a simple design principle for motor lattices that propagate waves along prescribed directions. As shown by the simulations in Figure~\ref{Fig5}b, we can control the propagation direction through smooth spatial variations of the motor speed: the speed gradients set the phase gradients, which in turn determines the direction of metachronal-wave propagation.

We confirm this design principle experimentally in Figure~\ref{Fig5}c, where a net linear speed gradient dominates over the uncontrolled disorder discussed above, and therefore controls the unidirectional propagation of  phase waves in the metamachine (see SI). 
Beyond this proof of concept, we can envision controlling the direction of propagation of phase waves  in space and time  by tuning locally the strength of the $\bf E$ field that powers the micromotors. In practice, this could be achieved through local and dynamical  modulations of the channel height, as explained in SI.\\} 

\noindent {\bf Conclusion and outlook.}
{\color{bleuf}
Our study demonstrates that lattices of interacting micromotors self-organize into a dynamical state combining antiferromagnetic order, robust phase coherence, and disorder-induced wave propagation.
From a fundamental perspective, our results show that disorder does not merely disrupt order but instead triggers propagating waves in ensembles of partially synchronized oscillators.
Such disorder-induced waves should therefore be expected to play a prominent role in a broad class of biological oscillators, from ciliated cells to cell tissues that are inevitably heterogeneous~\cite{fradique2023assessing,gilpin2020multiscale,brumley2015}.
Beyond the specifics of metachronal waves, we envision  micromotor lattices   as a powerful platform to investigate the transport of mechanical signals through geometrical design in active media and decentralized metamachines~\cite{nash2015topological,juraschek2025chiral,veenstra2025wave,bacconier2025}.}

\section*{Acknowledgments} We thank R. diLeonardo and G. Vizsnyiczai for help with preliminary experiments and P. Baconnier for useful discussions on coupled hysterons physics. 
 This project  received funding from the European Research Council (ERC) under the European Union’s Horizon 2020 research and innovation program (grant agreement No. [101019141]) (DB).

\section*{Author Contributions}
D. B. designed the project. R. B. performed the experiments. 
R. B., A. 
P. and A. M. performed the numerical simulations. 
R. B., A. P. and D. B. worked out the theory. All authors discussed the results and wrote the manuscript. 
Correspondence to \href{mailto:denis.bartolo@ens-lyon.fr}{Denis Bartolo}.

\section*{Competing Interest}
The Authors declare no competing interests.

\renewcommand{\thefigure}{Extended data Figure 1}
\begin{figure*}
\includegraphics[width=\textwidth]{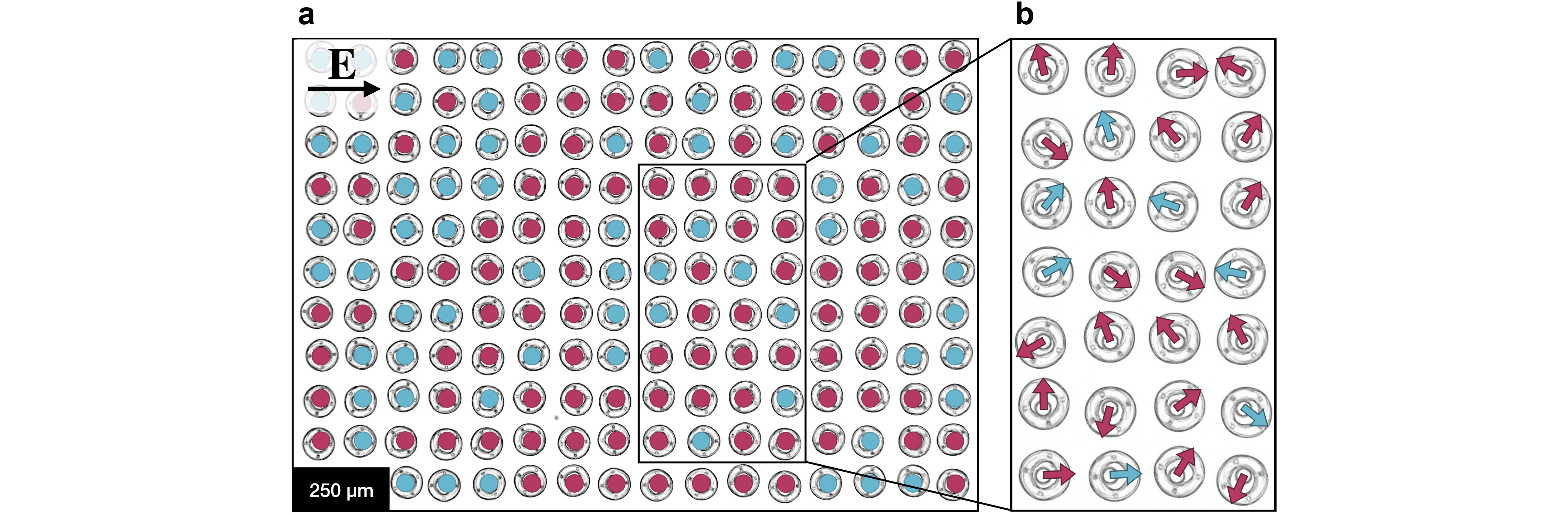}
\caption{{\bf Disorder in weakly interacting micromotors.}
When the lattice spacing of the metamachine is large, 
the motor rotations ({\bf a.}) and phases ({\bf b.}) are uncorrelated. Here $\ell=114$~$\mu$m, $E=1.1$~$\rm V\mu m^{-1}$, and $d=86$~$\mu m$.}
    \label{extended_data_1}
\end{figure*}

\bibliography{biblio}

@article{soni2019,
  title={The odd free surface flows of a colloidal chiral fluid},
  author={Soni, Vishal and Bililign, Ephraim S and Magkiriadou, Sofia and Sacanna, Stefano and Bartolo, Denis and Shelley, Michael J and Irvine, William TM},
  journal={Nature physics},
  volume={15},
  number={11},
  pages={1188--1194},
  year={2019},
  publisher={Nature Publishing Group UK London}
}

@article{ceron2023,
  title={Programmable self-organization of heterogeneous microrobot collectives},
  author={Ceron, Steven and Gardi, Gaurav and Petersen, Kirstin and Sitti, Metin},
  journal={Proceedings of the National Academy of Sciences},
  volume={120},
  number={24},
  pages={e2221913120},
  year={2023},
  publisher={National Academy of Sciences},
url={"https://www.pnas.org/doi/10.1073/pnas.2221913120"}
}

@article{chao2024,
  title={Selective excitation of work-generating cycles in nonreciprocal living solids},
  author={Chao, Yu-Chen and Gokhale, Shreyas and Lin, Lisa and Hastewell, Alasdair and Bacanu, Alexandru and Chen, Yuchao and Li, Junang and Liu, Jinghui and Lee, Hyunseok and Dunkel, Jorn and others},
  journal={arXiv preprint arXiv:2410.18017},
  year={2024},
url={"https://arxiv.org/abs/2410.18017"}
}

@book{kim2013microhydrodynamics,
  title={Microhydrodynamics: principles and selected applications},
  author={Kim, Sangtae and Karrila, Seppo J},
  year={2013},
  publisher={Butterworth-Heinemann}
}

@article{tan2022,
  title={Odd dynamics of living chiral crystals},
  author={Tan, Tzer Han and Mietke, Alexander and Li, Junang and Chen, Yuchao and Higinbotham, Hugh and Foster, Peter J and Gokhale, Shreyas and Dunkel, J{\"o}rn and Fakhri, Nikta},
  journal={Nature},
  volume={607},
  number={7918},
  pages={287--293},
  year={2022},
  publisher={Nature Publishing Group UK London},
url={"https://www.nature.com/articles/s41586-022-04889-6"}
}

@article{Zhang2022,
  title={Polar state reversal in active fluids},
  author={Zhang, Bo and Yuan, Hang and Sokolov, Andrey and de la Cruz, Monica Olvera and Snezhko, Alexey},
  journal={Nature Physics},
  volume={18},
  number={2},
  pages={154--159},
  year={2022},
  publisher={Nature Publishing Group UK London}
}

@article{Melcher1969,
  title={Electrohydrodynamics: a review of the role of interfacial shear stresses},
  author={Melcher, JR and Taylor, GI},
  journal={Annual review of fluid mechanics},
  volume={1},
  number={1},
  pages={111--146},
  year={1969},
  publisher={Annual Reviews 4139 El Camino Way, PO Box 10139, Palo Alto, CA 94303-0139, USA}
}

@Article{Kayaku_SU8,
  author    = {K. A. Materials},
  title     = {SU-8 Permanent Negative Epoxy Photoresist},
  year      = {2020},
}

@Article{Kayaku_omnicoat,
  author    = {K. A. Materials},
  title     = {Omnicoat},
  year      = {2020},
}

@article{quincke1896,
  title={Ueber rotationen im constanten electrischen felde},
  author={Quincke, G},
  journal={Annalen der Physik},
  volume={295},
  number={11},
  pages={417--486},
  year={1896},
  publisher={Wiley Online Library}
}

@Article{Bricard2013,
  author    = {Antoine Bricard and Jean-Baptiste Caussin and Nicolas Desreumaux and Olivier Dauchot and Denis Bartolo},
  journal   = {Nature},
  title     = {Emergence of macroscopic directed motion in populations of motile colloids},
  year      = {2013},
  month     = {nov},
  number    = {7474},
  pages     = {95--98},
  volume    = {503},
  doi       = {10.1038/nature12673},
  publisher = {Springer Science and Business Media {LLC}},
  url       = {https://doi.org/10.1038%2Fnature12673},
}

@article{Vilfan2006,
  title = {Hydrodynamic Flow Patterns and Synchronization of Beating Cilia},
  author = {Vilfan, Andrej and J\"ulicher, Frank},
  journal = {Phys. Rev. Lett.},
  volume = {96},
  issue = {5},
  pages = {058102},
  numpages = {4},
  year = {2006},
  month = {Feb},
  publisher = {American Physical Society},
  doi = {10.1103/PhysRevLett.96.058102},
  url = {https://link.aps.org/doi/10.1103/PhysRevLett.96.058102}
}

@article{Goldstein2009,
  title = {Noise and Synchronization in Pairs of Beating Eukaryotic Flagella},
  author = {Goldstein, Raymond E. and Polin, Marco and Tuval, Idan},
  journal = {Phys. Rev. Lett.},
  volume = {103},
  issue = {16},
  pages = {168103},
  numpages = {4},
  year = {2009},
  month = {Oct},
  publisher = {American Physical Society},
  doi = {10.1103/PhysRevLett.103.168103},
  url = {https://link.aps.org/doi/10.1103/PhysRevLett.103.168103}
}

@article{guirao2007,
  title={Spontaneous creation of macroscopic flow and metachronal waves in an array of cilia},
  author={Guirao, Boris and Joanny, Jean-Fran{\c{c}}ois},
  journal={Biophysical journal},
  volume={92},
  number={6},
  pages={1900--1917},
  year={2007},
  publisher={Elsevier}
}

@article{Uchida2010,
  title = {Synchronization and Collective Dynamics in a Carpet of Microfluidic Rotors},
  author = {Uchida, Nariya and Golestanian, Ramin},
  journal = {Phys. Rev. Lett.},
  volume = {104},
  issue = {17},
  pages = {178103},
  numpages = {4},
  year = {2010},
  month = {Apr},
  publisher = {American Physical Society},
  doi = {10.1103/PhysRevLett.104.178103},
  url = {https://link.aps.org/doi/10.1103/PhysRevLett.104.178103}
}

@article{pradillo2019quincke,
  title={Quincke rotor dynamics in confinement: rolling and hovering},
  author={Pradillo, Gerardo E and Karani, Hamid and Vlahovska, Petia M},
  journal={Soft matter},
  volume={15},
  number={32},
  pages={6564--6570},
  year={2019},
  publisher={Royal Society of Chemistry}
}

@article{bruot2016,
  title={Realizing the physics of motile cilia synchronization with driven colloids},
  author={Bruot, Nicolas and Cicuta, Pietro},
  journal={Annual Review of Condensed Matter Physics},
  volume={7},
  number={1},
  pages={323--348},
  year={2016},
  publisher={Annual Reviews}
}

@article{meng2021,
  title={Conditions for metachronal coordination in arrays of model cilia},
  author={Meng, Fanlong and Bennett, Rachel R and Uchida, Nariya and Golestanian, Ramin},
  journal={Proceedings of the National Academy of Sciences},
  volume={118},
  number={32},
  pages={e2102828118},
  year={2021},
  publisher={National Acad Sciences}
}

@article{box2015,
  title={Transitional behavior in hydrodynamically coupled oscillators},
  author={Box, S and Debono, L and Phillips, DB and Simpson, SH},
  journal={Physical Review E},
  volume={91},
  number={2},
  pages={022916},
  year={2015},
  publisher={APS}
}

@article{kotar2013,
  title={Optimal hydrodynamic synchronization of colloidal rotors},
  author={Kotar, Jurij and Debono, Luke and Bruot, Nicolas and Box, Stuart and Phillips, David and Simpson, Stephen and Hanna, Simon and Cicuta, Pietro},
  journal={Physical review letters},
  volume={111},
  number={22},
  pages={228103},
  year={2013},
  publisher={APS}
}

@article{Peters2005,
	title = {Experimental Observation of {{Lorenz}} Chaos in the {{Quincke}} Rotor Dynamics},
	author = {Peters, Fran{\c c}ois and Lobry, Laurent and Lemaire, Elisabeth},
	year = {2005},
	month = mar,
	journal = {Chaos: An Interdisciplinary Journal of Nonlinear Science},
	volume = {15},
	number = {1},
	pages = {013102},
	issn = {1054-1500, 1089-7682},
	doi = {10.1063/1.1827411},
	urldate = {2024-03-27},
	langid = {english}
}

@book{Smale2003,
	title = {Differential Equations, Dynamical Systems, and an Introduction to Chaos},
	author = {Smale, Stephen and Hirsch, Morris W and Devaney, Robert L},
	year = {2003},
	publisher = {{Elsevier Science}}
}

@article{hsu2005charge,
  title={Charge stabilization in nonpolar solvents},
  author={Hsu, Ming F and Dufresne, Eric R and Weitz, David A},
  journal={Langmuir},
  volume={21},
  number={11},
  pages={4881--4887},
  year={2005},
  publisher={ACS Publications}
}

@article{smith2015celebrating,
  title={Celebrating Soft Matter's 10th Anniversary: Influencing the charge of poly (methyl methacrylate) latexes in nonpolar solvents},
  author={Smith, Gregory N and Hallett, James E and Eastoe, Julian},
  journal={Soft Matter},
  volume={11},
  number={41},
  pages={8029--8041},
  year={2015},
  publisher={Royal Society of Chemistry}
}

@article{Uchida2011,
  title = {Generic Conditions for Hydrodynamic Synchronization},
  author = {Uchida, Nariya and Golestanian, Ramin},
  journal = {Phys. Rev. Lett.},
  volume = {106},
  issue = {5},
  pages = {058104},
  numpages = {4},
  year = {2011},
  month = {Feb},
  publisher = {American Physical Society},
  doi = {10.1103/PhysRevLett.106.058104},
  url = {https://link.aps.org/doi/10.1103/PhysRevLett.106.058104}
}

@article{brumley2015,
  title={Metachronal waves in the flagellar beating of Volvox and their hydrodynamic origin},
  author={Brumley, Douglas R and Polin, Marco and Pedley, Timothy J and Goldstein, Raymond E},
  journal={Journal of the Royal Society Interface},
  volume={12},
  number={108},
  pages={20141358},
  year={2015},
  publisher={The Royal Society}
}

@article{Acebron2005,
  title = {The Kuramoto model: A simple paradigm for synchronization phenomena},
  author = {Acebr\'on, Juan A. and Bonilla, L. L. and P\'erez Vicente, Conrad J. and Ritort, F\'elix and Spigler, Renato},
  journal = {Rev. Mod. Phys.},
  volume = {77},
  issue = {1},
  pages = {137--185},
  numpages = {0},
  year = {2005},
  month = {Apr},
  publisher = {American Physical Society},
  doi = {10.1103/RevModPhys.77.137},
  url = {https://link.aps.org/doi/10.1103/RevModPhys.77.137}
}

@inproceedings{kuramoto1975,
  title={Self-entrainment of a population of coupled non-linear oscillators},
  author={Kuramoto, Yoshiki},
  booktitle={International symposium on mathematical problems in theoretical physics: January 23--29, 1975, kyoto university, kyoto/Japan},
  pages={420--422},
  year={1975},
  organization={Springer}
}

@book{kuramoto1984,
  title={Chemical turbulence},
  author={Kuramoto, Yoshiki and Kuramoto, Yoshiki},
  year={1984},
  publisher={Springer}
}

@article{Metti2026,
	Author = {Liu, Zemin and Wang, Che and Ren, Ziyu and Wang, Chunxiang and Wang, Wenkang and Ko, Jongkuk and Song, Shanyuan and Hong, Chong and Chen, Xi and Wang, Hongguang and Hu, Wenqi and Sitti, Metin},
	Da = {2026/01/14},
	Doi = {10.1038/s41586-025-09944-6},
	Id = {Liu2026},
	Isbn = {1476-4687},
	Journal = {Nature},
	Title = {3D-printed low-voltage-driven ciliary hydrogel microactuators},
	Ty = {JOUR},
	Url = {https://doi.org/10.1038/s41586-025-09944-6},
	Year = {2026}}

@article{keim2019,
  title={Memory formation in matter},
  author={Keim, Nathan C and Paulsen, Joseph D and Zeravcic, Zorana and Sastry, Srikanth and Nagel, Sidney R},
  journal={Reviews of Modern Physics},
  volume={91},
  number={3},
  pages={035002},
  year={2019},
  publisher={APS}
}

@article{veenstra2025wave,
  title={Wave coarsening drives time crystallization in active solids},
  author={Veenstra, Jonas and Binysh, Jack and Seinen, Vito and Naber, Rutger and Robledo-Poisson, Damien and Hunt, Andres and van Saarloos, Wim and Souslov, Anton and Coulais, Corentin},
  journal={arXiv preprint arXiv:2508.20052},
  year={2025},
url={"https://arxiv.org/pdf/2508.20052"}
}

@article{nash2015topological,
  title={Topological mechanics of gyroscopic metamaterials},
  author={Nash, Lisa M and Kleckner, Dustin and Read, Alismari and Vitelli, Vincenzo and Turner, Ari M and Irvine, William TM},
  journal={Proceedings of the National Academy of Sciences},
  volume={112},
  number={47},
  pages={14495--14500},
  year={2015},
  publisher={National Academy of Sciences}
}

@article{juraschek2025chiral,
  title={Chiral phonons},
  author={Juraschek, Dominik M and Geilhufe, R Matthias and Zhu, Hanyu and Basini, Martina and Baum, Peter and Baydin, Andrey and Chaudhary, Swati and Fechner, Michael and Flebus, Benedetta and Grissonnanche, Gael and others},
  journal={Nature Physics},
  pages={1--9},
  year={2025},
  publisher={Nature Publishing Group UK London},
 url ={"https://www.nature.com/articles/s41567-025-03001-9.pdf"}
}

@article{kruse2011spontaneous,
  title={Spontaneous mechanical oscillations: implications for developing organisms},
  author={Kruse, Karsten and Riveline, Daniel},
  journal={Current topics in developmental biology},
  volume={95},
  pages={67--91},
  year={2011},
  publisher={Elsevier},
url={"https://www.sciencedirect.com/science/article/abs/pii/B9780123850652000037"}
}

@article{fradique2023assessing,
  title={Assessing motile cilia coverage and beat frequency in mammalian in vitro cell culture tissues},
  author={Fradique, Ricardo and Causa, Erika and Delahousse, Clara and Kotar, Jurij and Pinte, Laetitia and Vallier, Ludovic and Vila-Gonzalez, Marta and Cicuta, Pietro},
  journal={Royal Society Open Science},
  volume={10},
  number={8},
  pages={230185},
  year={2023},
  publisher={The Royal Society}
}

@article{byron2021metachronal,
  title={Metachronal motion across scales: current challenges and future directions},
  author={Byron, Margaret L and Murphy, David W and Katija, Kakani and Hoover, Alexander P and Daniels, Joost and Garayev, Kuvvat and Takagi, Daisuke and Kanso, Eva and Gemmell, Bradford J and Ruszczyk, Melissa and others},
  journal={Integrative and comparative biology},
  volume={61},
  number={5},
  pages={1674--1688},
  year={2021},
  publisher={Oxford University Press},
url={"https://academic.oup.com/icb/article/61/5/1674/6287617"}
}

@article{gilpin2020multiscale,
  title={The multiscale physics of cilia and flagella},
  author={Gilpin, William and Bull, Matthew Storm and Prakash, Manu},
  journal={Nature Reviews Physics},
  volume={2},
  number={2},
  pages={74--88},
  year={2020},
  publisher={Nature Publishing Group UK London},
url={"https://www.nature.com/articles/s42254-019-0129-0"}
}

@article{Gu2025,
  title={Emergence of collective oscillations in massive human crowds},
  author={Gu, Fran{\c{c}}ois and Guiselin, Benjamin and Bain, Nicolas and Zuriguel, Iker and Bartolo, Denis},
  journal={Nature},
  volume={638},
  number={8049},
  pages={112--119},
  year={2025},
  publisher={Nature Publishing Group UK London}
}

@article{Ferrante2013,
  title = {Elasticity-Based Mechanism for the Collective Motion of Self-Propelled Particles with Springlike Interactions: A Model System for Natural and Artificial Swarms},
  author = {Ferrante, Eliseo and Turgut, Ali Emre and Dorigo, Marco and Huepe, Cristi\'an},
  journal = {Phys. Rev. Lett.},
  volume = {111},
  issue = {26},
  pages = {268302},
  numpages = {5},
  year = {2013},
  month = {Dec},
  publisher = {American Physical Society},
  doi = {10.1103/PhysRevLett.111.268302},
  url = {https://link.aps.org/doi/10.1103/PhysRevLett.111.268302}
}

@article{Woodhouse2018,
  title = {Autonomous Actuation of Zero Modes in Mechanical Networks Far from Equilibrium},
  author = {Woodhouse, Francis G. and Ronellenfitsch, Henrik and Dunkel, J\"orn},
  journal = {Phys. Rev. Lett.},
  volume = {121},
  issue = {17},
  pages = {178001},
  numpages = {6},
  year = {2018},
  month = {Oct},
  publisher = {American Physical Society},
  doi = {10.1103/PhysRevLett.121.178001},
  url = {https://link.aps.org/doi/10.1103/PhysRevLett.121.178001}
}

@article{xi2024,
  title={Emergent behaviors of buckling-driven elasto-active structures},
  author={Xi, Yuchen and Marzin, Tom and Huang, Richard B and Jones, Trevor J and Brun, P-T},
  journal={Proceedings of the National Academy of Sciences},
  volume={121},
  number={45},
  pages={e2410654121},
  year={2024},
  publisher={National Academy of Sciences}
}

@article{Zheng2023,
  title = {Self-Oscillation and Synchronization Transitions in Elastoactive Structures},
  author = {Zheng, Ellen and Brandenbourger, Martin and Robinet, Louis and Schall, Peter and Lerner, Edan and Coulais, Corentin},
  journal = {Phys. Rev. Lett.},
  volume = {130},
  issue = {17},
  pages = {178202},
  numpages = {7},
  year = {2023},
  month = {Apr},
  publisher = {American Physical Society},
  doi = {10.1103/PhysRevLett.130.178202},
  url = {https://link.aps.org/doi/10.1103/PhysRevLett.130.178202}
}

@article{baconnier2022,
  title={Selective and collective actuation in active solids},
  author={Baconnier, Paul and Shohat, Dor and L{\'o}pez, C Hern{\'a}ndez and Coulais, Corentin and D{\'e}mery, Vincent and D{\"u}ring, Gustavo and Dauchot, Olivier},
  journal={Nature Physics},
  volume={18},
  number={10},
  pages={1234--1239},
  year={2022},
  publisher={Nature Publishing Group UK London},
url={"https://www.nature.com/articles/s41567-022-01704-x"}
}

@article{Hernandez2024,
  title = {Model of Active Solids: Rigid Body Motion and Shape-Changing Mechanisms},
  author = {Hern\'andez-L\'opez, Claudio and Baconnier, Paul and Coulais, Corentin and Dauchot, Olivier and D\"uring, Gustavo},
  journal = {Phys. Rev. Lett.},
  volume = {132},
  issue = {23},
  pages = {238303},
  numpages = {6},
  year = {2024},
  month = {Jun},
  publisher = {American Physical Society},
  doi = {10.1103/PhysRevLett.132.238303},
  url = {https://link.aps.org/doi/10.1103/PhysRevLett.132.238303}
}

@article{Xia2024,
  title = {Biomimetic Synchronization in Biciliated Robots},
  author = {Xia, Yiming and Hu, Zixian and Wei, Da and Chen, Ke and Peng, Yi and Yang, Mingcheng},
  journal = {Phys. Rev. Lett.},
  volume = {133},
  issue = {4},
  pages = {048302},
  numpages = {6},
  year = {2024},
  month = {Jul},
  publisher = {American Physical Society},
  doi = {10.1103/PhysRevLett.133.048302},
  url = {https://link.aps.org/doi/10.1103/PhysRevLett.133.048302}
}

@article{boudet2021,
  title={From collections of independent, mindless robots to flexible, mobile, and directional superstructures},
  author={Boudet, Jean-Fran{\c{c}}ois et al},
  journal={Science Robotics},
  volume={6},
  number={56},
  pages={eabd0272},
  year={2021},
  publisher={American Association for the Advancement of Science},
url={"https://www.science.org/doi/abs/10.1126/scirobotics.abd0272"}
}

@misc{ball2021,
  title={Animate materials},
  author={Ball, Philip},
  year={2021},
  publisher={Springer},
url={"https://royalsociety.org/news-resources/projects/animate-materials/"}
}

@article{sakaguchi1987local,
  title={Local and grobal self-entrainments in oscillator lattices},
  author={Sakaguchi, Hidetsugu and Shinomoto, Shigeru and Kuramoto, Yoshiki},
  journal={Progress of Theoretical Physics},
  volume={77},
  number={5},
  pages={1005--1010},
  year={1987},
  publisher={Oxford University Press}
}

@article{bacconier2025,
  title = {Self-aligning polar active matter},
  author = {Baconnier, Paul and Dauchot, Olivier and D\'emery, Vincent and D\"uring, Gustavo and Henkes, Silke and Huepe, Cristi\'an and Shee, Amir},
  journal = {Rev. Mod. Phys.},
  volume = {97},
  issue = {1},
  pages = {015007},
  numpages = {26},
  year = {2025},
  month = {Mar},
  publisher = {American Physical Society},
  doi = {10.1103/RevModPhys.97.015007},
  url = {https://link.aps.org/doi/10.1103/RevModPhys.97.015007}
}

@article{Tierno2021,
  title = {Arrested phase separation in chiral fluids of colloidal spinners},
  author = {Massana-Cid, Helena and Levis, Demian and Hern\'andez, Ra\'ul Josu\'e Hern\'andez and Pagonabarraga, Ignacio and Tierno, Pietro},
  journal = {Phys. Rev. Res.},
  volume = {3},
  issue = {4},
  pages = {L042021},
  numpages = {6},
  year = {2021},
  month = {Nov},
  publisher = {American Physical Society},
  doi = {10.1103/PhysRevResearch.3.L042021},
  url = {https://link.aps.org/doi/10.1103/PhysRevResearch.3.L042021}
}

@article{petroff2015,
  title = {Fast-Moving Bacteria Self-Organize into Active Two-Dimensional Crystals of Rotating Cells},
  author = {Petroff, Alexander P. and Wu, Xiao-Lun and Libchaber, Albert},
  journal = {Phys. Rev. Lett.},
  volume = {114},
  issue = {15},
  pages = {158102},
  numpages = {6},
  year = {2015},
  month = {Apr},
  publisher = {American Physical Society},
  doi = {10.1103/PhysRevLett.114.158102},
  url = {https://link.aps.org/doi/10.1103/PhysRevLett.114.158102}
}

@article{scholz2018,
  title={Rotating robots move collectively and self-organize},
  author={Scholz, Christian and Engel, Michael and P{\"o}schel, Thorsten},
  journal={Nature communications},
  volume={9},
  number={1},
  pages={931},
  year={2018},
  publisher={Nature Publishing Group UK London}
}

@article{jakli2008,
  title={Colloidal micromotor in smectic A liquid crystal driven by DC electric field},
  author={J{\'a}kli, Antal and Senyuk, Bohdan and Liao, Guangxun and Lavrentovich, Oleg D},
  journal={Soft Matter},
  volume={4},
  number={12},
  pages={2471--2474},
  year={2008},
  publisher={Royal Society of Chemistry}
}

@article{Pannacci2007,
  title = {How Insulating Particles Increase the Conductivity of a Suspension},
  author = {Pannacci, N. and Lobry, L. and Lemaire, E.},
  journal = {Phys. Rev. Lett.},
  volume = {99},
  issue = {9},
  pages = {094503},
  numpages = {4},
  year = {2007},
  month = {Aug},
  publisher = {American Physical Society},
  doi = {10.1103/PhysRevLett.99.094503},
  url = {https://link.aps.org/doi/10.1103/PhysRevLett.99.094503}
}

@article{brandenbourger2019,
  title={Non-reciprocal robotic metamaterials},
  author={Brandenbourger, Martin and Locsin, Xander and Lerner, Edan and Coulais, Corentin},
  journal={Nature communications},
  volume={10},
  number={1},
  pages={4608},
  year={2019},
  publisher={Nature Publishing Group UK London}
}

@article{veenstra2025,
  title={Adaptive locomotion of active solids},
  author={Veenstra, Jonas and Scheibner, Colin and Brandenbourger, Martin and Binysh, Jack and Souslov, Anton and Vitelli, Vincenzo and Coulais, Corentin},
  journal={Nature},
  pages={1--7},
  year={2025},
  publisher={Nature Publishing Group UK London}
}

@article{veenstra2024,
  title={Non-reciprocal topological solitons in active metamaterials},
  author={Veenstra, Jonas and Gamayun, Oleksandr and Guo, Xiaofei and Sarvi, Anahita and Meinersen, Chris Ventura and Coulais, Corentin},
  journal={Nature},
  volume={627},
  number={8004},
  pages={528--533},
  year={2024},
  publisher={Nature Publishing Group UK London}
}

@article{chen2024,
  title={Self-propulsion, flocking and chiral active phases from particles spinning at intermediate Reynolds numbers},
  author={Chen, Panyu and Weady, Scott and Atis, Severine and Matsuzawa, Takumi and Shelley, Michael J and Irvine, William TM},
  journal={Nature Physics},
  pages={1--9},
  year={2024},
  publisher={Nature Publishing Group UK London}
}

@article{liu2020,
  title={Oscillating collective motion of active rotors in confinement},
  author={Liu, Peng and Zhu, Hongwei and Zeng, Ying and Du, Guangle and Ning, Luhui and Wang, Dunyou and Chen, Ke and Lu, Ying and Zheng, Ning and Ye, Fangfu and others},
  journal={Proceedings of the National Academy of Sciences},
  volume={117},
  number={22},
  pages={11901--11907},
  year={2020},
  publisher={National Acad Sciences}
}

@article{liebchen2022,
  title={Chiral active matter},
  author={Liebchen, Benno and Levis, Demian},
  journal={Europhysics Letters},
  volume={139},
  number={6},
  pages={67001},
  year={2022},
  publisher={IOP Publishing}
}

@article{han2021,
  title={Fluctuating hydrodynamics of chiral active fluids},
  author={Han, Ming and Fruchart, Michel and Scheibner, Colin and Vaikuntanathan, Suriyanarayanan and De Pablo, Juan J and Vitelli, Vincenzo},
  journal={Nature Physics},
  volume={17},
  number={11},
  pages={1260--1269},
  year={2021},
  publisher={Nature Publishing Group UK London}
}

@article{martinet2025,
  title={Emergent dynamics of active elastic microbeams},
  author={Martinet, Q and Li, Y and Aubret, A and Hannezo, E and Palacci, J},
  journal={arXiv preprint arXiv:2508.20642},
  year={2025},
url={"https://arxiv.org/abs/2508.20642"}
}

@article{aubret2021,
  title={Metamachines of pluripotent colloids},
  author={Aubret, Antoine and Martinet, Quentin and Palacci, Jeremie},
  journal={Nature communications},
  volume={12},
  number={1},
  pages={6398},
  year={2021},
  publisher={Nature Publishing Group UK London},
url={"https://www.nature.com/articles/s41467-021-26699-6"}
}

@article{volpe2025roadmap,
  title={Roadmap for animate matter},
  author={Volpe, Giorgio and Ara{\'u}jo, Nuno AM and Guix, Maria and Miodownik, Mark and Martin, Nicolas and Alvarez, Laura and Simmchen, Juliane and Di Leonardo, Roberto and Pellicciotta, Nicola and Martinet, Quentin and others},
  journal={Journal of Physics: Condensed Matter},
  year={2025},
url={"https://iopscience.iop.org/article/10.1088/1361-648X/adebd3/meta"}
}

@article{aubret2018,
  title={Targeted assembly and synchronization of self-spinning microgears},
  author={Aubret, Antoine and Youssef, Mena and Sacanna, Stefano and Palacci, J{\'e}r{\'e}mie},
  journal={Nature Physics},
  volume={14},
  number={11},
  pages={1114--1118},
  year={2018},
  publisher={Nature Publishing Group UK London}
}

@article{Massana2021,
  title = {Arrested phase separation in chiral fluids of colloidal spinners},
  author = {Massana-Cid, Helena and Levis, Demian and Hern\'andez, Ra\'ul Josu\'e Hern\'andez and Pagonabarraga, Ignacio and Tierno, Pietro},
  journal = {Phys. Rev. Res.},
  volume = {3},
  issue = {4},
  pages = {L042021},
  numpages = {6},
  year = {2021},
  month = {Nov},
  publisher = {American Physical Society},
  doi = {10.1103/PhysRevResearch.3.L042021},
  url = {https://link.aps.org/doi/10.1103/PhysRevResearch.3.L042021}
}

\end{document}


\title{Phase coherence and disorder-induced wave propagation in micromotor arrays: 
 Supplementary Information
}

\author{Romane Braun}
\affiliation{Univ. Lyon, ENS de Lyon, Univ. Claude Bernard, CNRS, Laboratoire de Physique, F-69342, Lyon.}
\author{Alexis Poncet}
\affiliation{Univ. Lyon, ENS de Lyon, Univ. Claude Bernard, CNRS, Laboratoire de Physique, F-69342, Lyon.}
\author{Alexandre Morin}
\affiliation{Huygens-Kamerlingh Onnes Laboratory, Universiteit Leiden,
PO Box 9504, 2300 RA Leiden, the Netherlands}
\author{Denis Bartolo}
\affiliation{Univ. Lyon, ENS de Lyon, Univ. Claude Bernard, CNRS, Laboratoire de Physique, F-69342, Lyon.}
\email{denis.bartolo@ens-lyon.fr}

\maketitle
\setcounter{equation}{0}
\setcounter{figure}{0}
\setcounter{table}{0}
\setcounter{page}{1}
\renewcommand{\thefigure}{S\arabic{figure}}
\renewcommand{\thetable}{S\arabic{table}}
\renewcommand{\theequation}{${\rm S}$\arabic{equation}}
\renewcommand\thesubsection{\arabic{subsection}}
\renewcommand\thesubsubsection{\arabic{subsection}.\arabic{subsubsection}  }

\tableofcontents
\newpage 

\section{Description of the supplementary videos}
\renewcommand{\labelitemi}{{\bf-}}

\begin{itemize}
\item {\bf Movie 1.} 
A lattice of 48x44 motors self-organize into an antiferromagnetic and phase-coherent state. 
Our video shows $44\times34$ motors.
Rotor diameter: $d=86.4\,\rm \mu m$, lattice spacing: $\ell=100 \,\rm \mu m$ and field amplitude $E=0.62$~V/$\upmu$m. Video recorded at $90$ fps and played at $20$ fps.
\item {\bf Movie 2.}
Maps of the phase $\varphi$ measured in the experiment presented in Movie~1. 
The colors indicate the orientations of the rotors.
$d=86.4\,\rm \mu m$, lattice spacing: $\ell=100 \,\rm \mu m$ and field amplitude $E=0.62$~V/$\upmu$m. 
Data recorded at $90$ fps and played at $20$ fps.
\item {\bf Movie 3.}
Numerical simulations showing the phase ordering of a collection of interacting point rotors coupled by dipolar interactions (Eqs.~\ref{eq_dipolaire_2corps_1} and~\ref{eq_dipolaire_2corps_2}). 
In the absence of disorder, after a short transient  the rotors self-organize into a pristine phase-coherent state where the modified phase $\varphi^\star$ is uniform in space. Simulation parameters: $B=500$, $\ell=10$, $\sigma=0$.
\item {\bf Movie 4.}
The two videos represent the phase dynamics measured in the same experiment using the bare phase  ($\varphi$) in the left panel and the gauge-transformed phase ($\varphi^\star$) in the right panel. $\ell=100 \,\rm \mu m$ and field amplitude $E=0.62$~V/$\upmu$m. Video recorded at $90$ fps and played at $18$ fps.
\item {\bf Movie 5.}
Propagation of  phase waves in incompatible phase-coherent domains (Simulations). Simulation parameters: $B=500$, $\ell=10$, $\sigma=0.07$.
\item {\bf Movie 6.}
Phase waves emanate from a linear domain wall (Simulations). $B=500$, $\ell=10$, $\sigma=0.1$ ($\sigma$ here denotes the driving torque difference between the two domains). 
\end{itemize}

\section{Experimental Methods and Setup}
\subsection{3D Nanoprinting }
\begin{figure*}[h]
\includegraphics[width=\textwidth]{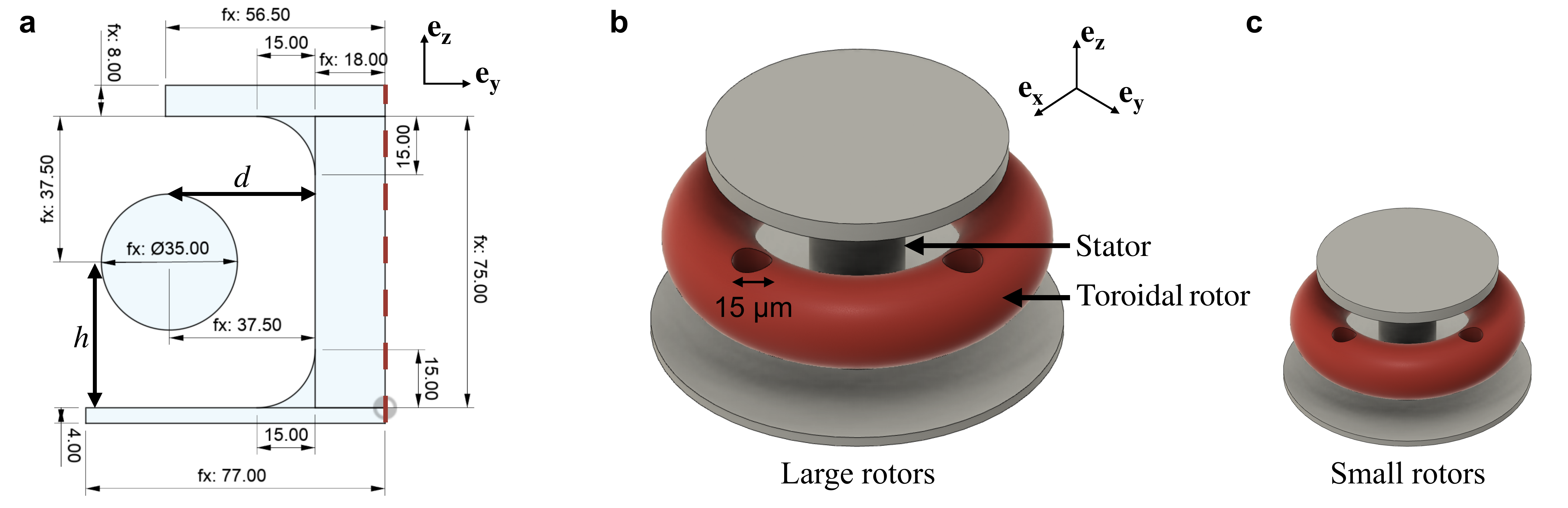}
\caption{{\bf Micromotor design.}
{\bf a. } Schematics of the micromotor geometry. All dimensions are in micrometers. 
{\bf b. } Large rotors: 3D rendering of the desired motor geometry. 
{\bf c. } Small rotors:  3D rendering of the desired motor geometry. (same scale as in {\bf b}).
}
    \label{schema_rotor}
\end{figure*}

We first design  3D models of the motors using the CAD software Fusion360 (Fig.~\ref{schema_rotor}). 
We then 3D-print micro-motor lattices using two-photon polymerization with a Nanoscribe Photonic Professional GT2 (PPGT2) 3D printer.
We print the motors in  photosensitive resin SU-8 100 spincoated on fused silica substrates, and enhance the adhesion of the resin by first spin-coating a thin layer of OmniCoat (Kayaku)~ \cite{Kayaku_omnicoat,Kayaku_SU8}.

We performed experiments on two different types of rotors having the same shape but different sizes (size ratio 0.6) (Fig.~\ref{schema_rotor}).
We provide the dimensions of the larger motors in Fig.~\ref{schema_rotor}a.
We  set the slicing and hatching distances to 0.7 $\mu m$, the hatching angle offset to 90°, and disable the splitting mode. 
We set the solidLaserPower and the baseLaserPower parameters to 42\%, and the solidScanSpeed and the baseScanSpeed parameters to 13000~$\upmu$m/s.

Following printing, we perform the post-exposure bake of the SU-8 according to the protocol detailed in \cite{Kayaku_SU8}. 
We then place the printed structures vertically into a beaker filled with SU-8 developer. 
The beaker is positioned on an orbital shaker (PSU-10i) rotating at 200 rpm for 20 minutes. 
Finally, we rinse the developed structures with fresh SU-8 developer and clean the device by agitating it in isopropyl alcohol for 2 minutes.

\subsection{Microfluidic device}
\begin{figure*}[h]
\includegraphics[width=\textwidth]{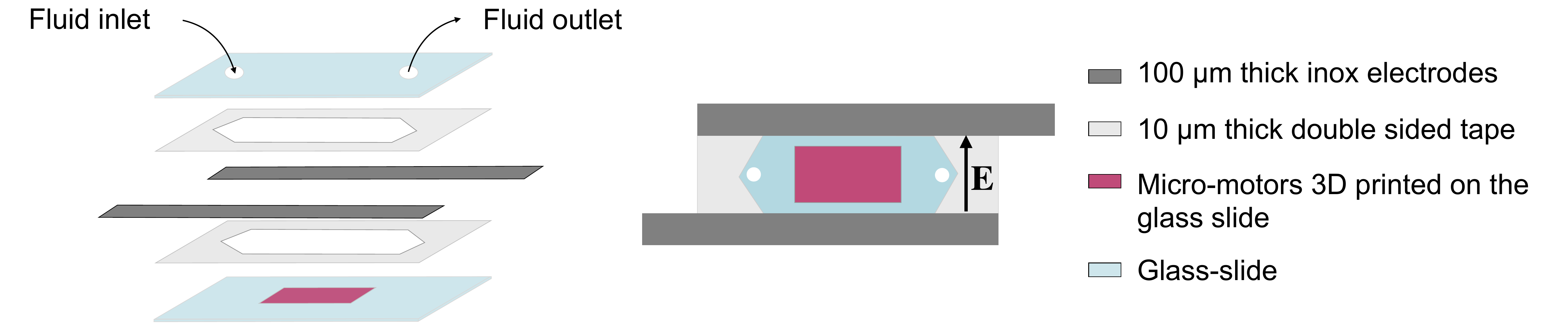}
\caption{{\bf Microfluidic chip.}
    {Our microfluidic channels are made of double sided tapes. The two electrodes are made of stainless steel.} 
}
    \label{schema_cellule}
\end{figure*}
We place the micromotors  in 1 cm wide microfluidic channels (Fig.~\ref{schema_cellule}).
We bond two electrodes  made of 100 $\mu m$ thick stainless steel plates to two glass slides using double-sided tapes with an approximate thickness of 10 $\mu m$. 
The total channel height is therefore approximately 120 $\mu m$. 
We seal the edges of the resulting microfluidic device with epoxy resin to ensure a leak-proof assembly.
For all the experiments described in this article, the distance between the electrodes was between 4~mm and 6~mm. 

Once the microfluidic device is assembled, we inject a 0.025 $\text{mol} \cdot \text{L}^{-1}$ Dioctyl sulfosuccinate sodium salt (AOT)/ hexadecane solution inside the microfluidic chamber. 
To release the mobile parts of the micromotors after SU-8 development,  we rely on ultrasonic shaking. 
We place the sealed microfluidic device in a glass box (Wheaton staining dishes) filled with a water-glycerol mixture containing 40\% glycerol by volume. 
We sonicate the microfluidic cell using an Electris UC317MD sonication bath in 2-second intervals.
After each sonication cycle, we check whether the 
rotors are free to move. 
The process is terminated once all mobile parts are functional, or when a rotor is damaged.

\subsection{Quincke motorization setup and principle}
\subsubsection{Electric setup}
To actuate our micromotors, we take advantage of the Quincke electrorotation instability, which we briefly recall in the next section~\cite{quincke1896,Melcher1969,Bricard2013}. 
Using a voltage amplifier (TREK 610E), we apply an electric field $\mathbf E$  transverse to the electrodes and parallel to one of the axes of the micromotor network, Fig.~\ref{schema_cellule}.
Above a critical field amplitude, the rotors are mechanically unstable, leading to the steady rotation of the rotors~\cite{Melcher1969}. 
At the end of each experiment, we turn off the electric field  and renew the solvent in the microchannel to reset the experiment and avoid unwanted memory effects~\cite{Zhang2022}. 

\subsubsection{Quincke electrorotation in a nutshell}
\begin{figure*}[h]
\includegraphics[width=\textwidth]{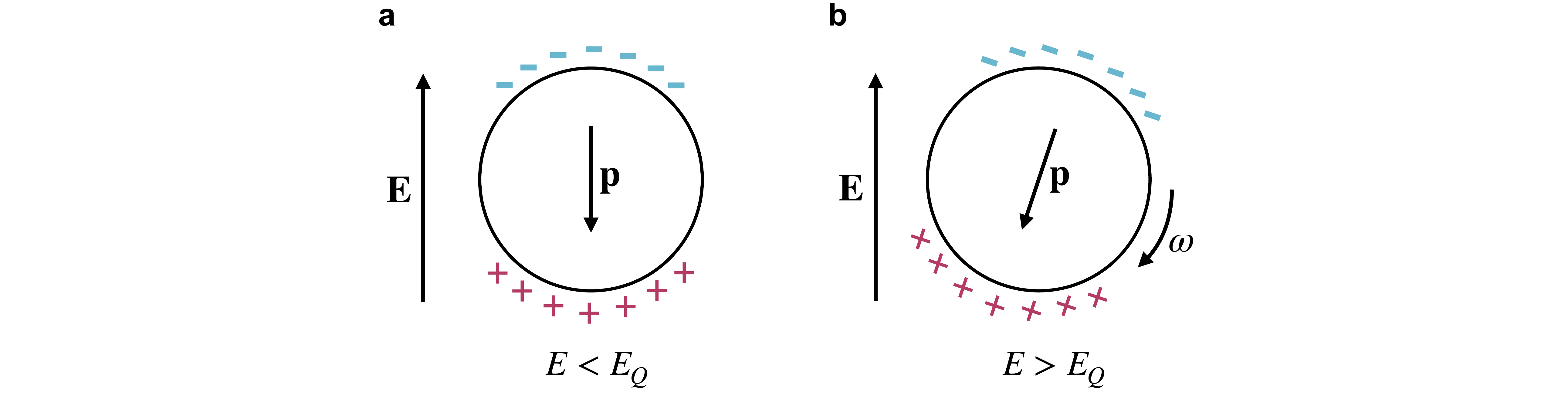}
\caption{{\bf Quincke electrorotation: principle.}
{\bf a.} Charge distribution around a  dielectric sphere immersed in a conducting fluid. 
Below $E_Q$, when the difference in electric permittivity between the fluid and the solid media is not too large, the Ohmic transport of the free ions in the solution induces the formation of a net electric dipole pointing in the direction opposite to $\mathbf E$.
{\bf b.} Above $E_Q$, the system reaches a different dynamic steady state where the particle spins at a constant rate $\omega$. 
This steady rotation is sustained by the competition between the electric and the viscous torques acting on the particle.
}
    \label{schema_quinke}
\end{figure*}

We provide a short qualitative description of the Quincke rotation mechanism, which we use to motorize our micromachines. 
A comprehensive description of this electrohydrodynamic instability can be found e.g. in~\cite{Melcher1969,Bricard2013}.

Consider an axisymmetric rigid body immersed in a conducting fluid, and subjected to a uniform DC electric field $\mathbf{E}$. 
After a transient period, the ohmic transport of the ions  under the action of the electric field results in the formation of a dipolar charge distribution at the surface of the solid particle. 
The resulting dipole moment $\mathbf{p}$ is oriented in the direction opposite to the $\mathbf{E}$ field (Fig.~\ref{schema_quinke}a).

The stability of this dynamical steady state is determined by the competition between two opposing effects. 
On one hand, $\mathbf{p}$ tends to align with $\mathbf{E}$, exerting an electric torque that amplifies any angular fluctuation of the particle orientation, and initiates  rotation. 
On the other hand, the Ohmic current in the electrolyte continuously redistributes the surface charges, stabilizing the dipole orientation in the direction opposite to $\mathbf{E}$. 
This competition leads to a supercritical instability: below the Quincke threshold $\mathbf{E_Q}$, the dipole orientation is stable and remains anti-aligned with $\mathbf{E}$. 
Above $\mathbf{E_Q}$, the configuration where $\mathbf E$ and $\mathbf p$ are collinear is unstable.
The amplification of any minute orientational fluctuation leads the system towards another dynamical steady state where $\mathbf p$ makes a finite angle with $\mathbf{E}$.
This charge distribution results in  a constant electric torque balanced by the viscous torque caused by the continuous rotation of the rigid body (Fig.~\ref{schema_quinke}b). 
The rotation speed $\omega$ is determined by the balance between the two torques: 
\begin{equation}
    \omega= \tau_{\rm M}^{-1}\left(1-E/E_{\rm Q}\right)^{1/2},
\end{equation}
where $\tau_{\rm M}$ is the so-called Maxwell-Wagner charge relaxation time. 
It is a material property and depends solely on the conductivity and electric permittivity of the materials. 
It is of the order of $10\,\rm ms$ in our experiments. 

A key aspect of the Quincke instability is that the charge distribution  spontaneously breaks the rotation symmetry. 
While the magnitude of the angular speed can be controlled by adjusting the magnitude of $E$, the direction of rotation remains only determined by the initial fluctuations: the axis of rotation of the particle can be in any direction in the plane perpendicular to $\mathbf{E}$.

\subsection{Measurements of the instantaneous phase and angular velocity}
\begin{figure*}[h]
\includegraphics[width=\textwidth]{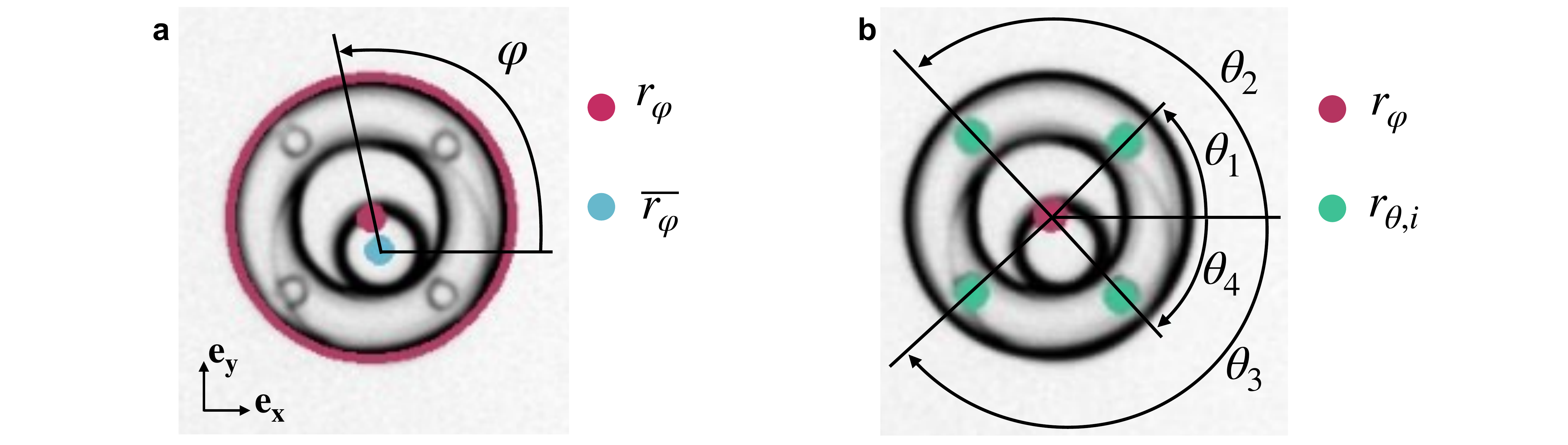}
\caption{
{\bf Tracking of the rotor phase and speed.}
{\bf a.} To measure the instantaneous phase $\varphi(t)$ we need to measure the instantaneous center of mass (pink dot) of the rotor and its time average (blue dot).
{\bf b.} We measure the instantaneous rotation angle $\theta$ by detecting the positions of the four marker holes on the rotor (green dots).
}
    \label{detection}
\end{figure*}
We observe the micro-motors using a Nikon AZ100 macroscope with a 2x objective (Nikon AZ Plan Fluor). 
A Ximea xiB-64 camera mounted on the macroscope via a DSC port records the images, introducing an additional 0.6x magnification factor. 
We adjust the magnification of the Nikon AZ100 macroscope according to experimental requirements, achieving a total magnification ranging from 3.6x to 8.4x.

Figure~\ref{detection}a and b illustrate how we measure the instantaneous phase and rotation speed of the motors. 
We use the HoughCircles function from the  OpenCV cv2 module to detect the rotors, and their four circular holes. 
We denote the rotor’s position as $\mathbf{r}_\varphi$, and the positions of the four holes as $\mathbf{r}_{\theta,i}$, where $i$ ranges from 1 to 4.

To measure the phase $\varphi$, we first compute the center of the rotor's trajectory: $\overline{\mathbf{r}_\varphi}=\left<\mathbf{r}_\varphi\right>_t$. 
We then define the phase  as: 
$\varphi$=arctan$((\mathbf{r}_\varphi-\overline{\mathbf{r}_\varphi}). \mathbf{e_y},(\mathbf{r}_\varphi-\overline{\mathbf{r}_\varphi}).\mathbf{e_x})$, where $\mathbf{e_x}$ and $\mathbf{e_y}$ are the unit vectors along and perpendicular to $\mathbf{E}$, respectively. 
To handle discontinuities in the trajectory caused by the periodic nature of the arctan2 function (confined to the interval [$-\pi$,$\pi$]), we add $\pm2\pi$ to the phase when we detect phase jumps with a magnitude larger than $\pi$.
We then compute the instantaneous rotation  rate $\dot \varphi$ using a finite difference method.

To measure the angle $\theta$, we calculate $\theta_i=$arctan2$((\mathbf{r}_{\theta,i}-\mathbf{r}_\varphi). \mathbf{e_y},(\mathbf{r}_{\theta,i}-\mathbf{r}_\varphi).\mathbf{e_x})$,  the angle of the $i$-th detected marker on the wheel. 
At each time, the angle $\theta$ is then defined as the mean of all $\theta_i$ modulo $\pi/2$, calculated for one rotor. As with $\varphi$, we correct the discontinuities in $\theta$ caused by its confinement to the interval [0,$\pi/2$]. 
If there is a jump in the trajectory greater than $0.42\frac{\pi}{2}$, we add $\pi/2$,  and if the jump is smaller than $-0.42\frac{\pi}{2}$, we subtract $\pi/2$.

\subsection{Experimental Protocols}
\noindent{\bf Antiferromagnetic order}
To characterize the emergence of antiferromagnetic order as the lattice spacing $\ell$ decreases, we conducted experiments both on lattices made of 20$\times$11 small motors, and on 13$\times$12 large motors. 
For each geometry, and each value of $\ell$, we repeated ten independent experiments.
The order parameter $\Omega$ plotted in Fig. 2a in the main text is averaged over these ten measurements.

\begin{figure*}[h]
\includegraphics[width=\textwidth]{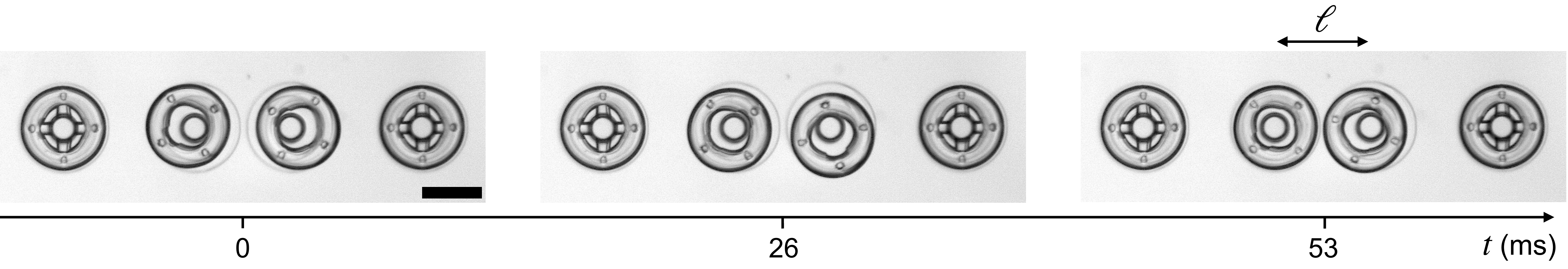}
\caption{{\bf Dynamics of a single pair of motors. }
    Three subsequent snapshots showing two functional motors surrounded by two immobile rotors. Scale bar: $100\,\rm \mu m$.
}
    \label{photo_manip_paire}
\end{figure*}

\noindent{\bf Dynamics of motor pairs}
To investigate the dynamics of the motor pairs presented in Fig.~2d of the main article, we printed series of four large motors as illustrated in Fig.~\ref{photo_manip_paire}. 
The leftmost and rightmost structures cannot rotate, as we added rigid links to immobilize the rotors.
The idea is to measure the dynamics of isolated pairs of motors feeling an electric field close to that of the motors in the bulk of the square lattices.
All measurements were performed on large motors.

{\color{bleuf}
\noindent{\bf Emergence of phase-coherence}
We investigate the phase coherence of the metamachines using  lattices of 48$\times$44 small micromotors with $\ell=99.6\,\rm{\mu m}$. 
Our observation window includes only the 36$\times$33 rotors at the center of the lattice.
We focus on this large bulk region to avoid boundary effects. 
Indeed, on the left and right edges of the lattice, the local electric field differs from the bulk, leading to lower Quincke thresholds and faster rotations speed as seen in Figure~\ref{FigureEdge}.
We can provide a qualitative explanation of this phenomenon. Whatever the E field magnitude, the motors locally hinder the Ohmic transport of the ionic current between the electrodes (they reduce the available space in the $z$ direction).
At large scales, this effect can be modeled as a localized reduction of the Ohmic conductivity $\sigma$. 
As captured by the leaky dielectric model of Taylor and Melchner~\cite{Melcher1969}, a local reduction of the conductivity translates into a dipolar perturbation to the $\mathbf E$ field centered on each motor.
The total $\mathbf E$ field that powers the Quincke rotation is the sum of the bare $\mathbf E$ field and of the dipolar perturbations caused by all the other rotors.
In a square geometry the sum of all these perturbations cancel out everywhere but at the left and right hand side of the lattice, where we indeed observed localized speed heterogeneities over short distances Figure~\ref{FigureEdge}. 
All the measurements reported in Figure 4 therefore correspond to bulk data only.
\begin{figure*}[h!]
\includegraphics[width=0.8\textwidth]{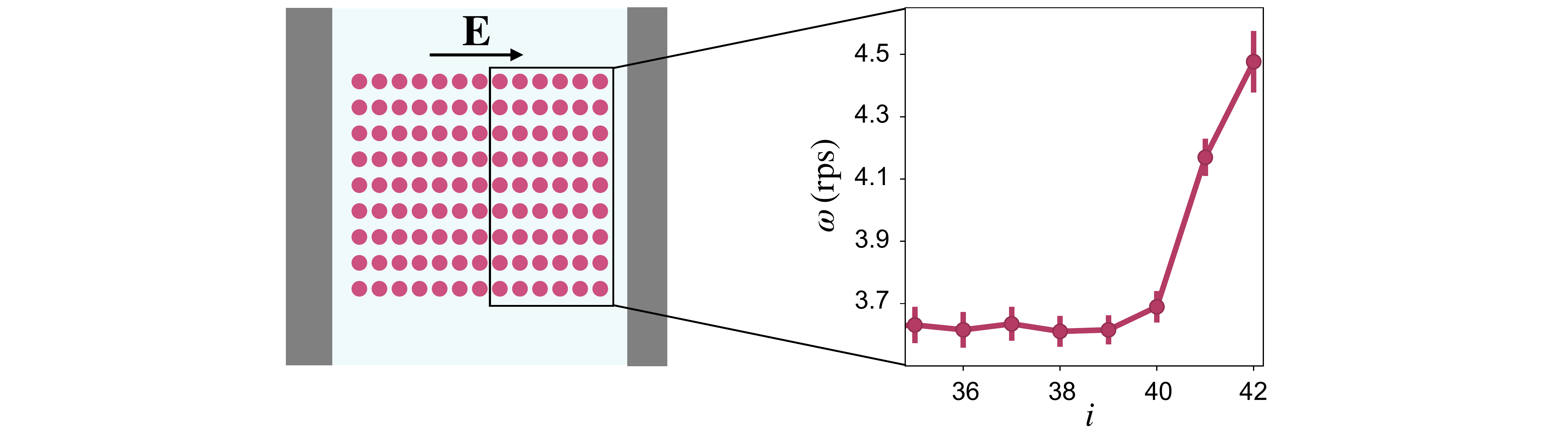}
\caption{{\bf Speed heterogeneities at the  edges of the lattice.}
Left: sketch of the experimental geometry. The two vertical electrodes induce a $\mathbf E$ field in the $x$ direction to power the Quincke motors. 
Right:
Map of the rotor speed measured close to the lattice edge. The speed of the motors is increased in a localized vertical strip. 
Experiments on small rotors, lattice spacing $\ell=101 \,\rm{\mu m} $, $E=0.66\, \rm V/\rm \mu m$
}
\label{FigureEdge}
\end{figure*}
}
{\color{bleuf}
\subsection{Geometric control of the motor speed}
We here describe a practical method to  control the spatial variations of the motor speed. 

The motors being powered by the Quincke instability, controlling their rotation speed amounts to controlling the  magnitude of the  electric field.
To achieve a local control of $E$ we take advantage of the Ohmic transport in the microchannel. 
The electric charge conservation imply that in steady state $\bm \nabla\cdot \mathbf j$=0, where $\mathbf j$ is the electric current. Combined with Ohm's law and integrating over the $z$ direction, we find that the magnitude of the $E$ field and the local height of the channel are related by 
\begin{equation}
\bm \nabla\cdot(\mathbf Eh)=0,
\label{equationcourant}
\end{equation}
where $h(x,y)$ is the local channel thickness. 
This simple relation readily tells us how to pattern the electric field in space.
Consider the simple case of a linear decrease of $h$ along the field direction $x$. In the limit of small gradients, Eq.~\ref{equationcourant} implies that $E$ increases linearly in the $x$ direction, thereby patterning the rotation speed of the motors. 

To demonstrate this effective design principle, we place a series of distant motors in a microfluidic cell, where the  left and right electrodes have different thicknesses. 
Figure~\ref{Figuregradient}a 
confirms that the motors spin at a higher rate as they are closer to the thinnest electrode.
For a more quantitative confirmation, we increase the voltage applied to the electrodes and measure its value $V_Q(x)$ when the motor located at $x$ reaches the Quincke threshold and start spinning. 
We find a linear relationship between $V_Q(x)$ and $h(x)$ in excellent agreement with our model that predicts $V_Q(x)=Ah(x)$, where $A=-E_Q\int_0^L\frac{{\rm d}x}{h(x)}$, see Figure~\ref{Figuregradient}b.

\begin{figure*}[h!]
\includegraphics[width=0.7\textwidth]{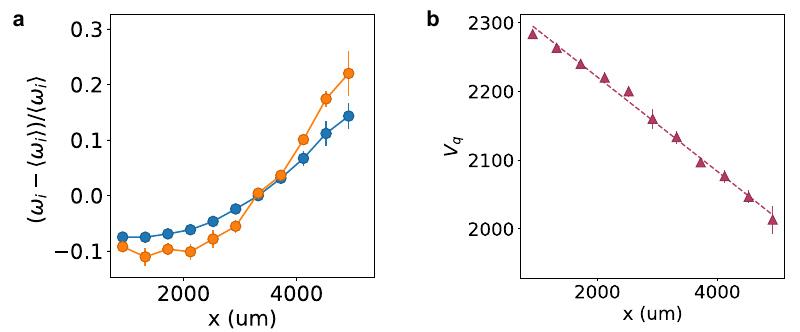}

\caption{{\bf Local control of the motor speed.}
{\bf a.} Plot of the motor speed normalized by their average value as a function of their position in the channel for two values of the imposed voltage V. Orange dots: $V=3000\, V$, Blue dots: $5000\, \rm V$. Distance between the electrodes: $6100 \,\rm \mu m$, height of the left electrode: $140 \,\rm \mu m$, height of the right electrode: $240 \,\rm \mu m$.
{\bf b.} Voltage $V_q$ needed to trigger the rotation of the motors at position $x$.  The experiments correspond to small motors separated by a distance of $400 \,\rm \mu m$.
}
\label{Figuregradient}
\end{figure*}
}

\newpage
\section{Single-motor dynamics}
\subsection{Quincke electrotation of micromotors: spin and precession}
\begin{figure*}[h!]
\includegraphics[width=\textwidth]{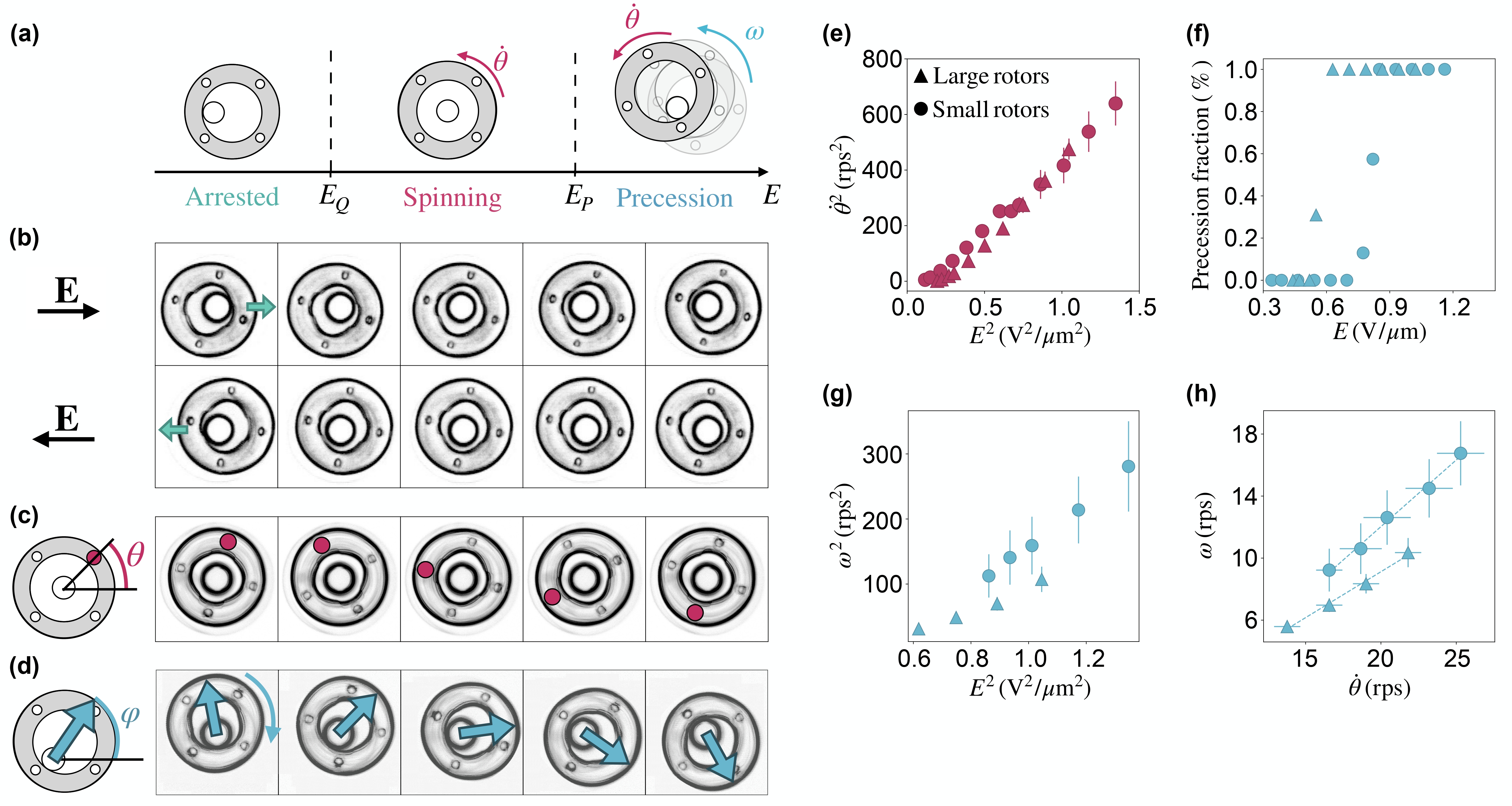}
\caption{{\bf a.} Sketch of the three states of the motor dynamics: Arrested, Spinning, Precession.
{\bf b.} $E=0.6 E_Q$: The motors do not rotate. Electrophoresis drives the rotors towards the stators.
{\bf c.} $E=1.3 E_Q$: Above the onset of electro-rotation, the rotors are centered and spin at a constant rate.
{\bf d.} $E=2.1 E_Q$: At large field amplitude, the rotors are not centered anymore. In contact with the stators, the rotors roll on the stator and precess. It hula-hoops around the stator. 
{\bf e.} Square of the Spinning rate, $\dot \theta^2$, plotted as a function of $E^2$.
{\bf f.} Fraction of rotors in the precession state.
{\bf g.} Square of the precession rate, $\omega^2=\dot \varphi^2$, plotted as a function of $E^2$.
{\bf h.} Precession rate $\omega$ plotted as a function of the spinning rate $\dot \theta$. A linear fit indicates that the rotors both roll and slip on the stators.
Circles: small rotors, triangles: large rotors.
}
    \label{dynamique_ind}
\end{figure*}
In the main text we focus on the collective dynamics of interacting motors powered by an electric field with a magnitude much higher than the Quincke threshold $E_Q$. 
In this section, we provide a quantitative characterization of the dynamics of isolated motors over a broader range of $E$ fields (Fig.~\ref{dynamique_ind}a).

At low field $E<E_Q$, the rotors do not spin. 
They are however driven towards the stator as a result of small electrophoretic forces (Fig.~\ref{dynamique_ind}b). The direction of motion indicates that the rotors have a positive charge.
As $E$ exceeds $E_Q$, right above the instability threshold,  the motors rotate at a constant rate $\dot \theta$, but do not precess. 
Instead the rotors are centered and spin around the stator (Fig.~\ref{dynamique_ind}c).
In this regime, $\dot\theta^2$ increases linearly with $E^2$ as predicted by Quincke-rotation theory~\cite{Melcher1969} (Fig.~\ref{dynamique_ind}e).
Further increasing the value of $E$, the motors abruptly change their dynamics (Fig.~\ref{dynamique_ind}f): 
the rotor contacts the stator thereby leading to a steady precession  akin to hula-hoop motion (Fig.~\ref{dynamique_ind}d).
In this regime, as reported in the main text,  the precession speed $\omega\equiv\dot\varphi$ increases with $E$ in agreement with Quincke-electrorotation theory~\cite{Melcher1969} (Fig.~\ref{dynamique_ind}g).
 The value of the slope of the linear relation between $\omega$ and $\dot\theta$  in Fig.~\ref{dynamique_ind}h implies that the rotor both slips and rolls on the stator.
{\color{bleuf}Three concluding comments are in order. Firstly, we stress that unlike in most active spinner experiments, see e.g.~\cite{soni2019,tan2022,ceron2023}, here  the parity symmetry of the dynamics is spontaneously broken: the motors spin, or precess, with equal probability in the clockwise and counterclockwise directions. Secondly, we  note that  the spinning and precession regimes do not coexist. Every single rotor is either in one state or the other unlike in the starfish embryos experiments of~\cite{chao2024}.
Thirdly, given the motor geometry we are using, the range of field amplitude over which all the motors rotate and remain in the spinning regime is too narrow to systematically investigate the collective organization of interacting pure spinners.   
}

\subsection{Modulation of the precession dynamics}\label{modulation_section}
\begin{figure*}[h]
\includegraphics[width=\textwidth]{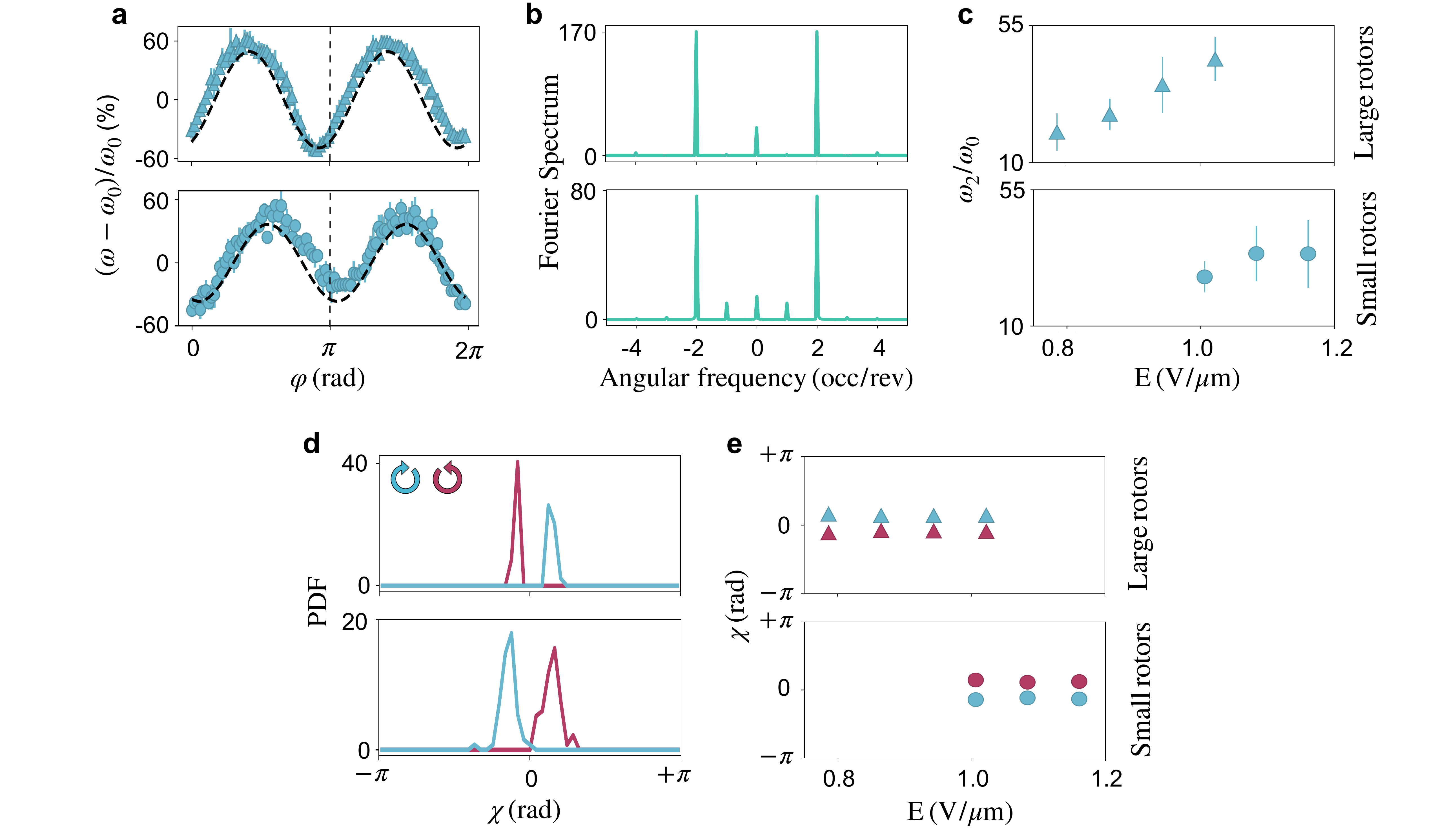}
\caption{{\bf Twofold modulation of the hula-hoop dynamics}
{\bf a.} Relative precession speed $(\omega-\omega_0)/\omega_0$ for one rotor plotted versus the orientation angle $\varphi$. 
We wrapped  the signal $\varphi(t)$ over the interval [0,2$\pi$] and  average $\omega(\varphi)$ over bins of width $2 \pi/100$. 
{\bf b.} Power spectra associated with the Fourier transform of the signals shown in {\bf a} (signal averaged over more than 50 different motors). 
{\bf c.} Evolution of the relative amplitude of the twofold modulation ($\omega_2/\omega_0$) with the field amplitude.
{\bf d.} Distribution of the twofold-modulation phase $\chi$ for motors spinning in the clockwise (blue) and counterclockwise (red) directions. 
{\bf e.} Twofold-modulation phase $\chi$ as a function of the field amplitude. $\chi$ hardly depends on $E$.
Top: Large rotors. Bottom: Small rotors.
}
    \label{modulation}
\end{figure*}

Before explaining this variety of dynamics, it is worth taking a closer look at the precession regime.
On average, the  precession angle $\varphi(t)$ increases linearly with time and is controlled by $E$ (Fig.~\ref{dynamique_ind}g). 
However the instantaneous value of $\omega=\dot\varphi(t)$ is not constant. 
Even when a rotor is isolated, $\omega$ fluctuates periodically around a constant value. $\omega$ depends on $\varphi$ (Fig.~\ref{modulation}a). 
To study this phenomenon, we conducted experiments on more than $50$ rotors. 
To minimize the interactions between the rotors, we set their distances to four diameters. We Fourier transform the  $\omega(\varphi)$ signals, and plot their power spectra in Fig.~\ref{modulation}b. 
We  find a clear twofold modulation of the precession speed regardless of the rotor size: the motors periodically accelerate and decelerate over half a rotation.
All our data are therefore well fitted by a function $\omega(\varphi)=\omega_0-\omega_2\cos(2\varphi+\chi)$, and Fig.~\ref{modulation}c shows that the relative magnitude of the modulation $\omega_2/\omega_0$  increases with $E$. 
More importantly, we find that the sign of the phase of the twofold modulation depends on the direction of rotation of the rotors. Running experiments with the large motors, we find that the sign of $\chi$ mirrors that of the rotation direction (Fig.~\ref{modulation}d). 
Rotors orbiting in the clockwise direction ($\omega$<0) are associated with a positive phase ($\chi>0$). 
Conversely, rotors orbiting in the counterclockwise direction ($\omega$>0) are associated with a negative phase ($\chi<0$). For the small rotors, it is the opposite: the sign of the phase is the same as the sign of the direction of rotation.
The value of $\chi$ does not depend on the field strength (Fig.~\ref{modulation}e).
We will show that the phase $\chi$ of the speed modulation will prove crucial to identify the origin of the precession dynamics, and of the phase coherence of motor lattices.

\subsection{Origins of the spinning and hula-hoop dynamics}\label{spinninghoolahoop_section}
Providing a quantitative and predictive theory of single motor  dynamics goes beyond the scope of our article.
On the sole basis of our measurements, we can however pinpoint the physical phenomena that rule the 
transition from spinning to hula-hoop motion with twofold speed modulation.
Before delving into our phenomenological reasoning, we summarize our main conclusions. 
In short,  at low electric field electrophoresis drives the rotor in the direction of the $\mathbf E$ field. 
Increasing the value of $E$, the Quincke instability powers rotation, and dielectrophoresis stabilizes a motion centered around the stator.
Further increasing the $E$ field, the dipolar interactions between the rotor and stator overcome the dielectrophoretic repulsion, thereby resulting in a modulated precession dynamics.

We now explain in more detail how our experimental observations hint towards this three-steps scenario:

-- At low $E$ field, below $E_{Q}$, the translation of the rotors in the direction of the field indicates that they acquire a positive charge. 
Surface charge dissociation in apolar solvent containing AOT salt is a well established phenomenon~\cite{hsu2005charge,smith2015celebrating}. 
The  drift of the rotors below $E_Q$ is a mere consequence of the net electrophoretic mobility of insulating bodies immersed in a solution of inverse micelles in an apolar solvent.

\begin{figure*}
\includegraphics[width=\textwidth]{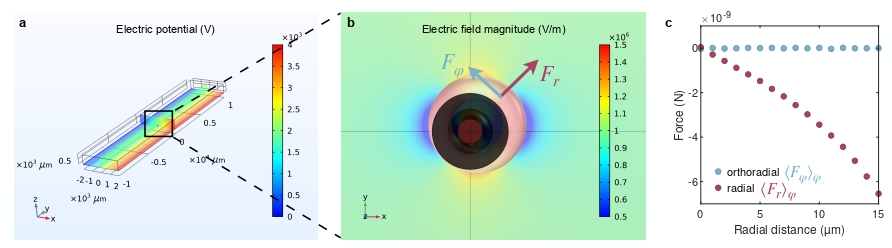}
\caption{{\bf Dielectrophoresis centers the rotor's position}
{\bf a} A single rotor immersed in hexadecane is modeled with the finite element software COMSOL under an applied voltage of $4\,\rm kV$ across $4\,\rm mm$ along the x-direction according to the experimental conditions. The relative permittivities of the SU8 rotor and hexadecane solution are $\epsilon_{\rm SU8} = 3.5$ and $\epsilon_{\rm hexa+AOT} = 2$ respectively. Their conductivities are $\sigma_{\rm hexa+AOT} = 10^{-9}\,\rm S/m \gg \sigma_{\rm SU8} = 10^{-17}\,\rm S/m$.
{\bf b} Close-up on the Quincke rotor showing heterogeneities in the electric field magnitude. This heterogeneities cause a net dielectrophoresis force on the toroidal rotor (white) when it is off-centered from the stator (black).
{\bf c} The radial component of the electric force is negative and tends to center the rotor. 
}
    \label{figureAlexandre}
\end{figure*}

-- At intermediate field magnitude, $E>E_Q$, 
another phenomenon competes with  electrophoresis  and centers the rotor position. 
The repulsion of Quincke rotors from solid surfaces has already been reported in Ref. \cite{pradillo2019quincke}.
In this article, Pradillo and co-workers show that 100 micron big Quincke spheres overcome both the effects of gravity and of electrophoresis, hovering over solid surfaces under the action of dielectrophoretic forces. 
To better establish the relevance of dielectrophoretic forces in our geometry, we solved the Maxwell equations and computed the $E$ field around an isolated  motor using a Finite Element Method (COMSOL), see (Fig. \ref{figureAlexandre}). 
We find that the rotor experiences a net radial dielectrophoretic force that stabilizes a centered spinning motion~\cite{pradillo2019quincke}. 

-- At high field amplitude, the rotors are not centered anymore, they contact the stator and undergo modulated hula-hoop motion. 
This additional transition must therefore involve a third interaction that can compete with  dielectrophoretic centering forces.
The twofold modulation of the orbiting speed   hints towards dipolar interactions.
Past the Quincke threshold, the conduction charges (the ions) accumulate at the rotor's surface and their spatial distribution forms a dipole, see Ref.~\cite{Melcher1969} and Fig.~\ref{schema_quinke}.  When the rotor spins, the charge dipole makes a finite angle with the electric field and its magnitude increases with $E$. 
The cylindrical stator also features a surface charge with a dipolar symmetry, but the orientation of the dipole remains exactly opposite to the $E$ field~\cite{Melcher1969}, see Fig.~\ref{schema_quinke}. 
In principle, increasing $E$, the magnitude of the dipole-dipole interaction between the rotor and   stator can overcome the dielectrophoretic force. 
In addition, the dipolar force depends on the relative orientation of the dipoles and on the displacement of the rotor. 
These variations are complex to derive exactly, but the dipolar symmetry implies a twofold modulation of the force with $\varphi$, which is consistent with the observed modulation of the precession speed in our experiments.

Going beyond this qualitative reasoning, would require modeling both the electrostatic and frictional interactions between the rotor and stator, taking into account their complex geometries. This  complex numerical investigation could help predicting and understanding the variations of the modulation phase with the motor size.

\newpage

  \section{Emergence of antiferromagnetic order: theory}
In this section, we investigate the coupled dynamics of Quincke rotors and explain how antiferromagnetic order emerges from transverse frictional interactions. 
We start by recalling the basic equations governing the motion of a single Quincke rotor within the leaky dielectric approximation~\cite{Melcher1969} actuated by a constant external torque. 
We then solve analytically the situation of two rotors coupled by frictional torques. 

\subsection{Quincke electro-rotation instability: a primer}
We first briefly recall how to model the dynamics of a single Quincke rotor, see for instance Ref.~\cite{Melcher1969,Peters2005} for more details.
We consider a rotor, an axisymmetric body, immersed in a conducting fluid. 
Both the solid body and the fluid are weak Ohmic conductors and have a finite electric  permittivity. 
We neglect all electrostatic double layer effects. 
Within this leaky dielectric approximation, when placed in a constant $\mathbf E$, the    electric polarization ${\bm p}$ and  rotation vector $\bm{\dot \theta}$ of the rotor obey two coupled equations:

\begin{align}
    \frac{d{\mathbf p}}{dt} &= \bm{ \dot \theta } \times ({\mathbf p} - \chi^\infty {\mathbf E}) - \frac{1}{\tau_{\rm M}} ({\mathbf p} - \chi^0 {\mathbf E}), \label{eq:QuickeP} \\
    I \frac{d\bm{ \dot \theta }}{dt} &= {\mathbf p}\times {\mathbf E} - \gamma \bm{ \dot \theta }, \label{eq:QuickeO}
\end{align}
where $\tau_M$ is the Maxwell-Wagner time, $\chi^0$ and $\chi^\infty$ are respectively the zero-frequency and high frequency susceptibilities, $I$ is the moment of inertia and $\gamma$ is a friction coefficient.
The first equation reflects the conservation of the electric charge at the surface of the particle, and the second  is the angular- momentum-conservation equation.   
In all that follows, the electric field points along the $x$ direction ${\bm E} = E \hat {\bm e}_x$, and the rotation vector  along the $z$ direction $\bm{ \dot \theta } = \dot \theta \hat {\bm e}_z$.  The polarization ${\bm p}$ vector lies in the $x$-$y$ plane.
It is convenient to recast Eqs.~\eqref{eq:QuickeP}-\eqref{eq:QuickeO}  into a dimensionless form by introducing the scaled variables
\begin{align} \label{eq:defXYZ}
    t^\ast &= \frac{t}{\tau_{\rm M}}, & \Pi^x &= \frac{\tau_{\rm M} E}{\gamma} (p_x - \chi^0 E), & \Pi^y &= -\frac{\tau_{\rm M} E}{\gamma} p_y, & \Omega &= \tau_{\rm M} \dot \theta
\end{align}
The dynamics of the Quincke rotor then take the compact form
\begin{align}
    \frac{d\Pi^x}{dt^\ast} &=\Omega\Pi^y - \Pi^x \label{eq:lorenz1} \\
    \frac{d\Pi^y}{dt^\ast} &= \Omega(\rho - \Pi^x) - \Pi^y\label{eq:lorenz2} \\
    \frac{d\Omega}{dt^\ast} &=\sigma (\Pi^y - \Omega) \label{eq:lorenz3}
\end{align}
$\sigma = \gamma\tau_M/I$ is the ratio of the Maxwell-Wagner time to the angular relaxation time, and $\rho= {\tau_M(\chi^\infty - \chi^0)E^2}/{\gamma}$ is the distance to the Quincke threshold. 
When $\chi^\infty>\chi^0$, a necessary condition for the Quincke instability, we  define $E_Q=\sqrt{\gamma/[\tau_M(\chi^\infty - \chi^0)]}$ and  $\rho \equiv (E/E_Q)^2$. 
Eqs.~\eqref{eq:lorenz1}-\eqref{eq:lorenz3} then form the well-known Lorenz system~\cite{Smale2003}. 
However, in the experimentally relevant regime, inertia plays a negligible role. 
The Maxwell-Wagner time $\tau_M$ is much larger than the angular relaxation timescale $I/\gamma$. 
In the limit $\sigma\gg 1$, we can therefore further simplify our model and replace the first dynamical equation by the identity $\Pi^y=\Omega$.
In this overdamped regime, the dynamical system that rules the dynamics of the Quincke rotors becomes two dimensional:
\begin{align}
    \frac{d\Pi^x}{dt^\ast} &= \Omega^2 - \Pi^x  \label{eq:QuinckeX}\\
    \frac{d\Omega}{dt^\ast} &= \Omega(\rho - 1 - \Pi^x) \label{eq:QuinckeZ} 
\end{align}
$\Omega=\tau_M\dot \theta$ quantifies the rotation of the rotor while $\Pi^x$ is the  perturbation of the electric polarization ${\bm p}$ along the $x$ direction. 
The stability analysis of Eq.~\eqref{eq:QuinckeX}-\eqref{eq:QuinckeZ} is straightforward. 
The quiescent stationary state $(\Pi^x, \Omega)=(0, 0)$ is stable when $\rho < 1$ and becomes unstable when $\rho > 1$. 
In this regime two stable stationary points exist at $(\Pi^x, \Omega)=(\rho-1,\pm\sqrt{\rho-1})$. This is the Quincke instability: when $\rho > 1$, i.e $E>E_Q$, the rotor is unstable to minute rotational perturbations and starts spinning spontaneously along any rotation axis transverse to the $\mathbf E$ field.

\subsection{Two coupled rotors}
\subsubsection{How do Quincke spinners respond to torques?}
\begin{figure*}[h]
\includegraphics[width=\textwidth]{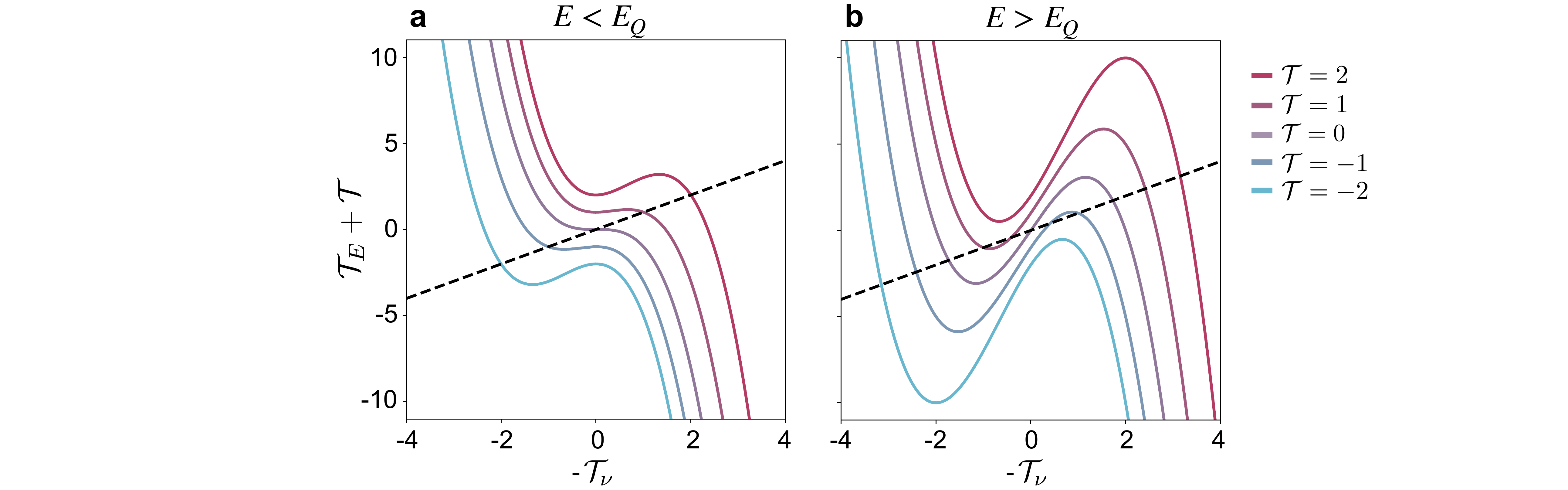}
\caption{Plots of the sum of torques $\mathcal{T}_E + \mathcal{T}$ as a function of the viscous torque $\mathcal{T}_V$, for different values of the imposed torque $\mathcal{T}$. The dashed line indicates the condition $\mathcal{T}_E + \mathcal{T} = \mathcal{T}_V$.
{\bf a.} Below the Quincke threshold $(\rho = 0.01)$, the motor rotates in the direction set by the imposed torque $\mathcal{T}$.
{\bf b.} Above the Quincke threshold $(\rho = 4)$, a sufficiently strong imposed torque can destabilize the rotation direction opposite to the one it promotes.
}
    \label{fig:Quincketorque}
\end{figure*}

Before addressing the  dynamics of coupled motors, we can gain some insight by considering a simpler situation where a single Quincke motor responds to an imposed constant torque.
The  dynamics then obeys the rescaled overdamped equations:
\begin{align}
    \frac{d\Pi^x}{dt^\ast} &= \Omega \Pi^y - \Pi^x  \label{eq:QuinckeXT}\\
    \frac{d\Pi^y}{dt^\ast} &= \Omega(\rho - \Pi^x) - \Pi^y \\
    0&=\Pi^y-\Omega+\mathcal T
    \label{eq:QuinckeZT} 
\end{align}
where $\mathcal T$ is the rescaled torque. 
The last equation represents the torque balance condition and it is worth keeping in mind that $\Pi^y\equiv\mathcal T_E$ is nothing else but the scaled electric torque, whereas $-\Omega\equiv\mathcal T_V$ is the scaled viscous torque. 
We can use the first two equations to express $\mathcal T_E$ as a function of $\Omega$ in steady state. It takes the cubic form
\begin{equation}
    \mathcal T_E=\Omega(\rho+\Omega\mathcal T-\Omega^2).
\end{equation}
The steady state and their stability are then readily deduced from the torque balance equation. In Fig.~\ref{fig:Quincketorque}, we plot the magnitude of $\mathcal T_V$ and of  $\mathcal T_E+\mathcal T$ for different values of $\mathcal T$. 
The motor dynamics is then readily inferred by a graphical analysis.
When $\mathcal T=0$, when $E>E_Q$, we recover the standard Quincke rotation result.  Two opposite values of the rotation rate satisfy the torque balance condition, and the  corresponding equilibrium positions are stable. The  situation where $\Omega=0$ corresponds to an unstable fixed point.
Applying an external torque bias the motor dynamics. Increasing $\mathcal T$ offsets the values of the two possible rotations rates, until the point where only one stable motor rotation is possible, in the direction set by the applied torque.
Below $E_Q$ applying an external torque always enable the electric actuation of the motor but solely in the direction imposed by the external torque.

This graphical analysis provides some intuition about the  dynamics of motors coupled by frictional forces. In this situation the external torque would be replaced by the interaction torque. 
When the coupling strength is vanishingly small, we of course expect the dynamics of the two motors to be decoupled. In the opposite limit where friction is very large, the first motor experiences a  large torque that promotes rotation only in the direction opposite to the neighboring motor.
However at intermediate friction values we can expect configurations where biased counter- and co-rotation states are possible.
This qualitative analysis hints towards a scenario where a finite coupling strength is required to achieve antiferromagnetic order, and that past this critical value all other states would be linearly unstable. This scenario is consistent with our experimental observations and confirmed by the detailed theory presented in the next section.

\subsubsection{Coupled dynamics of two Quincke motors: a dynamical system insight}

We now consider two Quincke rotors with polarizations ${\bm p_1}$ and ${\bm p_2}$, and rotation vectors $\bm {\dot \theta_1}$ and $\bm {\dot \theta_2}$.
We introduce a frictional  torque $-\gamma\mu\dot \theta_j$ acting on rotor $i$, where $\mu$ is a positive dimensionless number. 
This is a minimal ingredient mimicking a ``cogwheel'' effect that could be of  hydrodynamic or frictional origin. 
This minimal coupling will lead to rotations of the two rotors in opposite directions. In dimensional form, the equations of motion are
\begin{align}
        \frac{d{\bm p}_1}{dt} &= \bm {\dot \theta_1} \times ({\bm p_1} - \chi^\infty {\bm E}) - \frac{1}{\tau} ({\bm p_1} - \chi^0 {\bm E}) \notag \\
        \frac{d{\bm p}_2}{dt} &= \bm {\dot \theta_2} \times ({\bm p_2} - \chi^\infty {\bm E}) - \frac{1}{\tau} ({\bm p_2} - \chi^0 {\bm E})\notag \\
        I \frac{d\bm {\dot \theta_1}}{dt} &= {\bm p_1}\times {\bm E} - \gamma(\bm {\dot \theta_1}+\mu\bm {\dot \theta_2}) \notag\\
        I \frac{d\bm {\dot \theta_2}}{dt} &= {\bm p_2}\times {\bm E} - \gamma(\bm {\dot \theta_2}+\mu\bm {\dot \theta_1}) \notag
\end{align}
We can make these equations dimensionless by defining $(\Pi^x_1,\Pi^y_1,\Omega_1)$ and $(\Pi^x_2,\Pi^y_2,\Omega_2)$ according to Eq.~\eqref{eq:defXYZ}. 
In the limit $\sigma\gg 1$, we can eliminate $\Pi^y_1=\Omega_1+\mu \Omega_2$ and $\Pi^y_2=\Omega_2+\mu \Omega_1$. 
It is useful to recast the dynamical system in terms of the variables $\Pi^x_\pm = \frac{1}{2}(\Pi^x_1\pm \Pi^x_2)$ and $\Omega_\pm = \frac{1}{2}(\Omega_1\pm \Omega_2)$. 
\begin{align}
        \frac{d\Pi^x_+}{dt^\ast} &=(1+\mu) \Omega_+^2 + (1-\mu) \Omega_-^2 - \Pi^x_+  \label{eq:fc1}\\
        \frac{d\Pi^x_-}{dt^\ast} &= 2 \Omega_+ \Omega_- - \Pi^x_- 
        \label{eq:fc12}
        \\
            \frac{d\Omega_+}{dt^\ast} &=\frac{1}{1+\mu}\left[\left(\rho-(1+\mu)-\Pi^x_+\right)\Omega_+ - \Pi^x_- \Omega_-\right] \label{eq:fc21} 
            \\
            \frac{d\Omega_-}{dt^\ast} &= \frac{1}{1-\mu}\left[\left(\rho-(1-\mu)-\Pi^x_+\right)\Omega_- - \Omega_+ \Pi^x_-\right] \label{eq:fc2}
\end{align}
Solving for the stationary solutions yields :
\begin{align}
    \Pi^x_+&=(1+\mu) \Omega_+^2 + (1-\mu) \Omega_-^2 \label{eq:fp1}\\
    \Pi^x_-&=2\Omega_+ \Omega_-  \label{eq:fp2}\\
        0&=\Omega_+\left[\rho-(1+\mu)(1+\Omega_+^2) - (3-\mu) \Omega_-^2\right]  \label{eq:fp3} \\
        0&=\Omega_-\left[\rho-(1-\mu)(1+\Omega_-^2) - (3+\mu) \Omega_+^2\right]   \label{eq:fp4}
    \end{align}
which are satisfied if $\Omega_\pm = 0$ or if the bracket terms vanish. 
This yields four types of fixed points.
To study the linear statibily of a fixed point $(\Pi^x_+, \Pi^x_-, \Omega_+, \Omega_-)$ we need to compute  the eigenvalues of the dynamical matrix
\begin{equation}
\label{eq:Mstab}
		 M = 
            \begin{pmatrix}
                -1 & 0 & 2(1+\mu)\Omega_+ & 2(1-\mu)\Omega_- \\
                0 & -1 & 2\Omega_- & 2\Omega_+ \\
                -\frac{\Omega_+}{1+\mu} &-\frac{\Omega_-}{1+\mu} & \frac{\rho-\Pi^x_+}{1+\mu} -1& -\frac{\Pi^x_-}{1+\mu} \\
                -\frac{\Omega_-}{1-\mu} &-\frac{\Omega_+}{1-\mu} & -\frac{\Pi^x_-}{1-\mu} & \frac{\rho-\Pi^x_+}{1-\mu} -1
            \end{pmatrix}.
\end{equation}

\paragraph{\bf Non-rotating rotors.} The simplest solution of Eqs.~\eqref{eq:fp1}-\eqref{eq:fp2} is $(\Pi^x_+, \Pi^x_-, \Omega_+, \Omega_-) = (0, 0, 0, 0)$, corresponding to $(\dot \theta_1, \dot \theta_2) = (0, 0)$. 
The matrix $M$, Eq.~\eqref{eq:Mstab}, is then diagonal.
The system is stable  when $\rho < 1-\mu$ and  unstable  when $\rho>1-\mu$. 
Frictional coupling lowers the  Quincke threshold.

\paragraph{\bf Counter-rotating rotors.} If $\Omega_+ = 0, \Omega_-\neq 0$, the two rotors are counter-rotating.  
Eqs.~\eqref{eq:fp1}-\eqref{eq:fp2} then give two fixed points
\begin{equation}
        (\Pi^x_+, \Pi^x_-, \Omega_+, \Omega_-) = \left(\rho-(1-\mu), 0, 0, \pm \sqrt{\frac{\rho-(1-\mu)}{1-\mu}}\right) \notag
        \end{equation}
This type of solution exists when $\rho\geq1-\mu$, i.e. as soon as the non-rotating fixed point is unstable. This steady state configuration is always stable.

\paragraph{\bf Co-rotating rotors.} If $\Omega_- = 0, \Omega_+\neq 0$, the two rotors are co-rotating. There exists two such fixed points given by 
\begin{equation}
        (\Pi^x_+, \Pi^x_-, \Omega_+, \Omega_-) = \left(\rho-(1+\mu), 0, \pm \sqrt{\frac{\rho-(1+\mu)}{1+\mu}}, 0\right), \notag
        \end{equation}
when $\rho \geq 1+\mu$.
The four corresponding eigenvalues are 
\[
        \lambda_{1,\pm} = \frac{1}{2}\left[-1\pm \sqrt{1-8\frac{\rho-(1+\mu)}{1+\mu}}\right], 
        \quad \lambda_{2,\pm} = \frac{1}{2}\left[-\frac{1-3\mu}{1-\mu}\pm \sqrt{\left(\frac{1-3\mu}{1-\mu}\right)^2-8\frac{\rho-(1+\mu)^2}{1-\mu^2}}\right].
        \]
The fixed points are unstable for $\mu>1/3$. When $0<\mu<1/3$, they are stable if $\rho>(1+\mu)^2$. When $\mu>1/3$, the negative eigenvalue is associated with an eigenvector of the form $(\Pi^x_+, \Pi^x_-, \Omega_+, \Omega_-) =(0, a, 0, b)$.

\paragraph{\bf Unstable fixed points.} If $\Omega_\pm \neq 0$, Eqs.~\eqref{eq:fp1}-\eqref{eq:fp2} lead to four fixed points:
\begin{align}
        (\Pi^x_+, \Pi^x_-, &\Omega_+, \Omega_-) =\notag \\
        &\frac{1}{2}\left( \rho-(1-\mu)^2, \sqrt{[\rho-(1-\mu)^2][\rho-(1+\mu)^2]},\pm \sqrt{\rho-(1-\mu)^2}, \pm \sqrt{\rho-(1+\mu)^2}\right)\notag
        \end{align}
They exist only when $\rho \geq (1+\mu)^2$. 
The determinant of $M$, Eq.~\eqref{eq:Mstab}, is $\det M = -\frac{2}{1-\gamma^2}[\rho-(1-\gamma)^2][\rho-(1+\gamma)^2] < 0$. This implies that the four eigenvalues of $M$ cannot all have a negative real part. In other words, the four fixed points are always unstable.

 The stability analysis of the  two coupled motors is summarized in Fig.~2c (main text). 
 When the  frictional coupling constant  exceeds the critical value $\mu > 1/3$, and  above the Quincke threshold ($\rho > 1-\mu$), the only stable solution corresponds to counter-rotating motors. 
 All other steady states are unstable.
 This nonlinear dynamics correctly accounts for the emergence of pristine antiferromagnetic order below a finite lattice-spacing value.

 In the main text (Figure 2), we exemplify the dynamical states of the two coupled rotors, solving Eqs.~\eqref{eq:fc1}, \eqref{eq:fc12}, \eqref{eq:fc21}  and \eqref{eq:fc2} with $\rho=1.5$, $\mu=0.05$ (Low friction) and $\rho=1.5$, $\mu=0.35$ (Large friction). 

\newpage

\section{Identification of the interactions responsible for phase coherence: a minimal model}

In this section, we explain the emergence of phase coherence in our metamachines, and single out the classes  of interactions that can organize the phases of the interacting motors in space and time.

\subsection{A Minimal Model: coupled phase oscillators}
\label{section_model_theo}

Our goal is not to  model in detail all the specifics of our experiments. 
Rather, we seek to highlight the  types of dynamics and interactions that can give rise to a collective phase organization. 
To this end, we adopt a minimal approach inspired by models developed to study the synchronization of beating cilia in biological systems \cite{box2015,uchida2011,uchida2010,meng2021,kotar2013,brumley2015}.
These models have the advantage of simplifying the analysis while retaining the fundamental features of motor dynamics.

\begin{figure*}[h!]
\includegraphics[width=\textwidth]{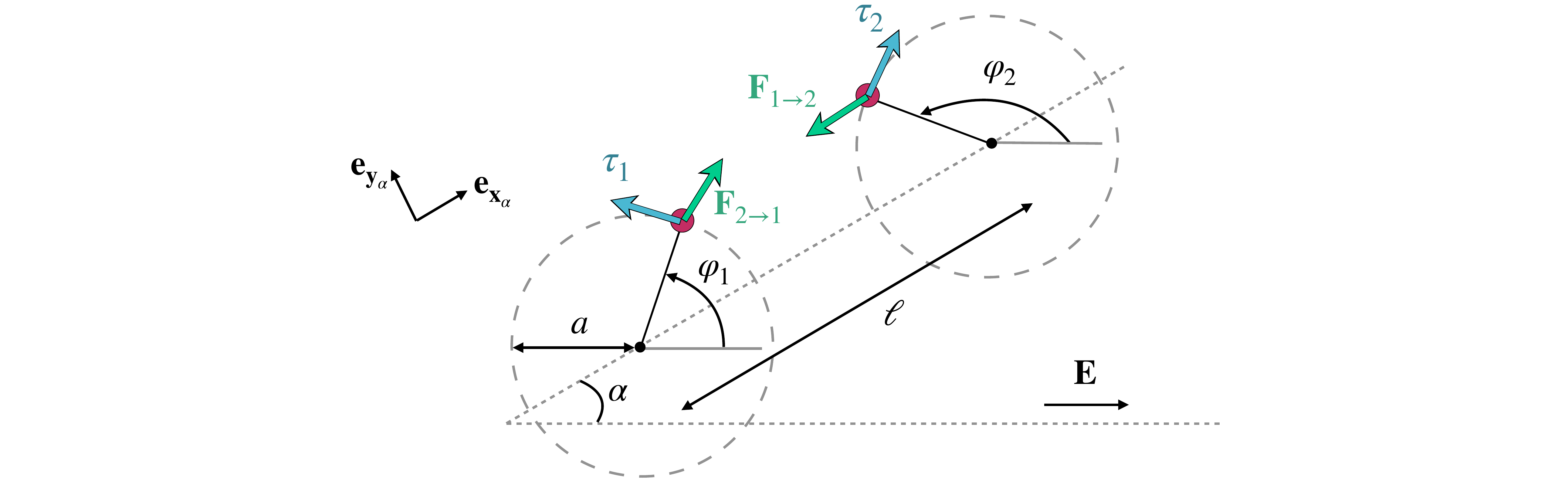}
\caption{{\bf A minimal model for interacting motors.}
    {Sketch of the minimal model used to describe the rotor dynamics. 
    The rotors are reduced to point particles orbiting along circular trajectories of radius $a$. 
    They are driven by constant torques $\tau_{1,2}$ and interact via forces $\mathbf F_{j\to i}$. The forces induce torques $\tau_{ji}$ on the particles, possibly accelerating or slowing down their orbital motion to find phase-coherent steady states.  } 
}
    \label{schema_modele}
\end{figure*}

First, we simplify the geometry of the motors and their precession dynamics by representing each rotor as a point particle following a circular orbit of radius $a$ under the action of a driving torque, see Fig.~\ref{schema_modele}. 
We then couple the particles by considering, one at a time, all possible interaction torques relevant to our experiments.  
In the overdamped limit, the torque balance equations govern the rotors' dynamics:
\begin{equation}
\label{bilan_force_1eq}
0 = -\gamma a \dot\varphi_i + \tau_{i} + \sum_{j\neq i}\tau_{ji}
\end{equation}
where
$\gamma$ is a friction coefficient,  $\tau_i$ is  the  driving torque, and $\tau_{ij}$ is the torque exerted by rotor $i$ on rotor $j$. In all that follows, we focus on nearest neighbor interactions.
We also note $\mathbf{r}_i$ the position of rotor $i$, and $\ell$ the lattice spacing. 

\subsection{Precession at constant speed: Dipolar interactions induce  phase coherence}
\begin{figure*}[h!]
\includegraphics[width=\textwidth]{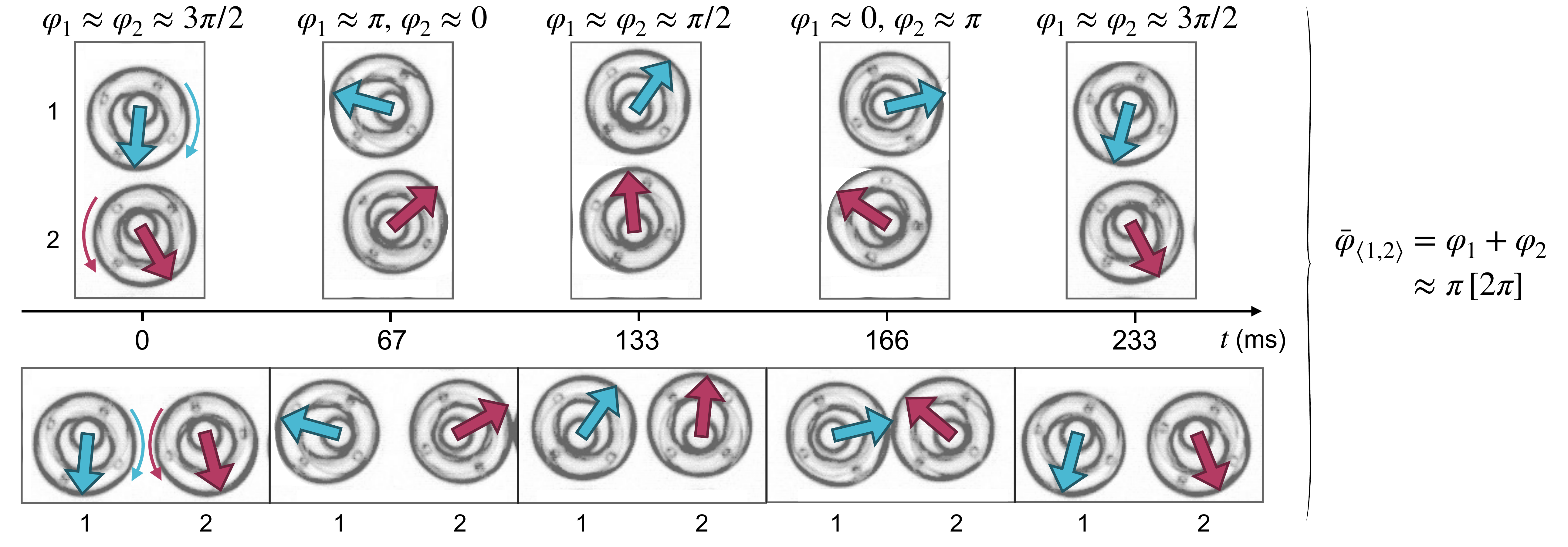}
\caption{{\bf Kinematics of  a pair of rotors}
Two series of snapshots showing the  rotor configurations when $\mathbf r_{12}$ points  in the direction transverse (top row)  and parallel (bottom row)  to $\mathbf E$.
In both cases $\bar{\varphi}$ is equal to $\pi$ but the rotor kinematics are markedly different.  }
\label{cinematique_experience}
\end{figure*}

We first disregard the modulation of the precession speed measured for isolated motors and take $\tau_1=-\tau_2=\tau_0$.
In our experiments  the rotors are in principle coupled by three  types of interactions:
(i) radial repulsive forces: these include both contact interactions, and electrostatic repulsion due to the net charge carried by the rotors.
(ii) hydrodynamic interactions.
 (iii) dipole-dipole interactions: these are induced by the electric dipoles carried by the motors, and which drive the Quincke instability.
All three types of interactions could organize the phase dynamics. 
However, in our system, the sum of the phases $\bar{\varphi}_{\langle k,l \rangle}$ is uniform and equal to $\pi$ in both lattice directions.
Crucially, a uniform $\bar \varphi$ corresponds to different kinematics in the horizontal and vertical directions  (Fig.~\ref{cinematique_experience}). 
We therefore conclude that  the interaction responsible for phase organization must be anisotropic. 

Radial forces, in the absence of modulation, are isotropic. They cannot account for the distinct behaviors observed along the principal directions of the lattice. Similarly, hydrodynamic interactions—mediated by a homogeneous and isotropic solvent—respect rotational symmetry and thus cannot explain the experimentally observed anisotropic dynamics.
The only interactions that are explicitly anisotropic are the dipole-dipole interactions. 
 Dipolar interactions are attractive along the $E$-field direction and repulsive in the perpendicular direction (Fig.~\ref{figure_dipolaire_2corps}a). 
 They are therefore the only interactions capable of producing direction-dependent kinematics. They naturally emerge as the leading candidate to explain our experimental observations.
To confirm this reasoning, we simulate our  model.

\subsubsection{Dipolar Interactions}

\begin{figure*}[h!]
\includegraphics[width=\textwidth]{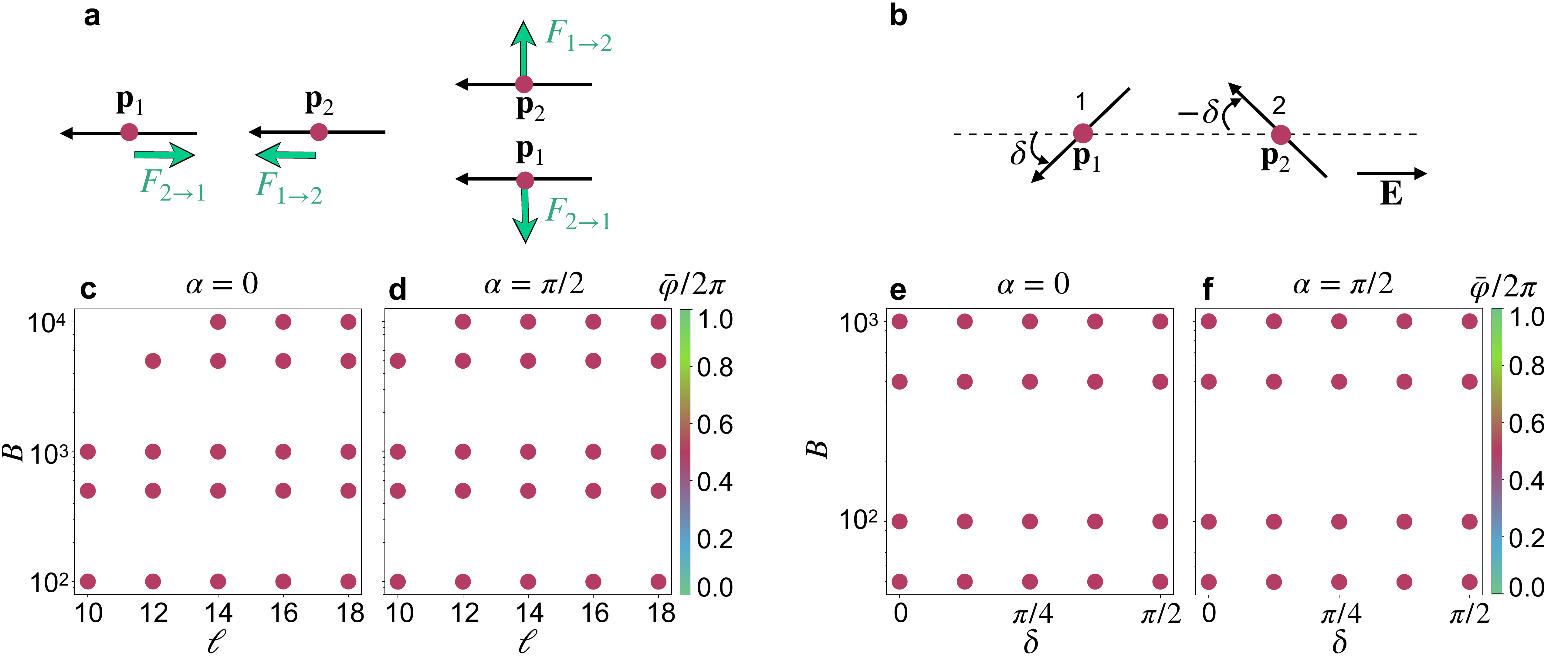}
\caption{{\bf Phase coherence induced by dipolar interactions.}
{\bf a.} Sketch of the anisotropic interactions between two point dipoles. Two identical dipoles attract when placed on the polarization axis and repel each other in the transverse direction. 
{\bf b.} Definition of the $\delta$ angle.
{\bf c-f.} Plot of the phase value $\bar{\varphi}$ computed for a pair of point rotors interacting via dipolar interactions (Eqs.~\ref{eq_dipolaire_2corps_1} and ~\ref{eq_dipolaire_2corps_2}). $\bar{\varphi}$ is averaged over the last 500 time steps in steady state, using four different initial conditions. 
When $\mathbf r_{12}$ points {\bf c.} along the direction of the dipole $\mathbf p$ ($\alpha=0$), {\bf d.} transverse to the dipole $\mathbf p$ ($\alpha=\pi/2$) the interactions promote a phase-coherent state with $\bar{\varphi}=\pi$ for all values of $B$ and $\ell$.
{\bf e-f.} Same observations for {\bf e.} $\alpha=0$ and {\bf f.} $\alpha=\pi/2$, varying the dipole orientation $\delta$ and magnitude $B$.
{\bf c-d.} $\delta=0.34$~rad, {\bf e-f.} $\ell=10$.
}
\label{figure_dipolaire_2corps}
\end{figure*}

In order to confirm our reasoning based on symmetry considerations, we focus on a simple setting where a pair of pointwise rotors orbit in opposite directions and interact via dipolar interactions.
Above the Quincke threshold, the rotor dipoles make a finite angle $\delta$ with the electric field $\mathbf{E}$, and  the sign of $\delta$ sets the rotation direction of the particle \cite{Bricard2013} (Fig.~\ref{figure_dipolaire_2corps}~{\bf b}). 
Accordingly, we assume that the first micromotor, rotating counterclockwise, carries a dipole $\mathbf{p_1}$ with angle $\delta > 0$, whereas the second micromotor, rotating clockwise, carries a dipole $\mathbf{p_2}$ with angle $-\delta$, with $\lvert \mathbf{p_1} \rvert = \lvert \mathbf{p_2} \rvert$.
The interaction potential between the two particles located at $\mathbf{r_1}$ and $\mathbf{r_2}$ is then given by:
\begin{equation}
U \propto -\frac{1}{r_{21}^3} \left[ 3(\mathbf{p_1} \cdot \hat{\mathbf{r}}_{21})(\mathbf{p_2} \cdot \hat{\mathbf{r}}_{21}) - \mathbf{p_1} \cdot \mathbf{p_2} \right],
\end{equation}
where  $\mathbf{r}_{21} = \mathbf{r}_2 - \mathbf{r}_1$.
Deriving the interaction forces and torques, and choosing units so that  $\tau_0/\gamma=1$ and $a=1$, we find the  equations of motion:
\begin{align}
\label{eq_dipolaire_2corps_1}
\dot{\varphi}_1 &= 1 - \frac{B}{r_{21}^{ 4}} \Big[ (\hat{\mathbf{p}}_2 \cdot \hat{\mathbf{r}}_{21}) \hat{\mathbf{p}}_1 + (\hat{\mathbf{p}}_1 \cdot \hat{\mathbf{r}}_{21}) \hat{\mathbf{p}}_2 \notag \\
&\quad - 5 (\hat{\mathbf{p}}_2 \cdot \hat{\mathbf{r}}_{21})(\hat{\mathbf{p}}_1 \cdot \hat{\mathbf{r}}_{21}) \hat{\mathbf{r}}_{21} + (\hat{\mathbf{p}}_1 \cdot \hat{\mathbf{p}}_2) \hat{\mathbf{r}}_{21} \Big] \cdot \hat{\mathbf{e}}_{\varphi_1}
\end{align}
\begin{align}
\label{eq_dipolaire_2corps_2}
\dot{\varphi}_2 &= -1 + \frac{B}{r_{21}^{4}} \Big[ (\hat{\mathbf{p}}_2 \cdot \hat{\mathbf{r}}_{21}) \hat{\mathbf{p}}_1 + (\hat{\mathbf{p}}_1 \cdot \hat{\mathbf{r}}_{21}) \hat{\mathbf{p}}_2 \notag \\
&\quad - 5 (\hat{\mathbf{p}}_2 \cdot \hat{\mathbf{r}}_{21})(\hat{\mathbf{p}}_1 \cdot \hat{\mathbf{r}}_{21}) \hat{\mathbf{r}}_{21} + (\hat{\mathbf{p}}_1 \cdot \hat{\mathbf{p}}_2) \hat{\mathbf{r}}_{21} \Big] \cdot \hat{\mathbf{e}}_{\varphi_2}.
\end{align}
Here, $B$ is a constant that sets the strength of the dipolar interactions.
We  numerically solve these equations of motion for particle pairs aligned either along the horizontal ($\alpha = 0$) or vertical ($\alpha = \pi/2$) directions. 
Varying the values of the coupling strength $B$, the interparticle distance $\ell$ and dipole angle $\delta$, we systematically find that the dipolar interactions drive the dynamics of the rotor towards a phase-coherent state where $\bar{\varphi}=\pi$, in agreement with our experimental observations (Figs.~\ref{figure_dipolaire_2corps}~{\bf c–f}).

This already suggests that dipole-dipole interactions govern the phase dynamics of the motors. 

\subsubsection{Radial Interactions}
\begin{figure*}[h!]
\includegraphics[width=\textwidth]{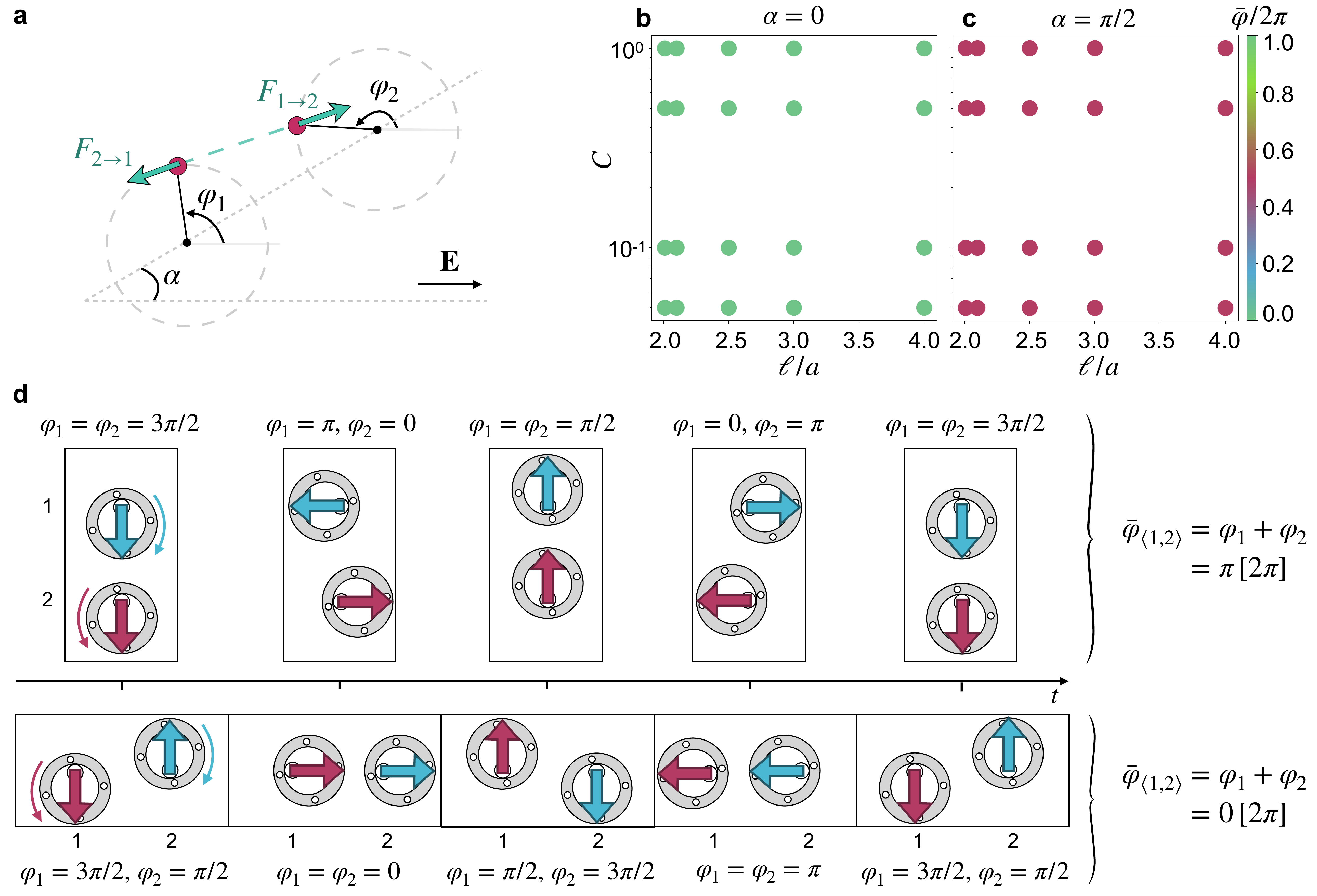}
\caption{{\bf Phase coherence induced by repulsive radial interactions.}
{\color{bleuf}{\bf a.} Geometry of the interacting particles and of the repulsive forces $\mathbf F_{i\rightarrow j}$ leading to the interaction torques $\tau_{ij}$.
{\bf b, c.} Phase diagrams showing the steady-state value of $\bar{\varphi}$ as a function of $\ell/a$ and $C$.
Panel {\bf b.} corresponds to $\alpha = 0$ (horizontal configuration), while panel {\bf c.} corresponds to $\alpha = \pi/2$ (vertical configuration).
For each diagram, the value of $\bar{\varphi}$ is averaged over the last 500 data points, using four different initial conditions.}
{\bf d.} Sketch of the rotor configurations that would result from repulsive interactions, for a horizontal  ($\bar{\varphi} = 0$) and a vertical pair ($\bar{\varphi} = \pi$).
In both cases, the rotor kinematics are identical.}
\label{figure_radiale_2corps}
\end{figure*}

In this section we show that different phase-coherence patterns can arise from repulsive interactions between the rotors (Fig.~\ref{figure_radiale_2corps}a).
As discussed earlier,  radial interactions are isotropic, and therefore cannot account for the anisotropic dynamics observed experimentally (for unmodulated drives). 
However, we show below that repulsive forces can induce phase coherence. 
While they are not dominant in our system, they could play a significant role in other realizations of metamachines.
{\color{bleuf}The interaction torque $\tau_{12}$ takes  generic form $\tau_{12}=f(r_{12})(\mathbf r_1\times \hat{\mathbf r}_{21})$, where the magnitude of the force $f$ is a decaying function of the distance $r_{21}$. 
In our simulations $f(r_{21})\propto1/r_{21}^2$.
We find that the equations of motion  take the form:
\begin{align}
\dot\varphi_1 &= 1 - \frac{C}{r_{21}'^3}\Big(\frac{\ell}{a}\sin{(\alpha-\varphi_1)}+\sin{(\varphi_2-\varphi_1)\Big)} \label{radial1} \\
\dot\varphi_2 &= -1 + \frac{C}{r_{21}'^3}\Big(\frac{\ell}{a}\sin{(\alpha-\varphi_2)}+\sin{(\varphi_2-\varphi_1)\Big)} \label{radial2},
\end{align}
where $C$ a positive constant controlling the amplitude of the interactions, and 
\begin{align}
r_{21}' &= \Bigl(\big[\frac{\ell}{a}\cos{\alpha}+(\cos \varphi_2-\cos \varphi_1)\big]^2+\big[\frac{\ell}{a}\sin{\alpha}+(\sin \varphi_2-\sin \varphi_1)\big]^2\Big)^{1/2}.
\label{eq_r12'}
\end{align}
 We numerically solve these two equations, and find that for all  values of $C$ and $\ell/a$ the rotors self-organize their phases with steady-state values so that $\bar{\varphi} = 0$ for  horizontal pairs ($\alpha = 0$) and $\bar{\varphi} = \pi/2$ for  vertical pairs ($\alpha = \pi/2$), see Fig.~\ref{figure_radiale_2corps}b and ~\ref{figure_radiale_2corps}c.
We  conclude that radial forces can induce phase coherence. However, unlike in our experiments, they would result in identical kinematics for both horizontal and vertical rotor pairs, as illustrated in Fig.~\ref{figure_radiale_2corps}d. }

\subsubsection{Hydrodynamic interactions}
Guided by earlier works on hydrodynamic synchronization in driven colloid lattices, we have systematically explored the effect of the far-field flow on phase coherence~\cite{box2015,uchida2011,uchida2010,meng2021,kotar2013,brumley2015}. 
Using the same approximations as in the previous section, we find that neither the Stokeslet nor the rotlet singularities can induce any form of phase coherence when the driving torque is constant~\cite{uchida2011}. 
{\color{bleuf} We have also used a minimal model for near-field hydrodynamics interactions, that again showed no synchronisation phenomena when the driving torques are constant. We do not report these negative results here, but detail the effect of near-field forces below, when we address the role of drive- modulations on hydrodynamic interactions (Section 3).}

\subsection{Twofold-modulated drive: impact of hydrodynamic interactions on phase coherence}
{\color{bleuf}Earlier studies have shown that hydrodynamic interactions can effectively synchronize the phase of driven colloids and beating microcilia when their oscillation pattern is modulated~\cite{box2015,uchida2011,uchida2010,meng2021,kotar2013,brumley2015}.
In section \ref{modulation_section} we have shown that the fluctuations of the rotation speed  has a two-fold symmetry.
In order to determine whether this modulated dynamics can impact the phase coherence of our motors we resort to our numerical model. 
We address the effect of a twofold modulation of the driving torque $\tau_0$ considering the effect of all possible interactions separately.}


In short, in the following sections, we show that the fluid flows induced by the modulated precession of our motors could self-organize their phases. 
However, comparisons between our numerical results and our experiments performed on different rotor sizes will rule out hydrodynamic interactions as the primary source of phase ordering. 
The conclusion of our detailed analysis remains unchanged: the phase coherence of our metamachines should be chiefly controlled by the dipolar interactions between the Quincke motors.

{\color{bleuf}Solving the dynamical equations for $\varphi_1$  and $\varphi_2$ when the forces are either radial  or dipolar, we find no effect of the drive modulation on the phase coherence. We therefore do not report these negative results and focus on  hydrodynamic interactions only (rotlet, Stokeslet and near-field  flows).
We investigate  the impact of  drive modulations on one type of hydrodynamic interaction at a time.}

\subsubsection{Twofold modulation of the driving torques}

We consider twofold modulation of the driving torques informed by the characterization of the single-rotor dynamics reported in Section~\ref{modulation_section}. 
The modulation of the mean torque $\tau_0$ is characterized by its amplitude $A_2$ and phase $\chi$. For a rotor pair:
\begin{align}
\tau_{1} &= \tau_0\left[1 - A_2  \cos{(2\varphi_1 + \chi_1)}\right] \label{modulation_1} \\
\tau_{2} &= -\tau_0\left[1 - A_2  \cos{(2\varphi_2 + \chi_2)}\right]  \label{modulation_2}
\end{align}
where the modulation phases are opposite: $\chi_1=-\chi_2$.
In our experiments we found that the sign of $\chi$ can change as the motor size increases. For small rotors ($d=86\,\rm \mu m$) $\chi_1=0.41\pm0.04$~rad, for large rotors  ($d=144\,\rm \mu m$) $\chi_1 =-0.37\pm0.12$~rad.
In all that follows, we set $A_2 = 0.25$ and $\lvert \chi_1 \rvert=0.44$~rad. We checked that none of our conclusions depend on this arbitrary choice.

\subsubsection{Hydrodynamic interactions: Rotlet flows}

\begin{figure*}[h!]
\includegraphics[width=\textwidth]{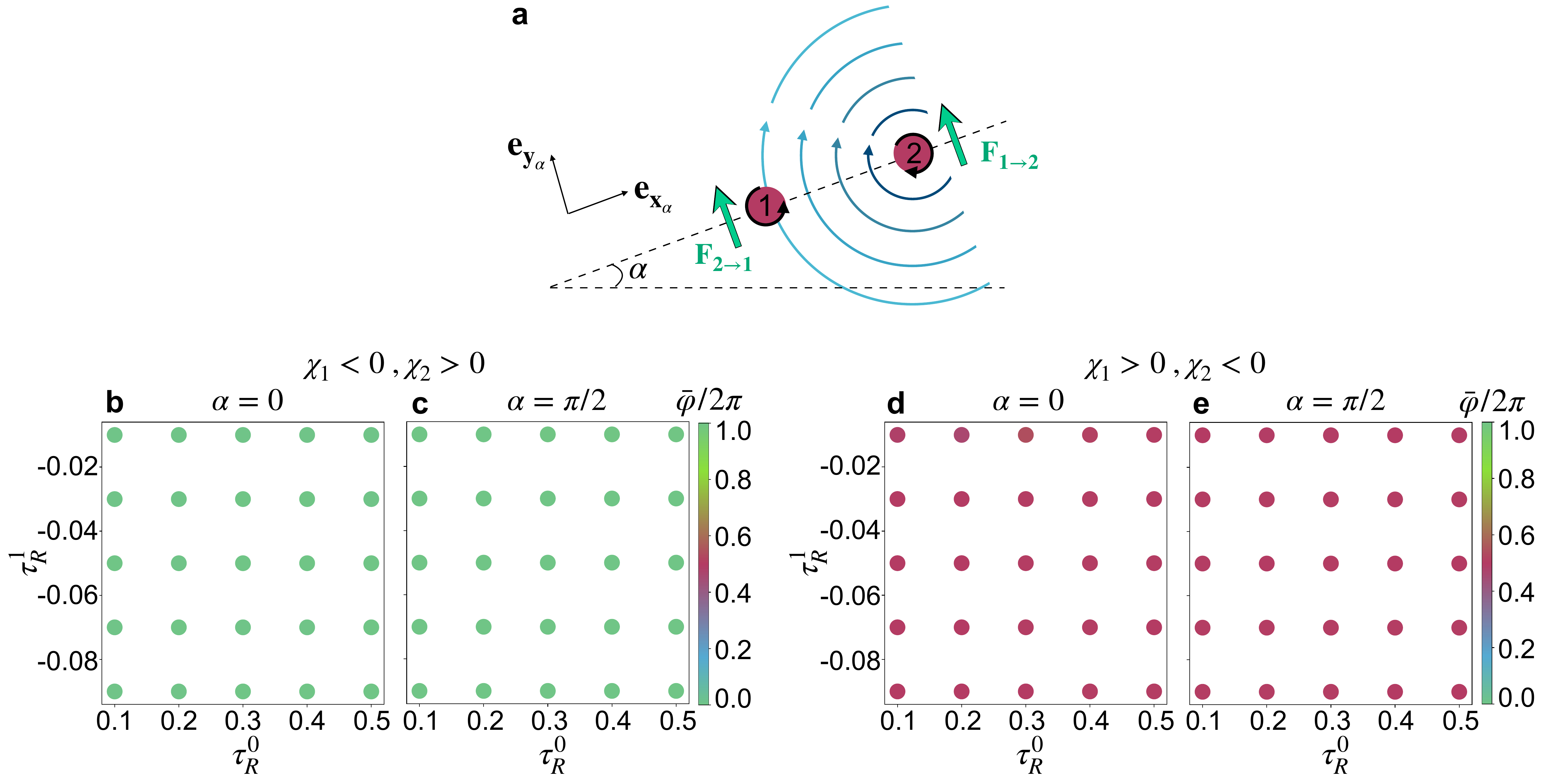}
\caption{{\bf Results of Rotlet-type hydrodynamic interactions with modulated driving forces.}
{\bf a.} Sketch of the rotlet flows and forces $\mathbf F_{i\rightarrow j}$ leading to the interaction torques $\tau_{ij}$.
{\bf b-e.} Phase diagrams of the steady-state value of $\bar{\varphi}$ as a function of $f_R^0$ and $f_R^1$.
Panels {\bf b} and {\bf d} correspond to $\alpha = 0$, and panels {\bf c} and {\bf e} to $\alpha = \pi/2$.
In each diagram, the value of $\bar{\varphi}$ is averaged over 500 time points and four different initial conditions.}
\label{figure_rotlet_2}
\end{figure*}

As the rotors spin and orbit, in the far field they are modeled by two essential singularities: a rotlet and a Stokeslet. 
We inspect the role of these two flows separately starting with the rotlet.
Within a standard far-field approximation ($\ell/a\gg1$) the rotlet flow points in the direction transverse to $\mathbf r_{12}$ (Fig.~\ref{figure_rotlet_2}a).
At leading orders, the interaction torque $\tau_{21}$ takes the  form $\tau_{21}(\varphi_1,\varphi_2,\alpha)=\left(\tau_R^0+\tau_R^1\delta r\right)\cos(\alpha-\varphi_1)$, with 
\begin{equation}
\delta r=|\mathbf r_{12}-\ell|=\left[(\cos\varphi_2 - \cos\varphi_1)\cos(\alpha)
+ (\sin\varphi_2 - \sin\varphi_1)\sin(\alpha)\right],
\label{eq:r12}
\end{equation} 

The resulting equations of motion hence thus read:
\begin{align}
\dot\varphi_1 &= 1 - A_2 \cos{(2\varphi_1 + \chi_1)} \notag \\
&+ \left\{ \tau_R^0 + \tau_R^1 \left[ (\cos\varphi_2 - \cos\varphi_1)\cos(\alpha) + (\sin\varphi_2 - \sin\varphi_1)\sin(\alpha) \right] \right\} \cos{(\alpha - \varphi_1)} \label{rotlet1} \\
\dot\varphi_2 &= -1 + A_2 \cos{(2\varphi_2 + \chi_2)} \notag \\
&+ \left\{ \tau_R^0 + \tau_R^1 \left[ (\cos\varphi_2 - \cos\varphi_1)\cos(\alpha) + (\sin\varphi_2 - \sin\varphi_1)\sin(\alpha) \right] \right\} \cos{(\alpha - \varphi_2)} \label{rotlet2}
\end{align}

We solve these equations numerically and compute the time average of $\bar{\varphi}$ in steady state. 
Figures.~\ref{figure_rotlet_2}b-e show that the twofold modulation of the driving torques results in phase coherence.
However, unlike in our experiments, we find that the asymptotic value of $\bar\varphi$ strongly depends on the sign of the $\chi_1=-\chi_2$. 
This essential difference rules out the rotlet-mediated interactions as the cause of phase organization in our metamachines.

\subsubsection{Hydrodynamic interactions: Stokeslet flows}
\begin{figure*}[h!]
\includegraphics[width=\textwidth]{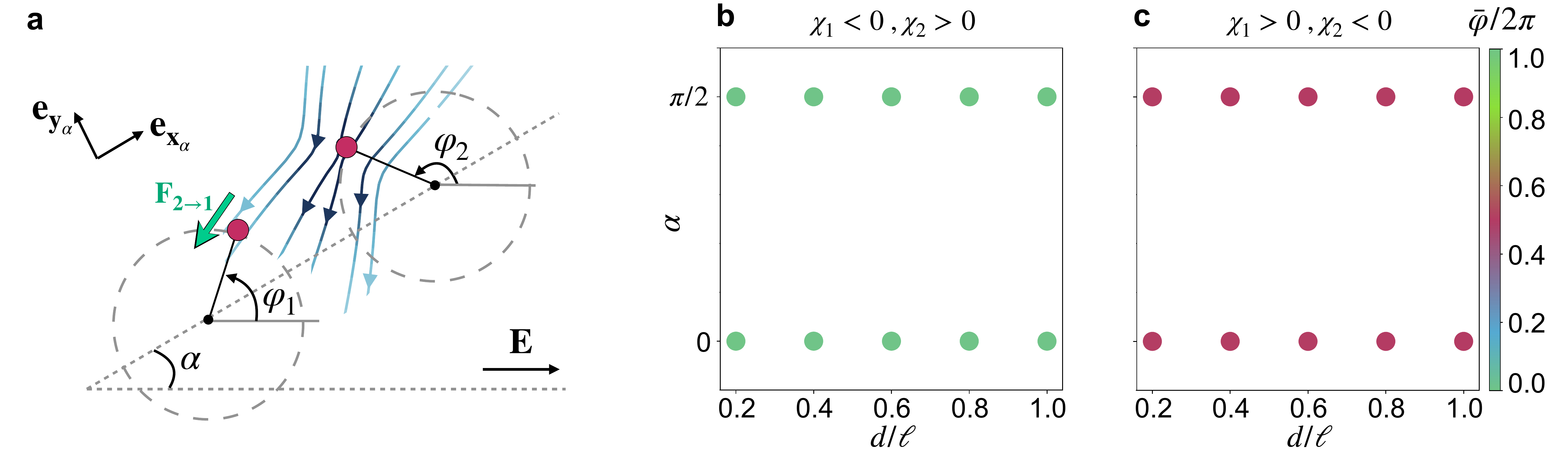}
\caption{{\bf Results of Stokeslet-type hydrodynamic interactions with modulated driving forces.}
{\bf a.} Sketch of the Stokeslet flows and forces $\mathbf F_{i\rightarrow j}$ leading to the interaction torques $\tau_{ij}$.
{\bf b-c.} Phase diagrams of the steady-state value of $\bar{\varphi}$ as a function of $d/\ell$ and $\alpha$.
In each diagram, the value of $\bar{\varphi}$ is averaged over 500 time points and four different initial conditions.}
\label{figure_stokeslet_2}
\end{figure*}

We now examine the role of Stokeslet flows (Fig.~\ref{figure_stokeslet_2}a) that have been extensively studied in previous works. 
Adapting the model detailed in Ref.~\cite{box2015} in the so-called rigid limit, we find the equations of motion for the two phase variables:
\begin{align}
\dot\varphi_1 &= 1 - A_2 \cos{(2\varphi_1 + 2\alpha + \chi_1)} \notag \\
&+ \frac{3d}{16\ell} \left( -1 + A_2 \cos{(2\varphi_2 + 2\alpha + \chi_2)} \right) \left( 3 \cos{(\varphi_1 - \varphi_2)} - \cos{(\varphi_1 + \varphi_2)} \right) \\
\dot\varphi_2 &= -1 + A_2 \cos{(2\varphi_2 + 2\alpha + \chi_2)} \notag \\
&+ \frac{3d}{16\ell} \left( 1 - A_2 \cos{(2\varphi_1 + 2\alpha + \chi_1)} \right) \left( 3 \cos{(\varphi_1 - \varphi_2)} - \cos{(\varphi_1 + \varphi_2)} \right).
\end{align}
As before, we solve these coupled nonlinear equations numerically and compute the steady-state value of $\bar\varphi$.
Our conclusions are identical to those drawn in the previous section (rotlet flows).
Stokeslet flows can stabilize a phase-coherent dynamics under modulated driving (Figs.~\ref{figure_stokeslet_2}b--c). 
We also studied in details the influence of the Stokeslet strength  and found that it merely affects the convergence time towards the phase-coherent states.
In all cases, we find that unlike in our experiment, the value of $\bar\varphi$ strongly depends on the sign of $\chi_1=-\chi_2$.
We therefore draw the same conclusions: Stokeslet-mediated interactions alone cannot explain the phase organization observed in our metamachines.

Since our experimental results indicate $\bar{\varphi} = \pi$ regardless of motor size, we conclude that although hydrodynamic interactions can induce phase coherence in the presence of modulation, they can only play a secondary role in the observed phase organization.

\begin{figure*}[h!]
\includegraphics[width=\textwidth]{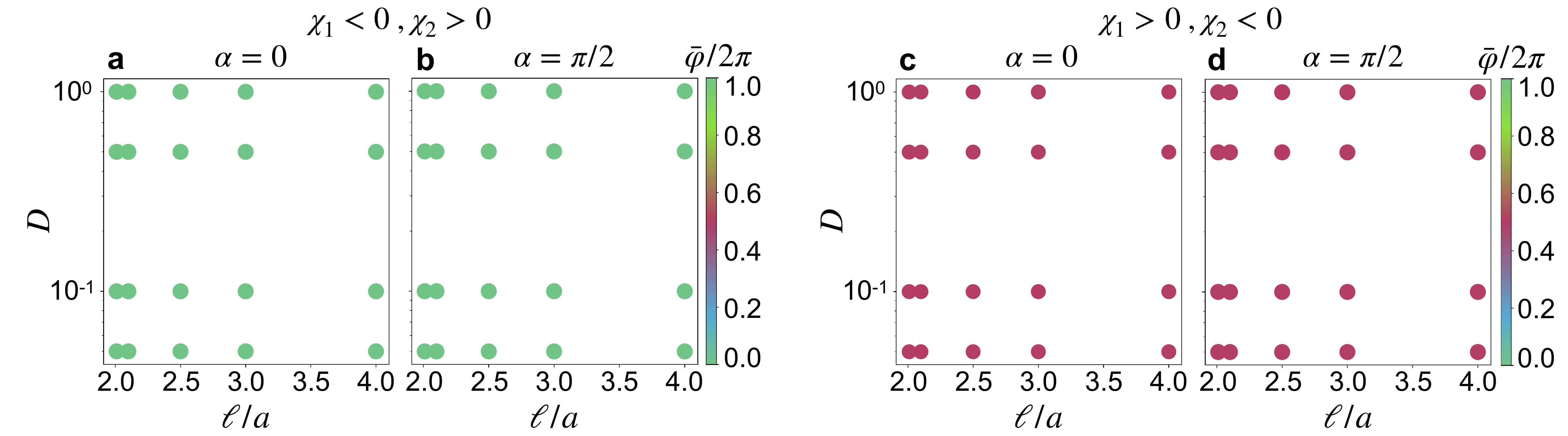}
\caption{{\color{bleuf}{\bf Results of Near-field hydrodynamic interactions with modulated driving forces.}
Phase diagrams of the steady-state value of $\bar{\varphi}$ as a function of $a/\ell$ and $D$. Panels {\bf a} and {\bf c} correspond to $\alpha=0$, and panels {\bf b} and {\bf d} to $\alpha=\pi/2$.
In each diagram, the value of $\bar{\varphi}$ is averaged over 500 time points and four different initial conditions.}}
\label{figure_near_field}
\end{figure*}

{\color{bleuf}\subsubsection{Near-field Hydrodynamic interactions}
In addition to the effect of the far field flows (rotlet and stokeslet), we also consider the effect of near-field hydrodynamics interactions. 
An accurate model of these lubrication interactions goes beyond the scope of our research.
However,  we can gain some valuable insight based on a minimal model where the near-field  lubrication forces are proportional to the velocity difference of the two rotors, with a drag coefficient $\xi$ that decays rapidly with the inter-rotor distance~\cite{kim2013microhydrodynamics}.
The resulting
interactions torques take the form $\tau_{21}=\xi( r_{21} ) (\dot \varphi_2\mathbf{e_{\varphi}}_2-\dot \varphi_1\mathbf{e_{\varphi}}_1). \mathbf{e_{\varphi}}_1$, where $\mathbf{e_{\varphi}}_i$ is the tangent vector of the trajectory for the i-th rotor. We take arbitrarily $\xi( r_{21} )\propto 1/r_{21}$.
The equation of motion for the two phase variable are then given by:

\begin{align}
    \dot \varphi_1\left(1+\xi( r_{21} )\right)&=1-A_2 \cos{(2\varphi_1+\chi_1)}+\xi( r_{21} )\dot \varphi_2\cos{(\varphi_2-\varphi_1)}\\
    \dot \varphi_2\left(1+\xi( r_{21} )\right)&=-1+A_2 \cos{(2\varphi_2-\chi_2)}+\xi( r_{21} )\dot \varphi_1\cos{(\varphi_2-\varphi_1)}
\end{align}

Then writing these equations for the variables $\bar \varphi=\varphi_1+\varphi_2$ and $\delta \varphi=\varphi_2-\varphi_1$, we obtain:
\begin{align}
    \dot{ \bar \varphi} \big[1+D/ r_{21}'-\cos{(\delta \varphi)}D/ r_{21}'\big]&=-2A_2 \sin{\big(\bar\varphi+(\chi_1+\chi_2)/2\big)}\sin{\big(\delta\varphi+(\chi_2-\chi_1)/2)\big)}\\
    \dot{ \delta \varphi} \big[1+D/ r_{21}'+\cos{(\delta \varphi)}D/ r_{21}'\big]&=-2+2A_2 \cos{\big(\bar\varphi+(\chi_1+\chi_2)/2)\big)}\cos{\big(\delta\varphi+(\chi_2-\chi_1)/2\big)}
\end{align}
where $r_{21}'$ is defined in Eq.\ref{eq_r12'} and $D$ is a positive constant that control the strength of the interactions.

As before, we solve these equations numerically and compute the steady-state value of $\bar\varphi$. Our conclusions are identical to those drawn for  far-field  interactions (rotlet and stokeslet). 
In all cases, we find that unlike in our experiment, the value of $\bar\varphi$ strongly depends on the sign of $\chi_1$ and $\chi_2$ (Figs.~\ref{figure_near_field}a--d). We therefore draw the same conclusions: our simplified model of near-filed hydrodynamics interactions alone cannot explain the phase organization observed in our metamachines.
}

\subsection{How to shape the phase-coherent states of phase-oscillator lattices}
\begin{figure*}[h!]
\includegraphics[width=\textwidth]{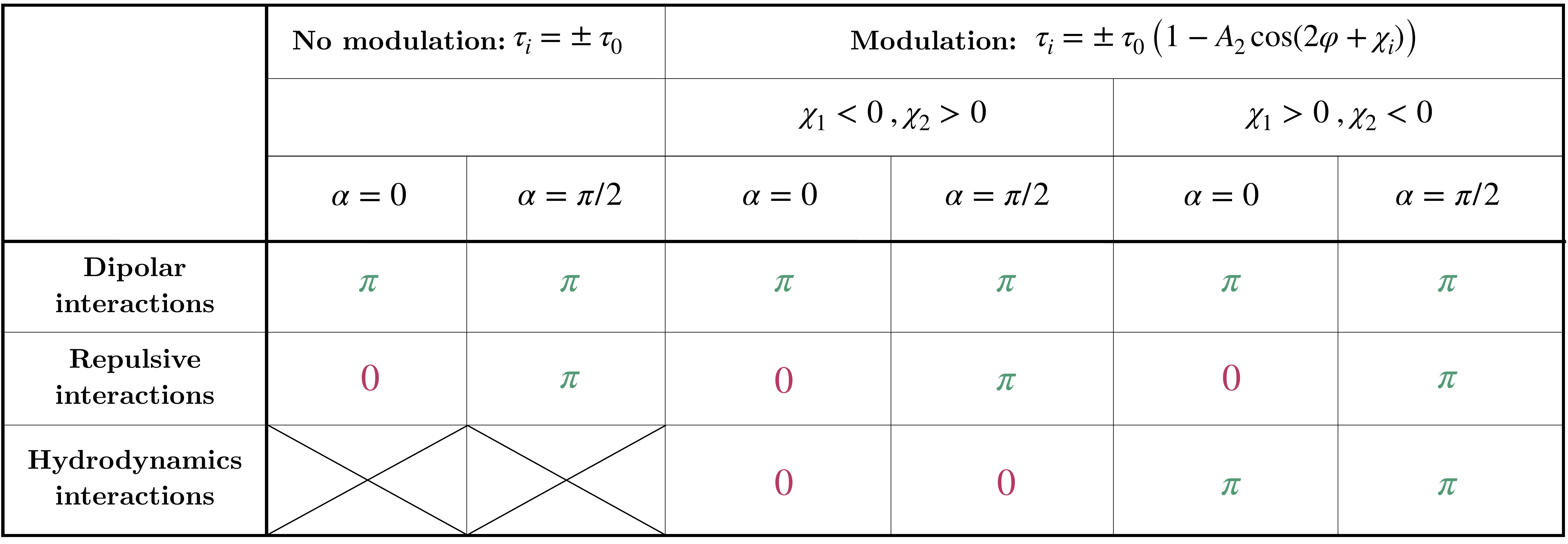}
\caption{{\bf Summary table of the effect of different interactions on phase coherence.}
This table summarizes, for each interaction type and parameter set considered, the steady-state value of $\bar{\varphi}$ reached in the stationary regime.}
\label{tableau_recap}
\end{figure*}

We close this section  with a more general discussion on the phase coherent states accessible to metamachines. 
We focus on coupled motors whose directions of rotation self-organize into an ordered antiferromagnetic state, as in our experiments. 
Beyond the specifics of our system, a broader question is: what type of phase-coherent state can be achieved when  phase oscillators interact.
Having microscopic systems in mind we have investigated the self-organization that can result from  longitudinal  repulsive forces, transverse forces (Rotlets), long-range hydrodynamic interactions (Stokeslet) and dipolar interactions. We have systematically considered the cases of constant and twofold-modulated drives. We summarize our main findings in the table Fig.~\ref{tableau_recap}.
We essentially find two types of phase coherent states. 
(i) States where $\bar\varphi$ is homogeneous. 
They correspond to different phase kinematics along the principal directions of the lattice. They require either anisotropic drives or interactions. In all cases we find that the only possible phase coherent states correspond to $\bar\varphi=0$ or $\bar\varphi=\pi$ thereby resulting in the kinematics illustrated in Figure~\ref{cinematique_experience}.
(ii) States where $\bar\varphi$ takes two different values along the $x$ and $y$ axes. In all our simulations we find that $\bar\varphi$ measured along the $x$ and $y$ axes can only take the values $(0,\pi)$. 
This leads to the isotropic phase kinematics depicted in Figure~\ref{figure_radiale_2corps}.

Our results suggest the possibility of directing the self-organization of metamachines into desired phase-coherent states by tuning the strength and symmetry of the dominant interactions.

\begin{figure*}[h!]
\includegraphics[width=\textwidth]{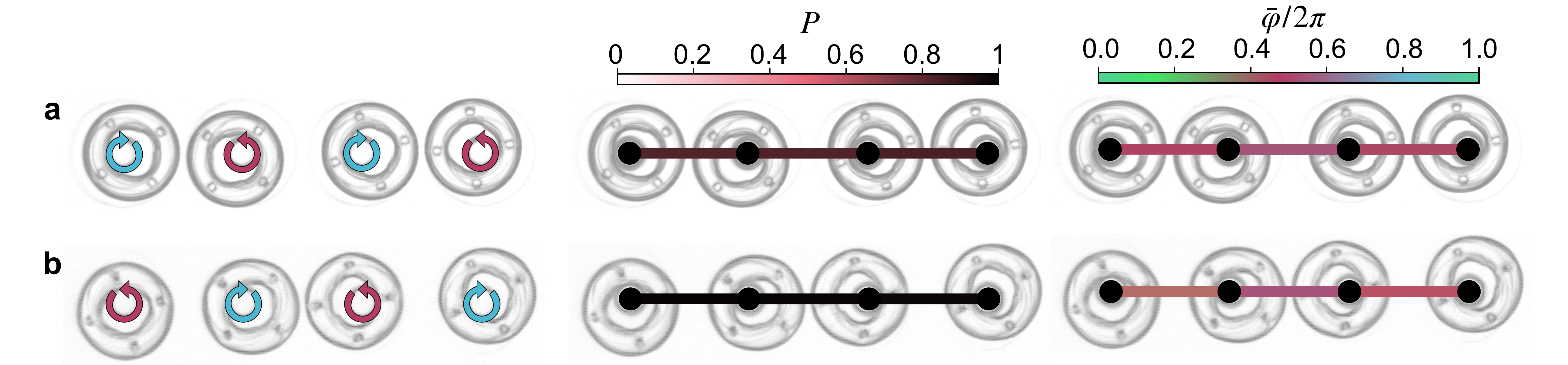}
\caption{{\color{bleuf}{\bf Phase coherence without physical contacts.} 
{\bf a.} Close up on four large rotors in an AF domain. The lattice spacing $\ell$ is larger that the radius of the precession orbit $R$ so that the rotors never contact. However the corresponding three links support a high level of phase coherence (middle panel), and the corresponding phase sum $\bar \varphi$ remains equal to $\pi$ as found deep in the ordered phase (right panel). Large rotors, E = 0.75 V/$\mu$m. 
{\bf b.} Same observations in an experiment with small motors, $E$ = 1.08 V/$\mu$m. }}
\label{figure_coherence_sans_contact}
\end{figure*}

{\color{bleuf}
\subsection{Synchronization without contact interactions}
We report in Fig.~\ref{figure_coherence_sans_contact} an additional  experimental observation that discards contact interactions as the primary source of phase coherence. 
When the lattice spacing is too large to yield pristine AF order, ordered domains remains (see Figure 2 in the main text). In those domains, we find that the motors precession remains phase coherent even though the rotor never contact. Contact interactions are therefore not the dominant interactions leading to phase coherence in our experiments.}

\begin{figure*}[h!]
\includegraphics[width=0.8\textwidth]{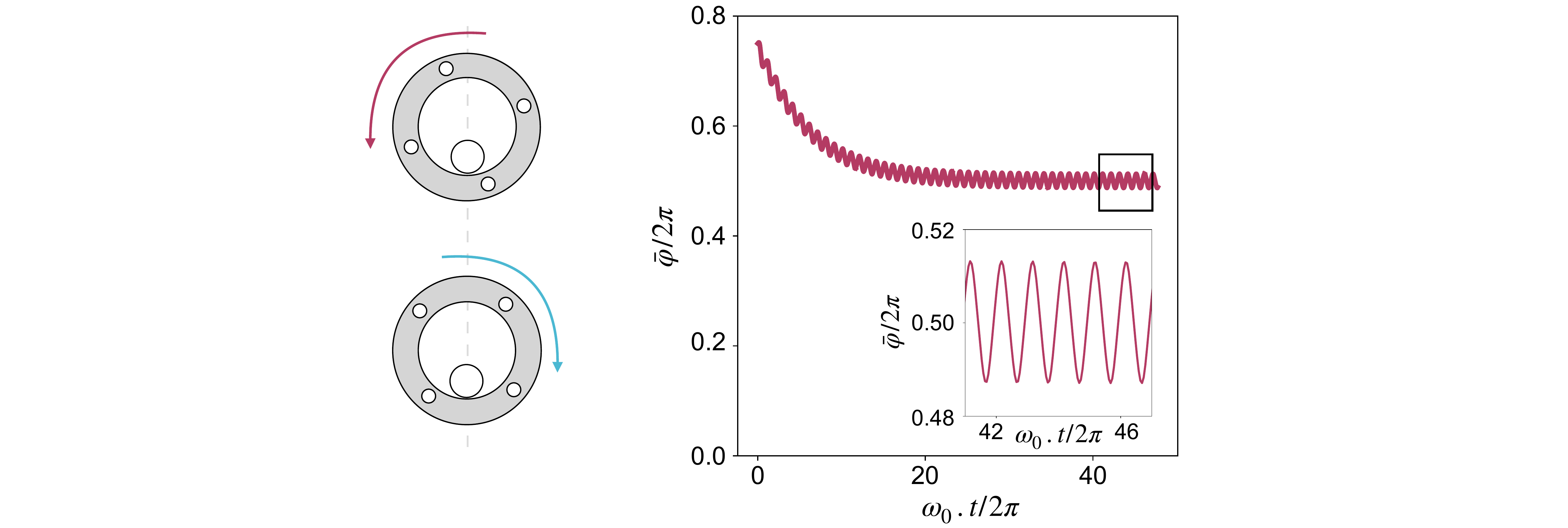}
\caption{ {\color{bleuf} {\bf Oscillations of the phase sum.} Numerical simulations of two rotors with constant natural frequencies $\pm \omega_0$ and coupled by dipolar forces. In steady state the phase sum oscillates around a constant value ($\pi$). $B=500$, $\ell=10$.}}
\label{figure_modulation paire}
\end{figure*}

{\color{bleuf}
\subsection{Oscillations of the phase sum in steady state}
Figure 3b in the main text shows that $\bar\varphi(t)$ oscillates around the mean value $\bar\varphi(t)=\pi$. 
These oscillations at the same frequency as the average rotation speed $\omega_0$ do not require the rotation speed to be modulated. 
To see this, we simulate the dynamics of a pair of rotors  with constant natural frequencies $\pm\omega_0$, and interacting through dipolar interactions. Figure~\ref{figure_modulation paire} shows that in steady state the phase sum oscillate at $\omega_0$ around the mean value $\bar \varphi=\pi$.}

\newpage
\section{Synchronization transition and phase waves: the role of disorder}
In this section, we explain the origin of the phase waves observed in the ordered state of the metamachines.

\subsection{Phase coherence in homogeneous motor lattices}
\begin{figure*}[h]
\includegraphics[width=\textwidth]{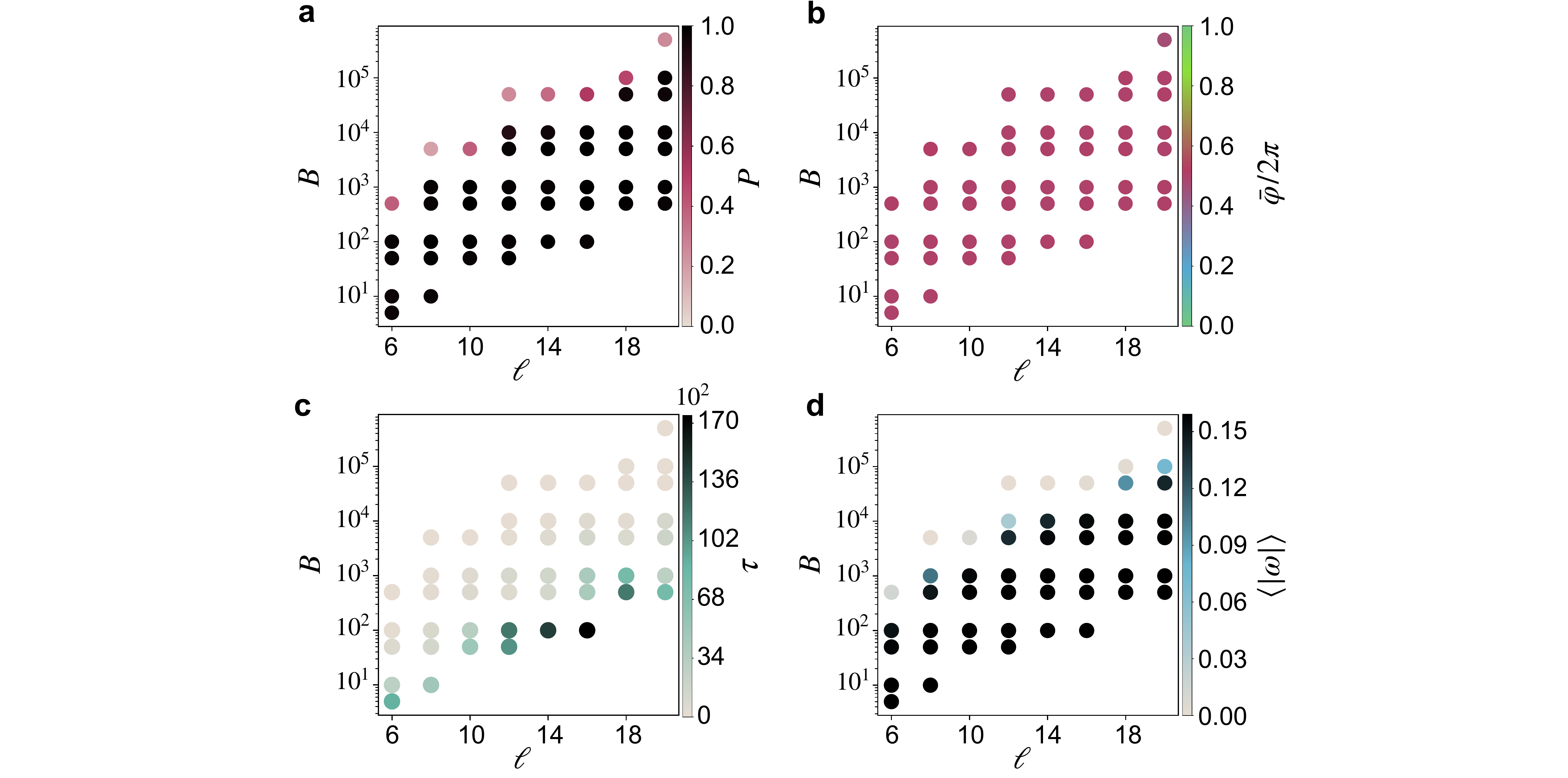}
\caption{{\bf Determination of the simulation parameters}
    {{\bf a.} Value of the order parameter $\lvert P \rvert$ for different values of $B$ and $\ell$.
    {\bf b.} Phase difference $\bar \varphi$ for different values of $B$ and $\ell$.
    {\bf c.} Convergence time $\tau$ for different values of $B$ and $\ell$. $\tau$ is defined as the time required for $P$ to reach within $\pm 5\%$ of its final value.
    {\bf d.} Average absolute motor speed $\langle \lvert\omega \rvert \rangle$ for different values of $B$ and $\ell$. All data correspond to  steady state measurements averaged over five independent realizations.} 
}
    \label{parametre_simu}
\end{figure*}

To guide our numerical investigation, we first inspect the role of the lattice spacing $\ell$ and dipole strength $B$ on the self-organization of $10\times10$ lattices (Figure~\ref{parametre_simu}). 
Informed by our previous analysis we focus on a minimal model where pointwise oscillators are only coupled via dipolar interactions.
We impose a perfect antiferromagnetic order for the rotation directions. 
We neglect the modulation of the rotation speed, and assume that all rotor dipoles point in the same direction, opposite to the electric field $\mathbf{E}$. This simplifying assumption is justified because both the modulation and dipole orientation have a negligible impact on dipolar interactions. 
Solving Eqs.~\ref{bilan_force_1eq} within a nearest neighbor approximation, we find that over a broad range of parameters  the system reaches an ordered steady state with a high degree of phase coherence where $P\simeq 1$ and  $\bar\varphi=\pi$ when averaged over all the edges. 
The only noticeable difference between our simulations is the length $\tau$ of the transient, which increases with $\ell$ and decreases with $B$, as expected. 
We also note that when the dipolar forces are much stronger than the individual drives, the rotors can freeze their dynamics (Figure~\ref{parametre_simu}d).
\label{section_dipole_onde}

\begin{figure*}[h!]
\includegraphics[width=\textwidth]{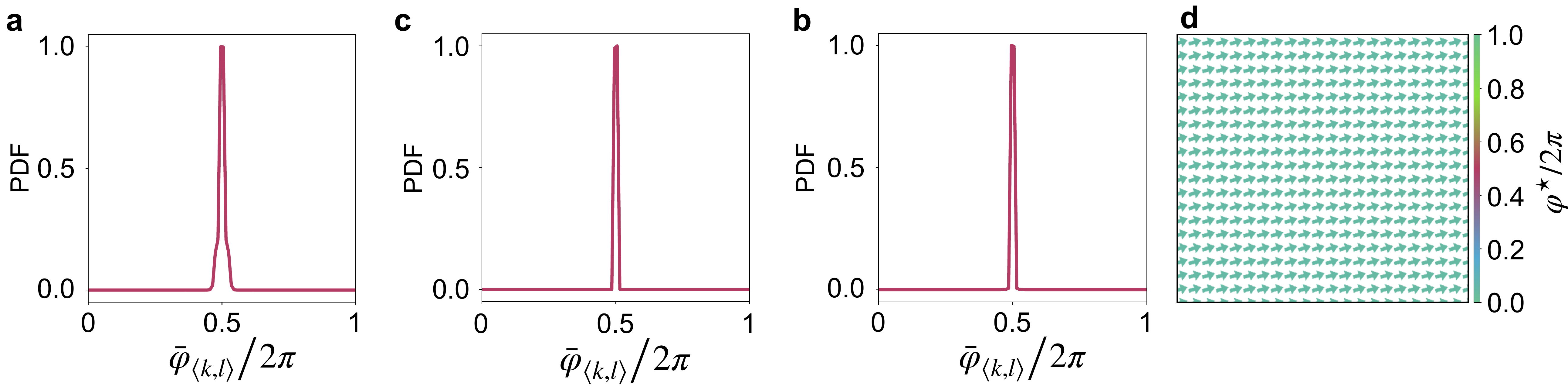}
\caption{{\color{bleuf}{\bf Boundary conditions do not alter phase coherence}
We plot the histogram of the phase sums $\bar \varphi$ measured on each link of the lattice in simulations corresponding to different boundary conditions. Phase coherence is not altered by the type of boundary conditions and $\langle\bar \varphi\rangle=\pi$ for
{\bf a.} Fixed boundary conditions, {\bf b.} free boundary conditions and {\bf c.} Periodic boundary conditions. 
{\bf d.} Instantaneous map of the  $\varphi^\star$ field in a simulation with periodic boundary conditions.
Simulation parameters: $B=500$, $\ell=10$.}}
\label{reseau_dipolaire}
\end{figure*}

{\color{bleuf}We performed simulations with open, fixed and periodic boundary conditions  (Fig.~\ref{reseau_dipolaire}). We find that  the  value of the phase sum $\bar{\varphi}$ always peak on the value  $\bar{\varphi}=\pi$ regardless of the type of boundary conditions in the phase-coherent state. 
To illustrate the nearly pristine organization of the phase sum, we show the results of simulations with periodic boundary conditions in Fig.\ref{reseau_dipolaire}d using the modified phase variable $\varphi^\star$ ($B = 500$ and $\ell = 10$).
 $\varphi^\star$  is uniform over the whole lattice at all times: no phase wave propagates in homogeneous motor arrays.
(Fig.~\ref{reseau_dipolaire}d).}  

\subsection{Disorder in rotation speeds as a source of waves}\label{section_desordre_vitesse_onde}

To assess the impact of disorder, 
we introduce noise in the driving torques of the motors.
The equations of motion take the same form:
\begin{equation}
0 = -\gamma a \dot\varphi_i + \tau_{i} + \sum_{j\neq i}\tau_{ji},
\end{equation}
but the $\tau_i$ are now quenched random variables. 
The sign of $\tau_i$ alternates on each site to account for antiferromagnetic ordering, but their amplitudes follow a uniform distribution centered at 1 with width $\sigma$.
We retain the same parameters as in the previous section: $B=500$ and $\ell=10$. 
However, we now perform simulations with open boundary conditions

\begin{figure*}[h!]
\includegraphics[width=470pt]{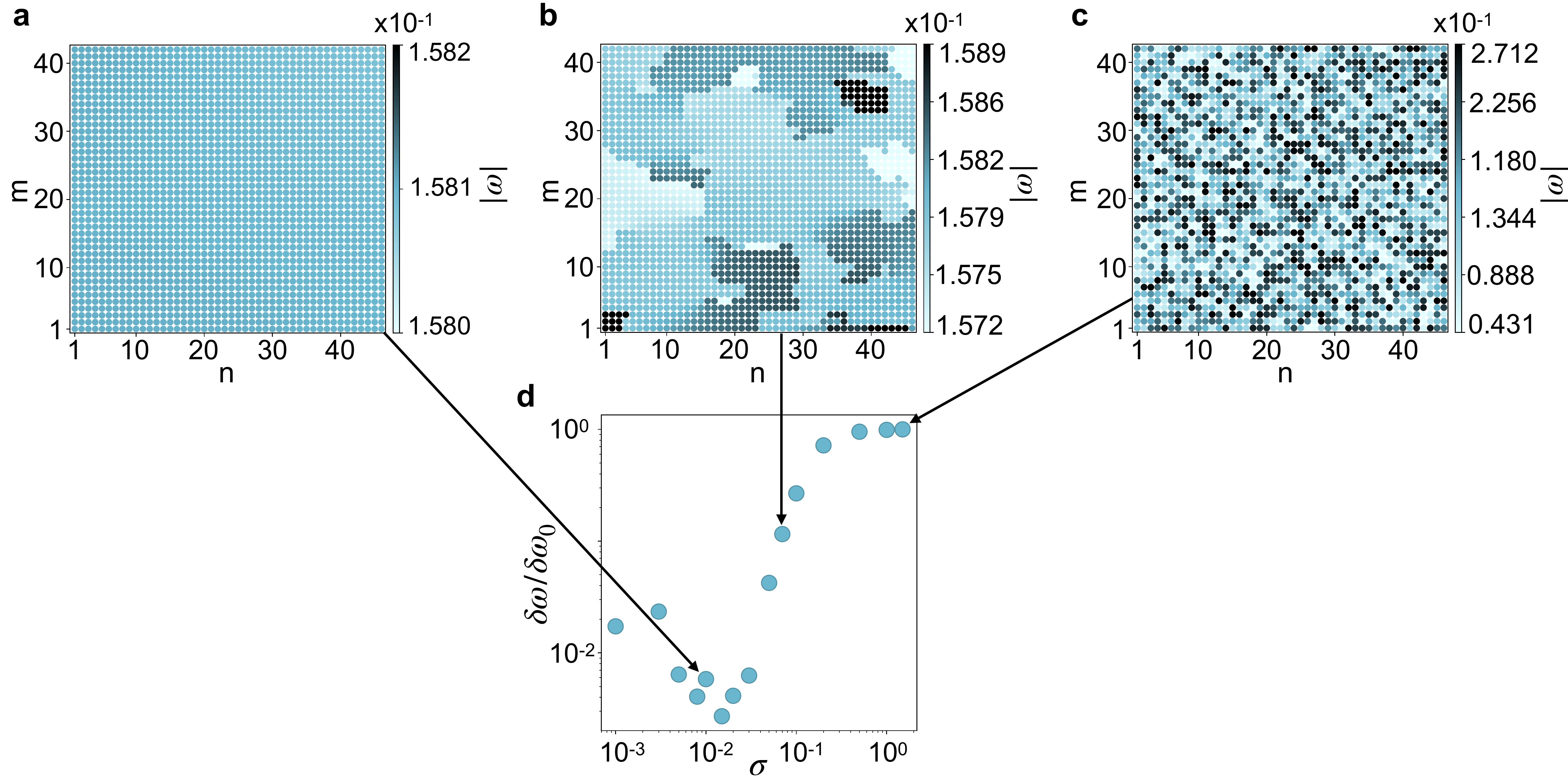}
\caption{{\bf Phase-locking transition.}
Maps of the absolute rotation speeds $\lvert \omega \rvert$ for {\bf a.} $\sigma=0.01$, {\bf b.} $\sigma=0.07$, and {\bf c.} $\sigma=1.5$.
{\bf d.} Standard deviation $\delta \omega$ of the absolute motor rotation speeds, normalized by the standard deviation $\delta \omega_0$ of a uniform distribution of width $\sigma$ ($\delta \omega_0 = \sigma / \sqrt{12}$), as a function of $\sigma$.}
\label{phase_locking_transition}
\end{figure*}
\subsubsection{A Kuramoto-like Phase-locking Transition}
When varying $\sigma$, we observe a phase-locking transition à la Kuramoto~\cite{kuramoto1975,kuramoto1984,Acebron2005}. 
For low values of $\sigma$ ($\sigma < 0.03$), the oscillators phase-lock. 
Despite differences in the driving torques, dipolar interactions organize their angular speeds, leading to a homogeneous state where all rotors orbit at a common rotation rate (Fig.~\ref{phase_locking_transition}a).
Increasing the noise strength above 
$\sigma \simeq 0.1$ all forms of phase-locking are lost and  $\delta \omega$ matches the standard deviation of the imposed torques (Fig.~\ref{phase_locking_transition}d). 
Interactions are not  strong enough to phase lock the rotors, leading to uncorrelated disorder in the rotation speeds (Fig.~\ref{phase_locking_transition}c).
At intermediate values of $\sigma$, ($0.03 < \sigma < 0.1$), the motors are not globally phase-locked but their angular speeds correlates over finite-size domains reminiscent of those seen in our experiments (Fig.~\ref{phase_locking_transition}b).
This phenomenology is the antiferromagnetic analogue of the (ferromagnetic) Kuramoto synchronization transition, see Ref.~\cite{Acebron2005} for a comprehensive review.

\subsubsection{Relationship between angular-speed disorder and phase coherence}
\begin{figure*}[h!]
\includegraphics[width=0.78\textwidth]{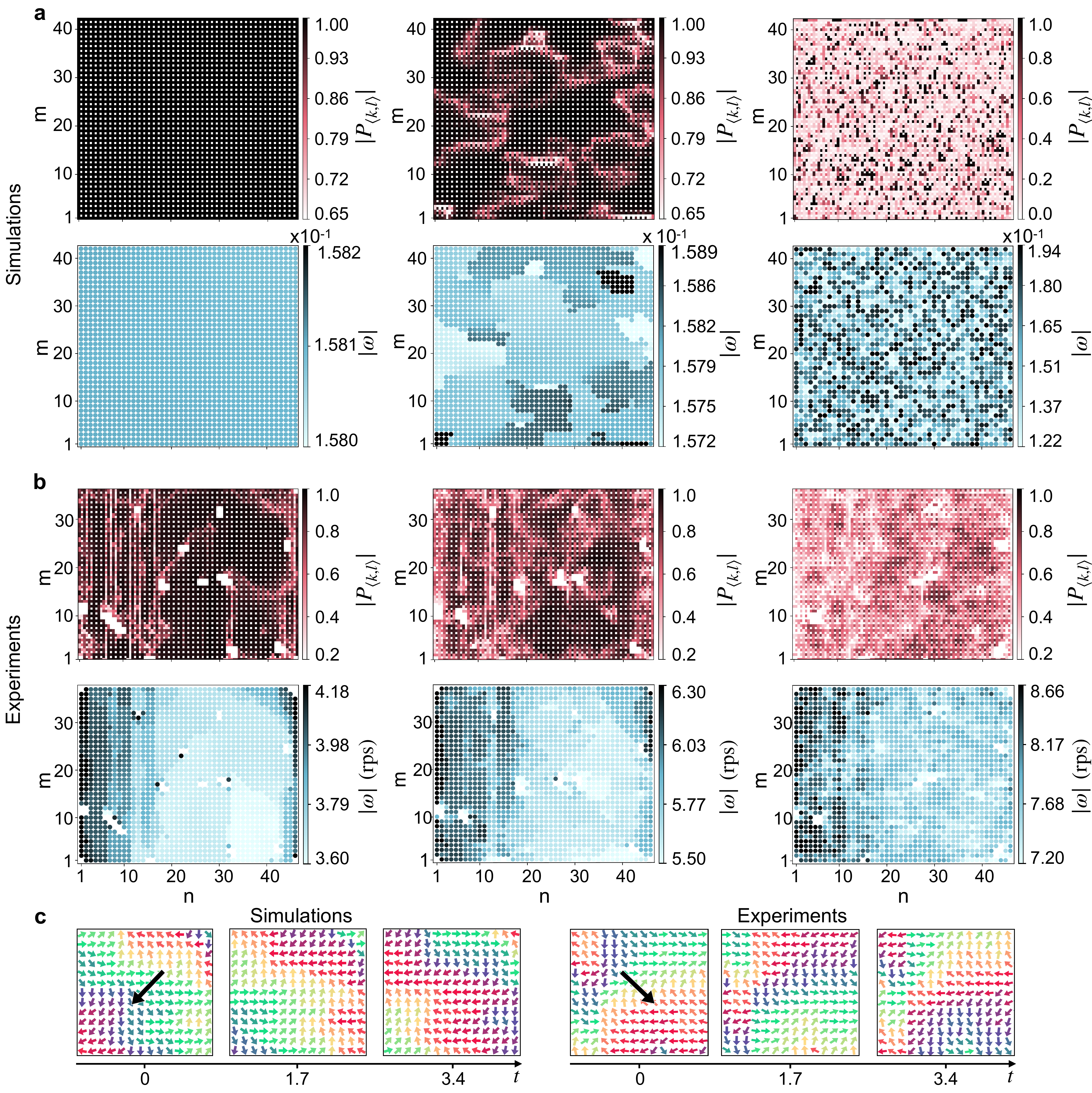}
\caption{{\bf Speed disorder and phase coherence.}
{\bf a–b.} Maps of the local order parameter $\lvert P_{\langle k,l\rangle} \rvert$ (top) and of the absolute rotation speeds $\lvert \omega \rvert$ (bottom).
{\bf a.} Simulations for different noise strengths: $\sigma = 0.01$ (left), $\sigma = 0.07$ (middle), and $\sigma = 0.5$ (right).
{\bf b.} Experimental results at increasing field strengths: $E = 1.2\,E_Q$ (left), $E = 1.5\,E_Q$ (middle), and $E = 1.8\,E_Q$ (right); with $E_Q = 0.52$~V/$\upmu$m.
{\bf c.} Phase maps $\varphi^\star$ at different times for simulations (left) and experiments (right), showing a wave propagating in the direction indicated by the black arrow.
}
\label{phase_coherence_transition}
\end{figure*}

In Figs.~\ref{phase_coherence_transition}a and ~\ref{phase_coherence_transition}b, we plot the map of the angular speed next to the map of the phase coherence order parameter $P$ for both our simulations and experiments.
Consistently, we find that the phase coherence and angular speed domains coincide. 
As expected from Kuramoto physics, their size decreases when disorder increases (see Fig. 4d in the main text). This reduction in domain size leads to a gradual decline in the global order parameter $P$ (Figs. 4c and 4e in the main text).

A remarkable consequence of the spatial heterogeneity in the angular speed is the emergence of phase waves propagating freely in all the phase-coherent domains (Fig.~\ref{phase_coherence_transition}c).

\begin{figure*}[h!]
\includegraphics[width=0.8\textwidth]{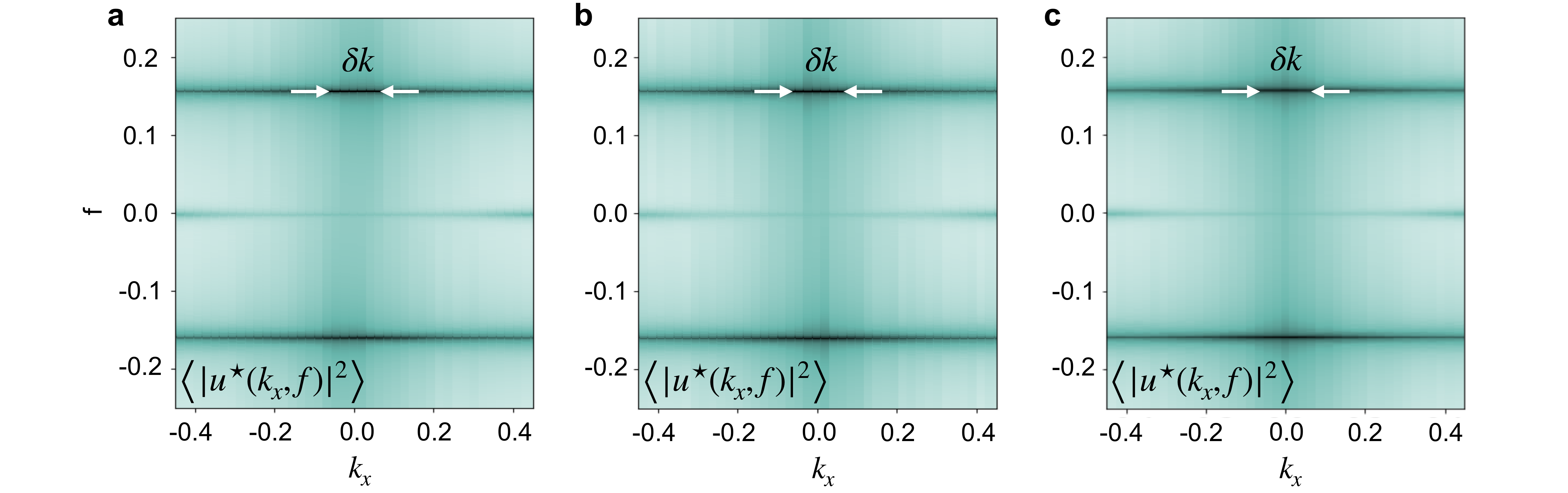}
{\color{bleuf}
\caption{{\bf Impact of  boundary conditions on the wave spectra.}
Same wave spectra as in Figure 4 (main text) but for three types of boundary conditions
{\bf a.} Open boundary conditions, as in the main text.
{\bf b.} Fixed boundary conditions. The rotors are fixed at the edge of the lattice.
{\bf c.} Periodic boundary conditions.}
}
\label{boundaryconditionsfig}
\end{figure*}

\newpage
{\color{bleuf}\subsection{The boundary conditions do not impact the wave spectra}
In order to assess whether the phase-wave patterns depends on the boundary conditions at the edges of the motor lattices, we performed three series of numerical simulations changing the type of boundary conditions and keeping all the other parameters identical. 
Figure~\ref{boundaryconditionsfig} shows that the wave spectra are identical for open, rigid and periodic boundary conditions.}

\subsection{Wave Generation Mechanism}
\label{mur_domaine}

\begin{figure*}[h!]
\includegraphics[width=0.8\textwidth]{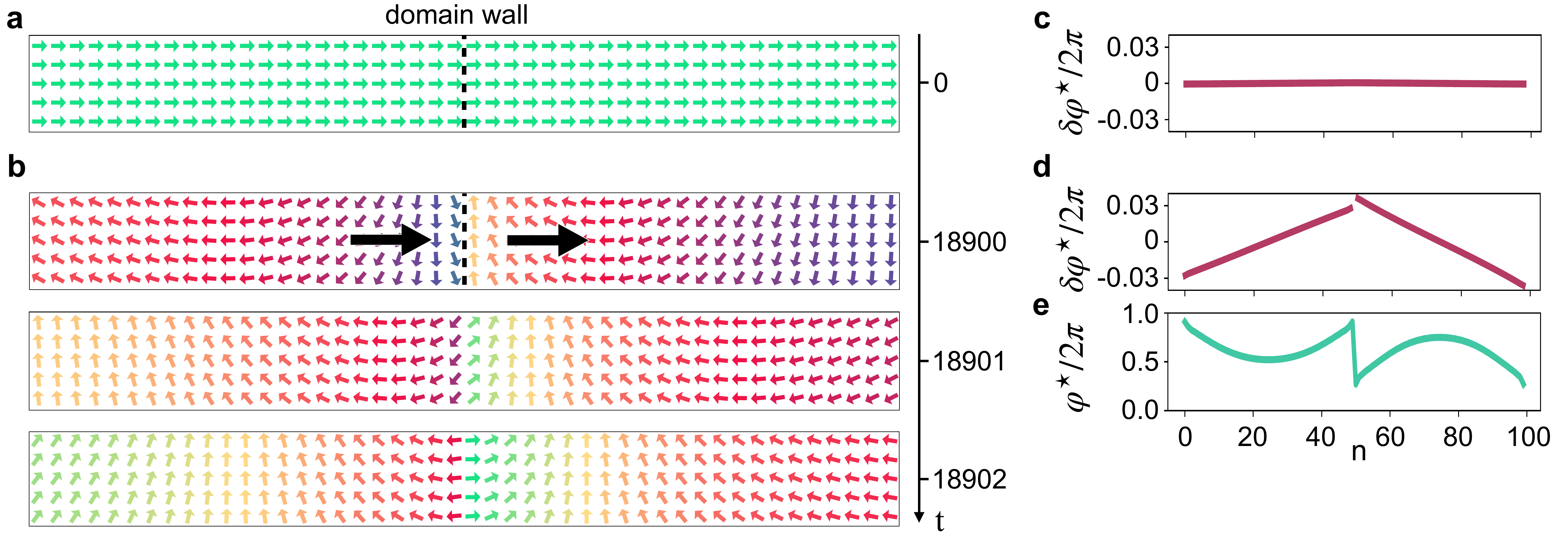}
\caption{{\bf Wave generation at domain walls.}
{\bf a, b.} Phase maps $\varphi^\star$ as a function of time. 
{\bf c, d.} Spatial profile of the horizontal phase difference $\bar{\varphi}^{\star}_{h}$.
{\bf e.} Phase $\varphi^\star$ plotted as a function of $n$.
{\bf a, c.} The system is initialized with a perfect phase coherence with a constant $\bar{\varphi}^{\star}_{h}$ equal to 0.
{\bf b.} After some time, waves appear in the system. They are associated with {\bf d.} a linear profile of $\bar{\varphi}^{\star}_{h}$ which induces {\bf e.} a quadratic profile for the phase $\varphi^\star$.
$\sigma = 0.1$, $B=500$ et $\ell=10$.
}
\label{domaine_vitesse_onde}
\end{figure*}

To elucidate the mechanism behind wave generation, we perform simplified one-dimensional simulations with dipolar interactions and periodic boundary conditions (Lattice size: 100×6).
We initialize the system in a perfectly coherent state and induce two distinct velocity domains by imposing two different driving torques in the halves of the lattice 
(Fig.~\ref{domaine_vitesse_onde}a and c).
Given the amplitude of the torque mismatch ($\sigma=0.1$), the Kuramoto transition cannot operate, leaving two domains where the angular speeds are distinct, thereby preventing the self-organization of a homogeneous phase coherent state (Note that we use the Greek letter $\sigma$  for the torque mismatch, but we stress that there is no quenched disorder).

 To quantify the deviations from the homogeneous situation where $\varphi^\star$ is uniform, we compute the phase difference $\delta{\varphi}^{\star}_{h}$ on all horizontal edges of the lattice.
 We find that the phase difference adopts a linear spatial profile in the two domains (Fig.~\ref{domaine_vitesse_onde}d).
This results in  quadratic spatial variations of $\varphi^{\star}$ across the domains, with sharp phase variations near the boundaries (Fig.~\ref{domaine_vitesse_onde}e). 
Since $\varphi^{\star}$ evolves periodically in time, the spatial gradients manifest as waves propagating freely across the velocity domains (Fig.~\ref{domaine_vitesse_onde}b).

\subsection{Long wave-length dynamics of $\delta \varphi^\star$: emergence of phase waves from domain walls}

To gain more insight into the propagation of phase waves from the boundaries of the ordered domains, 
we provide a long-wave-length approximation of the  minimal model developed in Section~\ref{section_model_theo}. 
Within this approximation, we will show that the modified  phase difference $\delta \varphi^\star$ is a slow variable whose dynamics is diffusive.
We will then use this central results to account for our experimental and numerical observations.

We consider an infinite chain of rotors aligned along the electric field, neglect the modulation of the rotation speed and focus on  dipolar interactions. 
For the sake of simplicity, we consider the limit $\ell\gg a$ where the dipolar forces  are predominantly longitudinal, and linearize the equations of motion Eqs.~\eqref{eq_dipolaire_2corps_1} and~\eqref{eq_dipolaire_2corps_2}  as in Eqs.~\eqref{radial1} and~\eqref{radial2}.
Given these simplifying hypotheses, the equations of motion take the compact form
\begin{align}
\dot \varphi_{n} = (-1)^n + \tau_D^1 \sin{\varphi_n} \left(\cos{\varphi_{n+1}} + \cos{\varphi_{n-1}} - 2\cos{\varphi_n} \right),
\end{align}
where the coupling torque $\tau_D^1$ is a positive constant that is set by the dipole strength and lattice spacing.
Assuming even-numbered rotors rotate counterclockwise and odd-numbered ones clockwise.
The dynamics of the modified phase field can then be expressed as:

\begin{align}
\partial_t \varphi^\star_n = 1 - \tau_D^1 \sin{\varphi^\star_n} \left( \cos{\varphi^\star_{n+1}} + \cos{\varphi^\star_{n-1}} + 2\cos{\varphi^\star_n} \right)
\end{align}

and that of the phase difference $\delta \varphi^\star_{n} = \varphi^\star_{n+1} - \varphi^\star_n$:

   \begin{align}    
        \partial_t\delta{  \varphi}^\star_{n}
        &=-\frac{\tau_D^1}{2}\left[-\sin{( \delta \varphi^\star_{n-1})}-2\sin{(2\varphi_n^\star)}-\sin{(\varphi_{n-1}^\star+\varphi_n^\star)}\right .\\
        &\left.\qquad \qquad +2\sin{(\delta \varphi^\star_{n})}+2\sin{(2\varphi_{n+1}^\star)}-\sin{(\delta \varphi^\star_{n+1})}+\sin{(\varphi_{n+1}^\star+\varphi_{n+2}^\star)}\right]\notag
    \end{align}

Assuming a high level of phase coherence, where $\delta \varphi^\star$ remains small, we can linearize the equation:

\begin{align}
\partial_t \delta \varphi^\star_{n} &= -\frac{\tau_D^1}{2} \left(-\delta \varphi^\star_{n-1} + 2 \delta \varphi^\star_{n} - \delta \varphi^\star_{n+1} \right) \\
&\quad -\frac{\tau_D^1}{2} \left[-2\sin(2\varphi_n^\star) - \sin(\varphi_{n-1}^\star + \varphi_n^\star) + 2\sin(2\varphi_{n+1}^\star) + \sin(\varphi_{n+1}^\star + \varphi_{n+2}^\star) \right]
\end{align}

Taking the continuum limit and assuming that $\delta \varphi^\star$ and $\varphi^\star$ vary slowly in space, we find:

\begin{align}
\partial_t \delta \varphi^\star(x,t) = D \Delta \delta \varphi^\star(x,t) - 4\tau_D^1 \delta \varphi^\star(x,t)\cos\left[2\varphi^\star(x,t)\right]
\end{align}

with $D = \frac{\tau_D^1}{2} \ell^2$.
We further simplify the equation by approximating $\varphi^\star(x,t)$ by its leading-order term
$\varphi^\star(x,t) \simeq \omega t$.
A spatial Fourier transform then yields:

\begin{align}
\partial_t \delta \varphi^\star_{q} = -D q^2 \delta \varphi^\star_{q} - 4\tau_D^1 \delta \varphi^\star_{q} \cos(2\omega t)
\end{align}

which we can readily integrate to find how each Fourier mode evolves in time:

\begin{align}
\delta \varphi^\star_{q} = B \exp\left(-D q^2 t - \frac{4\tau_D^1}{2\omega} \sin(2\omega t) \right)
\end{align}

In the limit of small $q$s, the term $\exp(-D q^2 t)$ varies slowly on the timescale of one period. We can therefore average the dynamics   over a  rotation period $T$. We then find 
\begin{align}
\left<\delta \varphi^\star_{q}\right>_T = B’ \exp(-D q^2 t)
\end{align}
where $B’ = \left< \exp\left( -\frac{4\tau_D^1}{2\omega} \sin(2\omega t) \right) \right>_T$.
In conclusion, in the long-wavelength limit, the time-averaged phase difference $\left< \delta \varphi^\star_h \right>_T$ obeys a diffusion equation:

\begin{align}
\partial_t \left<\delta \varphi^\star_h\right>_T = D \Delta \left<\delta \varphi^\star_h\right>_T
\end{align}
This diffusive dynamics explains the linear evolution of $\delta \varphi^\star$ in a domain as discussed in Section~\ref{mur_domaine}. Indeed, the Green’s function of the Laplacian in one dimension is a linear function. 
In steady state, linear variations of the modified phase difference are therefore the only possible solutions of the  dynamics.
This asymptotic behavior hence translates into quadratic variations of the modified phase away from one-dimensional domain walls in excellent agreement with our numerical simulations.

\bibliographystyle{abbrv}
\bibliography{bib_suplementary}